\documentclass[twocolumn, trackchanges,resetfootnote]{aastex701}

\usepackage{graphicx}
\usepackage{amsmath}
\usepackage{threeparttable}
\usepackage{wrapfig}
\usepackage{caption}
\usepackage{float}
\usepackage{hyperref}
\usepackage{enumitem}
\usepackage[table]{xcolor}  
\usepackage{booktabs}

\newcommand{\hl}[1]{{\setlength{\fboxsep}{1.5pt}\colorbox{yellow!30}{#1}}}

\newcommand{\lttab}{LTT~1445~A~b}
\newcommand{\lttac}{LTT~1445~A~c}
\newcommand{\rade}{$R_{\oplus}$}

\newcommand{\eureka}{\texttt{Eureka!}}
\newcommand{\helios}{\texttt{HELIOS}}

\newcommand{\stsci}{\affiliation{Space Telescope Science Institute, Baltimore, MD 21218, USA}}
\newcommand{\carnegie}{\affiliation{Earth and Planets Laboratory, Carnegie Institution for Science, Washington, DC 20015, USA}}
\newcommand{\umd}{\affiliation{Department of Astronomy, University of Maryland, College Park, MD 20742, USA}}
\newcommand{\PSUAA}{Department of Astronomy \& Astrophysics, The Pennsylvania State University, University Park, PA 16802, USA}
\newcommand{\PSUCEHW}{Center for Exoplanets and Habitable Worlds, The Pennsylvania State University, University Park, PA 16802, USA}

\begin{document}

\title{JWST Rocky Worlds DDT Observations of LTT~1445~A~b and c in Secondary Eclipse Using MIRI/F1500W}

\author[orcid=0000-0003-2775-653X,sname='Ih']{Jegug Ih}
\stsci
\carnegie
\email[show]{jih@stsci.edu}  

\author[orcid=0000-0001-8401-4300]{Shubham Kanodia}
\carnegie
\email{skanodia@carnegiescience.edu}

\author[orcid=0000-0002-0413-3308]{Nicholas Wogan}
\affiliation{NASA Ames Research Center, Moffett Field, CA 94035, USA}
\email{}

\author[orcid=0000-0002-8938-9715]{Mark Fortune} 
\affiliation{School of Mathematical \& Physical Sciences, Macquarie University, Macquarie Park, NSW 2113, Australia}
\email{}

\author[orcid=0000-0002-0832-710X]{Natalie H. Allen} 
\affiliation{Department of Physics and Trottier Institute for Research on Exoplanets, Universit\'{e} de Montr\'{e}al, Montr\'{e}al, QC, Canada}
\email{}

\author[orcid=0000-0001-8274-6639]{Hannah Diamond-Lowe}
\stsci
\email{}

\author[orcid=0000-0002-0371-1647]{Mike Greklek-McKeon} 
\carnegie
\email{mgreklek-mckeon@carnegiescience.edu}

\author[orcid=0000-0002-0931-735X]{Måns Holmberg} 
\stsci
\email{mholmberg@stsci.edu}

\author[orcid=0000-0002-2072-6541]{Jonathan Brande} 
\umd
\email{jbrande@umd.edu}

\author[orcid=0000-0001-8943-9148]{Reza Ashtari} 
\affiliation{Johns Hopkins University Applied Physics Laboratory, Laurel, MD 20723, USA}
\email{}

\author[orcid=0000-0002-9030-0132]{Katherine A. Bennett} 
\affiliation{Department of Earth \& Planetary Sciences, Johns Hopkins University, Baltimore, MD 21218, USA}
\email{}

\author[orcid=0000-0002-0746-1980]{Jacob Lustig-Yaeger} 
\affiliation{Johns Hopkins University Applied Physics Laboratory, Laurel, MD 20723, USA}
\email{jacob.lustig-yaeger@jhuapl.edu}

\author[orcid=0000-0003-4241-7413]{Megan Weiner Mansfield} 
\umd
\email{}

\author[orcid=0000-0002-9082-6337]{Andrea S.J.\ Lin}
\affiliation{Department of Astronomy, California Institute of Technology, Pasadena, CA 91125, USA}
\email{asjlin@caltech.edu}

\author[0000-0003-4835-0619]{Caleb I. Ca\~nas}
\affiliation{Southeastern Universities Research Association, Washington, DC 20005, USA}
\affiliation{Exoplanets and Stellar Astrophysics Laboratory, NASA Goddard Space Flight Center, Greenbelt, MD 20771, USA}
\email[]{c.canas@nasa.gov}

\author[orcid=0000-0002-7127-7643]{Te Han}
\affiliation{Kavli Institute for Astrophysics and Space Research, Massachusetts Institute of Technology, Cambridge, MA, USA}
\email{tehan@mit.edu}

\author[orcid=0000-0002-2990-7613]{Jessica Libby-Roberts}
\affiliation{\PSUAA}
\affiliation{\PSUCEHW}
\affiliation{Department of Physics and Astronomy, University of Tampa, Tampa, FL  33606}
\email{jlibbyroberts@ut.edu}

\author[orcid=0000-0001-9596-7983]{Suvrath Mahadevan}
\affiliation{\PSUAA}
\affiliation{\PSUCEHW}
\email{suvrath@astro.psu.edu}
\email[]{suvrath@astro.psu.edu}



\begin{abstract}


The JWST Rocky Worlds Director's Discretionary Time (RWDDT) program uses secondary eclipse observations to determine the conditions under which rocky exoplanets around M~dwarfs retain their atmospheres.  We present RWDDT MIRI/F1500W observations of LTT~1445~A~b and c, the transiting rocky siblings in the nearest such system (6.9\,pc), comprising two eclipses of each planet across three visits.  We perform three independent reductions and a joint fit that includes all existing JWST data and 220 radial velocities, improving the precision on the planetary radii, masses, and orbital parameters by factors of 1.6--2.3.  We measure eclipse depths of $157^{+18}_{-17}$\,ppm and $154^{+54}_{-37}$\,ppm for planets b and c, corresponding to 15\,$\mu$m brightness temperatures of $528^{+39}_{-29}$\,K and $625^{+109}_{-79}$\,K, both consistent with the dark, bare rock limit.  For planet b, combining these data with the archival MIRI/LRS spectrum in a Bayesian retrieval limits the CO$_2$ partial pressure to $<0.6$\,mbar and the surface pressure to $\lesssim0.1$\,bar, while forward models exclude thick aerosol layers that could mimic a bare surface in transmission.  We conclude that planet b has no substantial atmosphere.  For planet c, a potentially grazing eclipse and overlapping events leave the depth poorly constrained; the data rule out only thick, CO$_2$-bearing atmospheres ($\gtrsim1$\,bar with $\geq1\%$ CO$_2$).  LTT~1445~A~b is the planet closest to the canonical XUV-driven Cosmic Shoreline observed in eclipse so far. Its lack of a substantial atmosphere implies that the Cosmic Shoreline cannot lie much farther toward the airless side than nominally predicted.

\end{abstract}

\setcounter{footnote}{0}



\section{Introduction} 



JWST has enabled the detailed characterization of a growing sample of small ($\leq 1.8$ \rade), terrestrial planets beyond our Solar System \citep{kreidberg25_first, parkcoy25_population, monaghan26_uniform}. 
Studying rocky planets around M dwarfs in particular provides an advantageous opportunity due to the favorable signal size and the higher occurrence rate \citep{charbonneau07_dynamics, dressing_occurrence_2013, dressing_occurrence_2015}.  However, the harsh irradiation environment around M dwarfs is also expected to drive rapid atmospheric escape, likely stripping most observable rocky worlds of their atmospheres \citep{tian09, shields16, france16, chatterjee24, pass25}. 
The central open question at this time is identifying the conditions under which rocky planets around M dwarfs retain their atmospheres.
This is often framed around the so-called ``Cosmic Shoreline'', or a line separating planets with and without thick atmospheres along some relevant axes \citep{zahnle17, ji25_cosmic, berta-thompson_3d_2025}.   Escape speed and XUV irradiation history are most commonly adopted; this choice is informed by the Solar System \citep{zahnle17} but emphasizes the importance of greater cumulative XUV irradiation in M~dwarf systems \citep{pass25, ih26_do}. 

For identifying atmospheres, secondary eclipse observations of 1:1 tidally locked rocky planets offer a relatively efficient test.  Broadly speaking, the presence of an atmosphere can lower the brightness temperature of a planet's thermal emission by transporting heat away from the permanent dayside to the nightside, by molecular spectral absorption, or by forming reflective clouds \citep{mansfield19, koll19_identifying}.  A shallow eclipse depth relative to the ``dark, bare rock'' limit could hint at the presence of an atmosphere, though identifying both false positives and negatives, such as shiny surfaces or absorptive aerosol layers, is an active area of research \citep[e.g.][]{hammond2024, maurel25_constraints, ih26_do, lyu26_subsurface}. 

This observing technique has been applied to numerous JWST targets with both photometry and spectroscopy, including LHS~3844~b \citep{kreidberg19, zieba26_lhs3844}, TRAPPIST-1~b \citep{greene23, ducrot25_combined}, TRAPPIST-1~c \citep{zieba23}, GJ~486~b \citep{mansfield24}, GJ~1132~b \citep{xue24}, LTT~1445~A~b \citep{wachiraphan25_the}, TOI-1468~b \citep{meier2025_toi1468b}, LHS~1478~b \citep{august25_hot}, LHS~1140~c \citep{fortune25_lhs1140c},  LTT~3780~b \citep{allen25_hot}, GJ~3473~b \citep{holmberg26_hot}, and more using other instruments or facilities \citep[see][for a recent review]{monaghan26_uniform}.  Most targets observed so far, with the exception of LHS~1478~b, appear consistent with the dark, bare rock limit to varying limits on low-surface pressure atmospheres (not including ``lava worlds'' $T_\mathrm{irr}>2000$\,K).  This may be because they are all on the ``dry'' side of the nominal Shoreline often normalized to pass near Mars, as hotter targets are easier to observe in thermal emission.   Yet, the inferred shallow eclipses of LHS~1478~b, which also lies on the ``dry'' side, imply that there are confounding bare surface processes at play or that our understanding of atmospheric escape or replenishment motivating the Shoreline is too simplistic and does not consider the intrinsic stochasticity therein.

While the Cosmic Shoreline extrapolates from the Solar System and is most certainly an oversimplification \citep[e.g.][]{chatterjee24, pass25, august26_atmospheric}, it provides a testable hypothesis that motivates observational surveys.  To this end, because (a) the targets most expected to host atmospheres (cool equilibrium temperature and high gravity) are naturally the hardest to characterize and require large numbers of observing hours to achieve the needed signal-to-noise, and (b) the population-level question is most efficiently answered by a dedicated large survey rather than an accretion of piecemeal efforts \citep{ih26_do}, 500 hours has been allocated as the Rocky Worlds Director's Discretionary Time (RWDDT) to observe rocky planets around M stars in secondary eclipse using MIRI F1500W photometry, which targets a CO$_2$ absorption band \citep{redfield24_report, deming09}.

\begin{figure}
    \centering
    \includegraphics[width=0.99\linewidth]{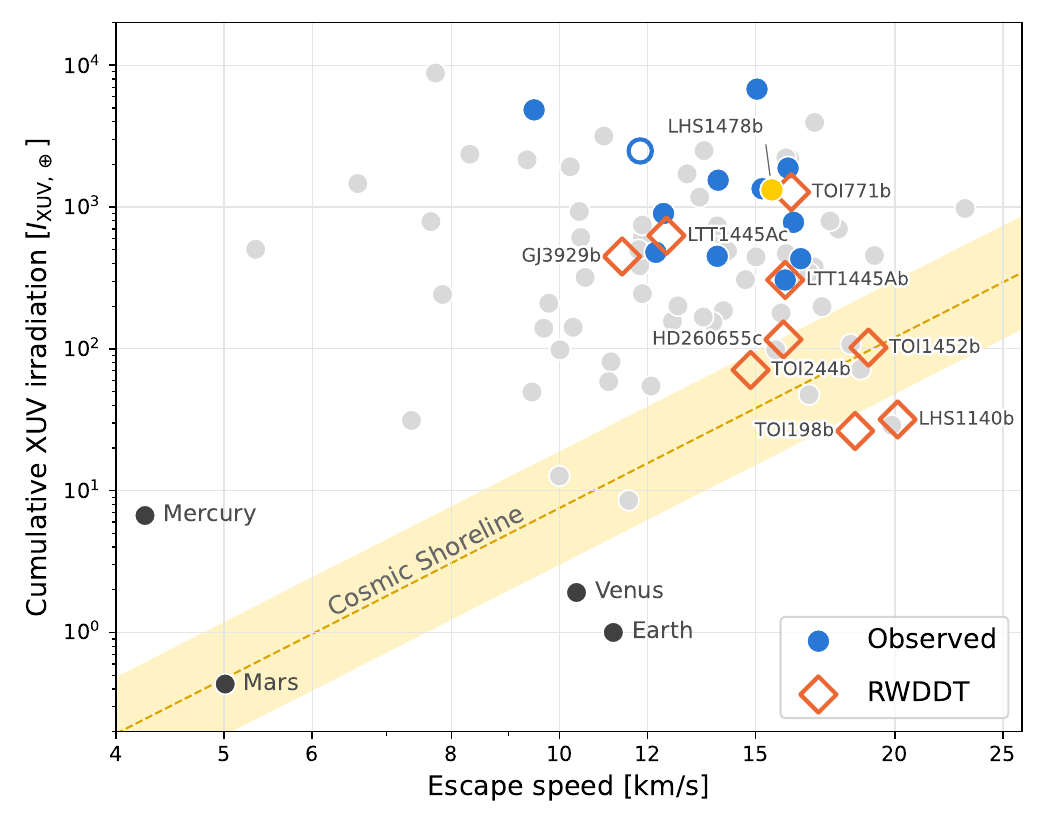}
    \caption{Escape speed and cumulative XUV irradiation of all rocky planets observed in secondary eclipse so far (blue circles) and selected RWDDT targets (orange diamonds).  Cumulative XUV irradiation is estimated using the methods in \citet{pass25}.  Observed ``lava worlds'' with $T_\mathrm{irr}>2000$\,K are not shown.  Known transiting rocky planets (``Targets Under Consideration'' from RWDDT) are shown in grey.  All targets have JWST observations except for GJ~1252~b (empty blue circle).  LHS~1478~b is shown in yellow.  The width of the Shoreline reflects a notional combined uncertainty in the estimated XUV flux and system parameters of $\pm0.4$ dex.}
    \label{fig:cosho}
\end{figure}

The RWDDT will extend the observed sample to cooler and thus more difficult targets\footnote{\url{https://rockyworlds.stsci.edu/rw-website-targets.html}}.  We show all observed targets to date and the chosen RWDDT targets in Figure~\ref{fig:cosho}.  The first RWDDT target observed, GJ~3929~b, was revealed to be consistent with a dark, bare rock but is also the furthest from the nominal Shoreline \citep{xue25_the, connors26_gj}. 

In this observational landscape, the sibling planets in the LTT~1445~A system stand out as prime targets for characterization, as they reside in the nearest M dwarf system with transiting rocky planets at a distance of 6.9 parsec ($J_{\rm mag}$=7.3, $K_{\rm mag}$=6.5).  The outer planet b was discovered in \citet{winters19_three} with TESS and MEarth and is a warm, terrestrial planet of $R_\mathrm{p}=1.31^{+0.07}_{-0.06} R_\oplus$ and $M_\mathrm{p}=2.87^{+0.26}_{-0.25} M_\oplus$ with an equilibrium temperature of $T_\mathrm{eq}=424\pm21$K ($A_\mathrm{B}=0$).  The \textit{inner} planet c was discovered in \citet{winters22_a} and found to potentially have a grazing transit, setting only a lower limit on its radius of $R_\mathrm{p}\geq 1.15 R_\oplus$, with $M_\mathrm{p}=1.54^{+0.20}_{-0.19} M_\oplus$ and $T_\mathrm{eq}=508 \pm 25$K.  While both planets are on the supposed ``dry'' side of the Shoreline, \lttab{}, being the closest target to the Shoreline observed in eclipse so far, is close enough to it that the uncertain evolution history of the host star could plausibly place it on either side.  Two planets that share a common XUV history provide an opportunity to characterize the Shoreline \textit{without} marginalizing over this major unknown \citep{pass25, ih26_do}. 

The LTT~1445~A system, particularly the cooler and non-grazing planet b, has thus received much community interest. 
\lttab{} was first observed spectroscopically from the ground with Magellan II/LDSS3C in the optical and found to not have a clear H/He-dominated atmosphere \citep{diamondlowe23_ground}.  It was then observed with HST/WFC3 in 0.2-1.65 $\mu$m and was also consistent with a mostly featureless spectrum, though retrievals weakly favored the presence of an atmosphere ($2\sigma$) \citep{bennett25_hst}.  

The community interest has continued in the JWST era.  \lttab{} was observed in secondary eclipse with MIRI/LRS; these observations revealed that its LRS-broadband brightness temperature was consistent with the dark, bare rock limit to within $<1\sigma$ but with a fairly large uncertainty that also allowed for moderate atmospheres scenarios, as the analysis was hampered by strong systematics that led to visit-to-visit variation \citep{wachiraphan25_the}.  Planet b was also observed in transit with NIRSpec/G395H as part of the JWST COMPASS program (GO2512, PI Batalha) and was revealed to lack any significant spectral features in its transmission spectrum even at JWST-level precision, ruling out a clear, $\leq500\times$ Solar metallicity ($\mu \leq 15$) when combined with the HST data \citep{batalha26_jwst}.  Planets b and c were both observed in transit with NIRISS/SOSS and NIRSpec/G395H as part of the JWST Cosmic Shoreline program (GO7073, PI Lustig-Yaeger), with the purpose of performing comparative planetology within a single system
(Brande \& Wiser et al. \textit{in prep.}, Ashtari et al. \textit{in prep.}).  Finally, planet b will be observed with NIRCam in Cycle 4 (GO 7251, PI Bennett), which has not yet been executed at the time of writing. 

Observations of \lttab{} testify to the difficulty of identifying and characterizing atmospheres on rocky exoplanets in both transmission and emission.  In transmission, high mean molecular weight (whose value is unknown a priori) renders the spectrum flat and therefore hard to distinguish from a bare rock or an opaque cloud deck, while stellar contamination can masquerade as absorption or scattering features \citep{millerricci09, rackham18, moran23, bennett_additional_2025, cadieux_transmission_2024}.  In emission, even before considering false positives or negatives, a thick atmosphere would lead to a \textit{shallow} eclipse and naturally make its detection more challenging.  This is especially true if the eclipse timing is not precisely known or if strong systematics (instrumental or astrophysical) are present in the light curve \citep[e.g.][]{wachiraphan25_the, august25_hot}.  The RWDDT employs a checkpoint strategy to account for the unknown eclipse timing, wherein observations are taken in two stages to first constrain the orbit with a long baseline and then follow up with more eclipses to build the S/N necessary to distinguish whether the eclipse is deep or shallow\footnote{\url{https://rockyworlds.stsci.edu/rw-website-schedule.html\#strategy}}. 

Observing the LTT~1445~A siblings is advantageous in that both goals can be achieved with very few eclipses (1 for b and 2 for c) of short durations and they were thus chosen as targets of the RWDDT\footnote{It is perhaps worth noting that \lttac{}, alongside GJ~3929~b, was chosen early on, separately from the rest of the chosen targets. It is unclear whether it would have been selected if the full list of targets was assembled together from the start \citep{ih26_do}.}.  Two eclipses of each planet were successfully observed with JWST in September of 2026\footnote{\url{https://rockyworlds.stsci.edu/downloads/LTT1445Ac_JWST_SchedulingReport_20260716.pdf}}\footnote{\url{https://rockyworlds.stsci.edu/downloads/LTT1445Ab_JWST_SchedulingReport_20260716.pdf}} \citep{espinoza_dd9234}.

In this work, we analyze those MIRI eclipse data and aim to characterize  the atmospheres and/or surfaces of the LTT~1445~A siblings and draw broader implications for the M~dwarf Cosmic Shoreline.  The paper is organized as follows.  In \S \ref{sec:ddt_obs}, we introduce the RWDDT observations analyzed in this work.  In \S \ref{sec:newglobal}, we present a new, initial joint fit that includes JWST transits.  In \S \ref{sec:datared}, we present our data reduction and analysis of the RWDDT observations. In \S \ref{sec:global_fit_results}, we present a final joint fit that includes the RWDDT eclipses.  In \S \ref{sec:models}, we interpret the measured eclipse depths using both Bayesian retrieval and a suite of forward models.  In \S \ref{sec:discuss}, we discuss the implications of our results for the system and MIRI observations.  In \S \ref{sec:conclusion}, we summarize our findings and share prospects for the RWDDT and observations of the LTT~1445~A system.

\section{DDT Observations of LTT~1445~A \lowercase{c} \& \lowercase{b} } \label{sec:ddt_obs}

As a part of the RWDDT program, JWST observed LTT~1445~A across three visits in two DD programs, \texttt{DD09234} and \texttt{DD12758} \citep{espinoza_dd9234}, using the MIRI imaging F1500W filter and the SUB64 subarray with FASTR1 readout.  In chronological order, visit \texttt{DD09234\_1} observed an eclipse of \lttac{} using one exposure of 7264 integrations with 22 groups per integration (3.9 hours); visit \texttt{DD12758} observed an eclipse of \lttab{} using two exposures of 6623 integrations each and 25 groups per integration (8.1 hours); and visit \texttt{DD09234\_2} observed the eclipses of both using three exposures of 7475 integrations each and 22 groups per integration (12.2 hours).  This observing strategy was intended to first confirm the orbit of each planet with unknown eclipse timings, and then capture the eclipses of both planets in one continuous light curve, according to the scheduling reports.  

Visual inspection of preliminary light curves confirms that both eclipse timings are consistent with nearly-circular orbits for both planets.  The eclipse of planet c coincided with a transit of planet b in \texttt{DD09234\_1} and again with an eclipse of planet b in \texttt{DD09234\_2}.  The former coincidence reduces the baseline for planet c by a factor of $\sim 3$ and makes the eclipse depth of c sensitive to limb darkening assumptions; the latter induces possible anti-correlation in the eclipse depths of the two planets.  We discuss the ramifications in \S \ref{sec:datared}.  

Interestingly, the repeated overlaps are possible because the siblings are very close to a 12:7 orbital commensurability (an integer ratio that does not imply resonance).  Planet c completed 6 orbits between the two \texttt{DD09234} eclipses while planet b completed exactly 3.5 orbits.  We test whether this is a true mean motion resonance in \S \ref{sec:discuss}.

\section{New Joint Fit including JWST transits} \label{sec:newglobal}

We first perform a new, initial joint fit to ascertain the planetary and orbital parameters of the LTT~1445~A system before analyzing the RWDDT eclipse observations.  
Determining the system parameters to the best precision achievable is essential to both inferring and interpreting the eclipse depths.
 
Previous studies have found a range of values.  Using the radius of planet b for illustration, \citet{winters22_a} found $R_{\mathrm{p,b}} = 1.30^{+0.07}_{-0.06}\,R_\oplus$, whereas \citet{lavie23_planetary} found a larger $1.43 \pm 0.09\,R_\oplus$ from TESS, and \citet{oddo23_characterization} found a smaller $1.18 \pm 0.06\,R_\oplus$ from TESS and CHEOPS.  Analysis in \citet{monaghan26_uniform} found that even such minor discrepancies make a precise interpretation of the MIRI LRS eclipse observations in \citet{wachiraphan25_the} challenging, as the uncertainty in the model due to uncertain parameters is comparable to that in the eclipse measurement itself.  

For planet c, large uncertainties in system parameters can leave whether the planet is eclipsing at all undetermined, especially as the eclipse impact parameter $b_\mathrm{ecl}$ depends not only on the transit impact parameter $b_\mathrm{tr}$ but also on $e\sin\omega$, where $e$ and $\omega$ refer to the planet's eccentricity and argument of periastron, respectively. The TESS light curve permits both grazing and non-grazing transits: \citet{winters22_a} could only place a lower limit of $1.15\,R_\oplus$ on its radius, while \citet{lavie23_planetary} found the transit to be grazing with 85\% probability, giving $R_{\mathrm{p,c}} = 1.60^{+0.67}_{-0.34}\,R_\oplus$.  \citet{pass23_hst} combined a partial transit of c in HST/WFC3 data with existing TESS and RV data and ruled out a grazing geometry at 97\% confidence, giving $1.07^{+0.10}_{-0.07}\,R_\oplus$.

Now that JWST has observed \textit{both} planets in transit multiple times (JWST GO 2512, PI Batalha; JWST GO 7073, PI Lustig-Yaeger; JWST DD 9234), we can characterize the orbits with far greater precision than before.  We therefore perform a new joint fit including the JWST transit observations. 

We perform two sets of joint fits, one with and one without including the RWDDT observations, in order to avoid double-counting the RWDDT data when we analyze them independently.  We use the posterior from the initial joint fit as priors in the independent analyses of the RWDDT data performed in \S \ref{sec:datared}, wherein we explore data reduction choices and systematics model assumptions.  We then fit all of the data again, now including the RWDDT observations in \S\ref{sec:global_fit_results} to produce self-consistent system parameters.  This posterior is reported as the final system parameters.  

\subsection{Included Data} \label{subsec:lc_rv}

\begin{table*}[ht]
\centering
\caption{Photometric and RV data included in the joint fits of LTT~1445~A~b and c.  The DDT visits and MIRI LRS eclipses are not included in the initial RV+LC joint fit.}
\label{tab:lc_rv_data}
\begin{tabular}{cllcll}
\hline
\multicolumn{6}{c}{Photometry} \\
\hline
Planet & Type & Instrument & $N_{\rm visits}$ & JWST PID & Reference \\
\hline
b & Eclipse & JWST MIRI/LRS        & 3 & GO 2708 & \citet{wachiraphan25_the} \\
 & Eclipse & JWST MIRI/F1500W     & 2 & DD 9234$^\ddagger$ & This work (\S \ref{sec:jegug_red}) \\
 &  &     &  &  DD12758 & This work (\S \ref{sec:jegug_red}) \\
 
 & Transit & JWST MIRI/F1500W     & 1 & DD 9234$^\dagger$ & This work (\S \ref{sec:jegug_red}) \\ 
 & Transit & JWST NIRSpec/G395H$^*$   & 2 & GO 2512$^{\#}$& \citet{batalha26_jwst} \\
 &  &  & &  GO 7073 &  R. Ashtari et al. \textit{in prep.} \\
 & Transit & JWST NIRISS/SOSS     & 2 & GO 7073 & R. Ashtari et al. \textit{in prep.} \\ 
 & Transit & Magellan II/LDSS3C   & 4 & ---        & \citet{diamondlowe23_ground} \\
\hline
c & Eclipse & JWST MIRI/F1500W     & 2 & DD 9234$^\dagger$$^\ddagger$ & This work (\S \ref{sec:jegug_red}) \\ 
 & Transit & JWST NIRSpec/G395H$^*$   & 2 & GO 7073 & J. Brande \& L. Wiser et al.\ \textit{in prep.} \\
 & Transit & JWST NIRISS/SOSS     & 2 & GO 7073 & J. Brande \& L. Wiser et al.\ \textit{in prep.} \\
 & Transit & ARC 3.5\,m/ARCTIC    & 1 & ---        & This work (Appendix~\ref{app:arctic}) \\
\hline
\end{tabular}
\begin{tablenotes}
\centering
\footnotesize
\item $^\dagger$ $^\ddagger$ Overlapping.  \quad $^*$ Two NRS orders fit separately. \quad $^{\#}$ Re-reduced in R. Ashtari et al. \textit{in prep}.
\end{tablenotes}
\vspace{1em}

\begin{tabular}{lll}
\hline
\multicolumn{3}{c}{Radial velocities} \\
\hline
Instrument & $N_{\rm RV}$ & Reference \\
\hline
ESPRESSO  & 84 & \citet{lavie23_planetary} \\
ESPRESSO  & 19 & \citet{winters22_a} \\
HARPS     & 38 & \citet{winters22_a} \\
HIRES     & 39 & \citet{winters22_a} \\
MAROON-X  & 25 & \citet{winters22_a} \\
PFS       & 15 & \citet{winters22_a} \\
\hline
Total     & 220 & \\ 
\hline
\end{tabular}

\end{table*}

For our joint fits, we include the photometric data from all existing JWST transits from GO 2512 and GO 7073.  We fit the two NRS detectors separately in NIRSpec data.   We also include a set of ground-based observations for longer baselines for each planet:  four Magellan II/LDSS3C transit observations of b \citep{diamondlowe23_ground} and one ARC 3.5\,m/ARCTIC transit of c (see Appendix \ref{app:arctic} for a description of the observations).  For the Magellan data, we use the light curves already de-trended rather than fitting for systematics again; this is sufficient to constrain the timing.  This amounts to a total of 17 light curves.  For RV, we include the combined set of archival data in \citet{winters22_a} and \citet{lavie23_planetary}, which do not share any epoch, for a total of 220 RVs. The included data and their references are summarized in Table \ref{tab:lc_rv_data}.

Other than the Magellan and ARCTIC transits, we do not include any other photometric data from \textit{e.g.} HST, TESS, CHEOPS, or MEarth.
Given that there are even more planned JWST observations with JWST using NIRCam within JWST Cycle 4 (JWST GO 7251, PI Bennett), we leave an exhaustively complete global fit (including stellar SED) for future work and consider the included dataset sufficient for the accurate analysis of the RWDDT data.  

We also exclude the MIRI LRS eclipse observations of planet b from \citet{wachiraphan25_the}; while the data could have been included to refine our priors, we found after fitting eclipses to the data that the dataset was ultimately not very constraining and made no difference to the eclipse timing in the prior.

\subsection{Initial Joint Fit Model Description} \label{subsec:fit_model}

To jointly analyze the photometric and spectroscopic observations for the LTT~1445 system, we use the \texttt{exoplanet} \citep{foreman-mackey_exoplanet-devexoplanet_2021} framework that performs posterior sampling using the Hamiltonian Monte Carlo (HMC) method via \texttt{PyMC3} \citep{salvatier_probabilistic_2016} and \texttt{Theano} \citep{the_theano_development_team_theano_2016}. We use the Keplerian solver to simultaneously fit the instrumental and astrophysical terms associated with the RVs and transits listed in \S \ref{subsec:lc_rv}. While converting the measured planetary parameters into derived parameters, we assume Gaussian distributions for the stellar mass, radius and effective temperature based on \citet{winters22_a} , and place Gaussian priors on the orbital periods and transit times.

Our fit included 220 archival RVs and 17 light curves (see Section \ref{subsec:lc_rv}), where we bin the JWST NIRSpec and NIRISS light curves to 60~s.

We use a uniform prior on the impact parameter between 0 and  $1+R_p/R_\star$ to allow for grazing transit configurations (but not non-transiting configurations), while the eccentricity and argument of periastron are sampled as $(\sqrt{e}\cos\omega, \sqrt{e}\sin\omega)$ on the unit disk, with a $\beta$ prior on $e$ \citep{kipping_parametrizing_2013}.  We use a quadratic limb-darkening law for all the photometric observations using the reparametrization of \citet{kipping_efficient_2013}, with separate coefficients for each instrument. Lastly, we also factor in a separate dilution term for each Magellan and ARCTIC light curve to account for any potential uncorrected transit depth contamination by LTT~1445 BC; we do this using a uniform prior between 0.1 and 1.5 \citep[explained in][]{kanodia_toi-5205b_2023}, which was ascertained in the joint fit using the \textit{true} depth from JWST for transits of planet b and c (i.e., no contamination in JWST light curves and the dilution term = 1). 

Each light curve has its own white jitter term that is added in quadrature to the propagated uncertainties, as well as a baseline trend. The baseline ranged from an offset (zeroth-order polynomial) to a quadratic, or an exponential.  For two NIRISS/SOSS transit light curves (planet b in GO 7073 visit 1, and planet c in GO 7073 visit 2), we use a stochastically driven, damped simple harmonic oscillator (SHO) Gaussian Process (GP) kernel, as implemented in \texttt{celerite2} \citep{foreman-mackey_fast_2017, foreman-mackey_scalable_2018} of timescale $\mathcal{LU}(10.3, 240)$\,min to account for the 86- and 158-minute oscillations seen in the data.

Each RV instrument is modeled with a separate RV offset term ($\gamma$) and a jitter term (white noise) that is added in quadrature to the formal uncertainties before combining the instruments.  The combined RV time series is fit including a linear and quadratic baseline to test for the presence of additional astrophysical signals.  We optimize the model in stages to find the maximum a posteriori solution, then run four chains of 5000 tuning and 2500 posterior steps each, with a target acceptance rate of 0.95. All parameters have $\hat{R}\le1.01$ \citep{vehtari_rank-normalization_2021}, which along with a large bulk and tail effective sample size conveys chain convergence. Instead of asserting or fixing the orbit to be circular we adopt a flexible floating eccentricity model. 

We show the results from the RV and transit-only joint fit in Table \ref{tab:jointfit} in Appendix \ref{app:global_posterior}.  These are used as priors in the next section.

\section{Analysis of RWDDT Observations} \label{sec:datared}

In this section, we describe three independent data reductions and analyses performed on the RWDDT data.  

\subsection{JI Reduction and Analysis}\label{sec:jegug_red}

We use \eureka{} \citep{bell22_eureka} to perform data reduction and fit light curves.  We also adopt this light curve in the final RV+LC joint fit in \S \ref{sec:global_fit_results}.   We begin from the Stage 2 calibrated integrations (\texttt{calints}) produced by the default STScI \texttt{jwst} pipeline, which applies last-frame rejection but neither first-frame rejection nor reset switch charge decay correction.  We see some non-linearity at the start of each integration of up to a few \% increase in slope but choose to not throw out any groups at the start, as the non-linearity did not vary over integrations during each visit, and any groups rejected would increase the photon noise.

\begin{figure}
    \centering
    \includegraphics[width=0.98\linewidth]{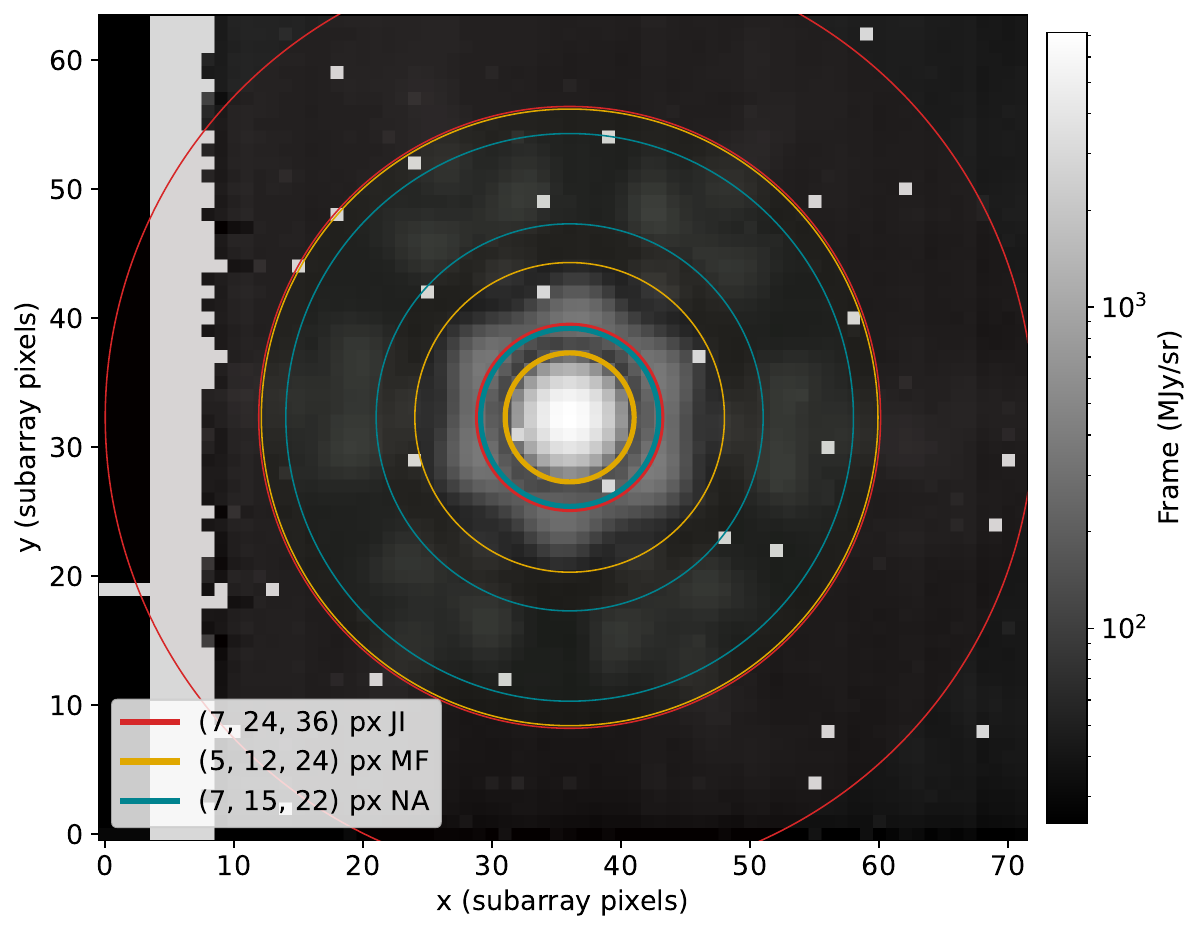}
    \caption{The PSF of LTT~1445~A on the \texttt{SUB64} subarray and the apertures used by the three reductions (circles).  The legend denotes aperture, background inner, and background outer radii, respectively.  The frame shown is the median frame of a portion of visit \texttt{DD09234\_1}.  Due to the small subarray, the background inevitably contains some of the snowflake of the PSF.  Note that the first 13 columns are flagged and not used. }
    \label{fig:psf}
\end{figure}

In Stage 3, we perform simple circular aperture photometry.  We first restrict each integration to the illuminated region of \texttt{SUB64} and also exclude row 0, which was not flagged but has roughly half the count levels of the other rows and $\sim 1.4$ times larger pixel-to-pixel scatter.  For the background region, we choose an annulus with inner and outer radii of 24 and 36 px.  Because the subarray is too small for an annulus beyond the PSF wings, inevitably the background contains some of the PSF (Figure~\ref{fig:psf}).  We deem this acceptable given the brightness of the star.  We also experimented with using a slightly larger inner radius of 28 px instead; this produced comparable or larger scatter.  We optimize the circular aperture size by performing a grid from 4-9 px in 0.5 px increments.  An aperture of 7 px produced the best tradeoff between point-to-point and binned (over 10-60 minute bins) scatter across visits.  In Stage 4, we mask outliers with an iterative 5$\sigma$ clip relative to an 80-integration moving mean.  This removes only 3, 11, and 14 integrations from \texttt{DD09234\_1}, \texttt{DD12758}, and \texttt{DD09234\_2}, respectively.


We perform both separate fits for each visit and a joint fit across all visits.  We choose to clip the first 300 integrations of each visit (9.8 min for \texttt{DD09234\_1,2} and 11.1 min for \texttt{DD12758}), during which the background level and PSF width are still settling; the remaining slower settling is modeled with an exponential ramp.  The eclipse depths change by less than 0.3 sigma when discarding 150 or 600 integration instead. 

\begin{figure*}[ht]
    \includegraphics[width=0.99\linewidth]{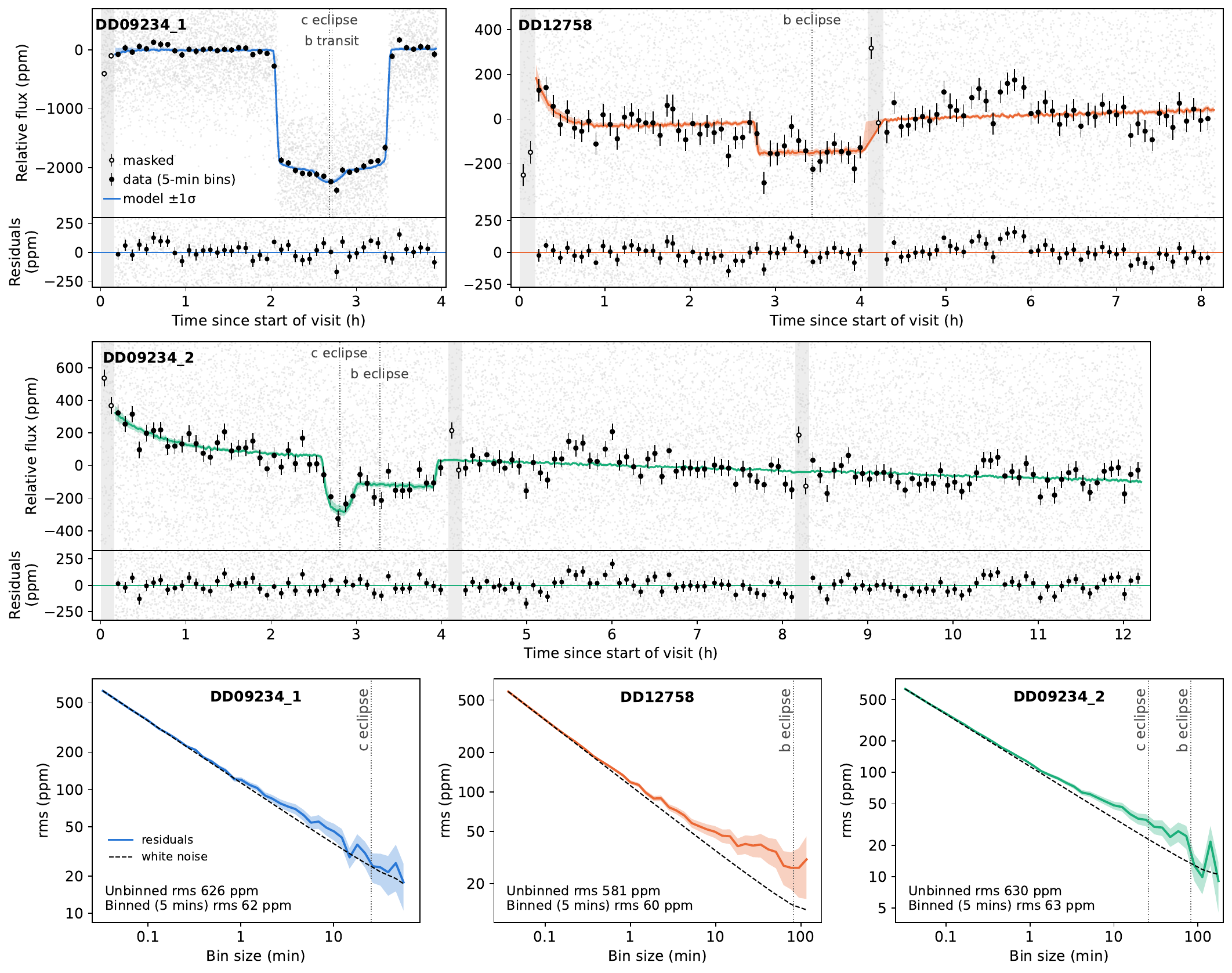}
    \caption{The individually fit light curves in the JI \eureka{} reduction without GPs (top panels) and bin size-residual plots for each visit (bottom).  \textit{Top}: the fit \eureka{} light curves (including systematics) and their 95\% uncertainties are shown as colored contours.  The data binned to 5 minute bins for visual clarity are shown as black points, while the unbinned data is shown as grey dots.  The masked regions at the start of each visit and exposure are shown as grey regions with white points.  The egress of planet b overlaps with a masked region due to the exposure break in visit \texttt{DD12758}.  The panels are shown in chronological order and share the same horizontal time scaling. \textit{Bottom}: the rms of the residuals at different bin sizes (color) compared to white noise expectation (dashed line).  The timescale for the eclipse duration of each planet is marked as vertical lines. }
    \label{fig:lightcurves}
\end{figure*}

Additionally, in \texttt{DD12758}, we notice a flare-like sharp increase in flux that is followed by a rapid decay (exponential timescale of $\sim$ 2 min).  While LTT~1445~A has been observed to flare in UV and X-ray \citep{diamond24, brown22_x}, and M star flares have been seen at 15 micron \citep{gillon_first_2025, holmberg26_hot}, the onset of the ``flare'' in the observed data coincides suspiciously within the $\sim$40 s gap between exposures.  We see a similar sharp increase and decay exactly at the start of each new exposure in \texttt{DD09234\_2} as well, albeit to lesser magnitudes (see \autoref{fig:lightcurves}).  While a flare model does fit these rises well, the odds of bona fide astrophysical events coinciding with new exposures three times (over total $>20$ hours) are rather small.  We argue that correlation does imply causation here and judge that these ``flares'' are in fact mini-detector settlings after exposure breaks as previously seen in \citet{zhang24} for MIRI/LRS.  We discuss this further in \S \ref{subsec:systematics}.  For this reason, we choose to also mask 300 integrations at the start of each new exposure.  When not fitting for systematics with Gaussian processes (GP), the masks affect the fit eclipse depths to less than 0.3 sigma (6 ppm).  We also tested fitting for offsets between exposures, which we discuss in \S \ref{subsec:systematics}.

\begin{table}
\begin{threeparttable}
\caption{Priors used in the JI \eureka{} light curve fits (\S \ref{sec:jegug_red}), based on the RV+transit-only joint fit in \S \ref{sec:newglobal} and tidal arguments on $e_c\sin\omega_c$.}
\label{tab:priors}
\small
\begin{tabular}{lc}
\hline\hline
Parameter & Prior \\
\hline
\multicolumn{2}{l}{\textit{LTT~1445\,A~b}} \\
$R_p/R_\star$                       & $\mathcal{N}(0.045,\ 0.001)$ (\texttt{DD09234\_1})  \\
$R_p/R_\star$                       & $0.044744 \pm 0.000068$$^{\,j}$               \\
$F_p/F_\star$ (ppm)                 & $\mathcal{U}(-1000,\ 1000)$                       \\
$t_0$ (BMJD$_\mathrm{TDB}$)         & $\mathcal{N}(61284.507907,\ 0.000017)$        \\
$P$   (d)                           & $\mathcal{N}(5.35876803,\ 1.1\times10^{-7})$  \\
$i$   ($^{\circ}$)                  & $89.694 \pm 0.130$$^{\,j}$                    \\
$a/R_\star$                         & $30.91 \pm 0.35$$^{\,j}$                      \\
$e\cos\omega$                       & $-0.015 \pm 0.026$$^{\,j}$; 0 (\texttt{DD09234\_1})                  \\
$e\sin\omega$                       & $0.0010 \pm 0.0136$$^{\,j}$                   \\
\hline
\multicolumn{2}{l}{\textit{LTT~1445\,A~c}} \\
$R_p/R_\star$                       & $0.037141 \pm 0.000284$$^{\,j}$               \\
$F_p/F_\star$ (ppm)                 & $\mathcal{U}(-1000,\ 1000)$                       \\
$t_0$  (BMJD$_\mathrm{TDB}$)        & $\mathcal{N}(61257.952088,\ 0.000055)$        \\
$P$     (d)                         & $\mathcal{N}(3.12389316,\ 6.1\times10^{-7})$  \\
$i$     ($^{\circ}$)                & $87.441 \pm 0.033$$^{\,j}$                    \\
$a/R_\star$                         & $21.57 \pm 0.25$$^{\,j}$                      \\
$e\cos\omega$                       & $0.009 \pm 0.022$$^{\,j}$                     \\
$e\sin\omega$                       & $\mathcal{N}(0,0.002)$$^{j,p}$ \\
\hline
\multicolumn{2}{l}{\textit{Star}} \\
$R_\star$  ($R_\odot$)              & 0.2673                                        \\
$u_1$                               & $\mathcal{N}(0.053,\ 0.005)$ (\texttt{DD09234\_1}) \\
$u_2$                               & $\mathcal{N}(0.089,\ 0.006)$ (\texttt{DD09234\_1}) \\
\hline
\multicolumn{2}{l}{\textit{Systematics and noise (each visit)}} \\
$c_0$                               & $\mathcal{N}(1,\ 0.005)$                      \\
$c_1$                               & $\mathcal{N}(0,\ 0.01)$                       \\
$r_0$                               & $\mathcal{U}(-0.05,\ 0.01)$                   \\
$r_1$                               & $\mathcal{LU}(5,\ 1500)$                      \\
$x_{\rm pos}$, $y_{\rm pos}$ (px$^{-1}$)        & $\mathcal{U}(-0.1,\ 0.1)$         \\
$\sigma_{\rm white}$                & $\mathcal{U}(0,\ 2000)$                       \\
\hline
\multicolumn{2}{l}{\textit{Mat\'ern-3/2 Gaussian process (GP fits only)}} \\
$\ln \sigma_{\rm GP}^2$             & $\mathcal{U}(-24,\ -14)$                      \\
$\ln \rho$  (d)                     & $\mathcal{U}(-7.27,\ -4.97)$                  \\
\hline
\end{tabular}
\begin{tablenotes}[flushleft]
\footnotesize
\item \textbf{Notes.} $\mathcal{N}(\mu, \sigma)$: normal; $\mathcal{U}(a, b)$: uniform; $\mathcal{LU}(a, b)$: log-uniform between $a$ and $b$. Priors for each planet are used only when a visit contains the planet.
\item $^{j}$ Joint multivariate normal prior over both planets based on the transit and RV joint fit, wherein mean and 1$\sigma$ are reported in the table.  In the order $(R_p/R_\star,\ i,\ a/R_\star,\ e\cos\omega,\ e\sin\omega)$ of b and of c, the correlation
matrix is
$\rho = \left(\begin{smallmatrix}
 1 & & & & & & & & & \\
 -0.64 & 1 & & & & & & & & \\
 -0.23 & 0.33 & 1 & & & & & & & \\
 0.01 & 0.00 & -0.02 & 1 & & & & & & \\
 -0.34 & 0.51 & -0.61 & 0.07 & 1 & & & & & \\
 -0.21 & 0.05 & -0.40 & 0.00 & 0.37 & 1 & & & & \\
 -0.23 & 0.32 & 0.99 & -0.02 & -0.60 & -0.41 & 1 & & & \\
 -0.23 & 0.33 & 1.00 & -0.02 & -0.61 & -0.40 & 0.99 & 1 & & \\
 -0.01 & 0.03 & -0.03 & 0.06 & 0.06 & -0.01 & -0.09 & -0.03 & 1 & \\
 0.01 & -0.04 & -0.14 & 0.01 & 0.09 & 0.01 & -0.26 & -0.14 & 0.00 & 1
\end{smallmatrix}\right)$, and the covariance is $\Sigma_{ij} = \rho_{ij}\,\sigma_i\,\sigma_j$; diagonal terms $\Sigma_{ii}$ are inflated by 1 ppm to make $\Sigma$ invertible. 
\item $^p$ This prior is set based on physical arguments; see \S \ref{sec:jegug_red}. 
\end{tablenotes}
\end{threeparttable}
\end{table}

We choose our priors for the system from the results of the initial RV and transit-only joint fit.  We adopt the posterior as a joint multivariate prior for $R_\mathrm{p}/R_\mathrm{s}$, $i$, $a/R_\mathrm{s}$, $e\cos\omega$, and $e\sin\omega$ for each planet.  In an eccentric orbit, the eclipse impact parameter $b_\mathrm{ecl}$ differs from the transit impact parameter $b_\mathrm{tr}$. However, the joint prior constrains $e\sin\omega$ of planet c only to $0.006 \pm 0.023$, which predicts the eclipse impact parameter $b_\mathrm{ecl}$ from a non-grazing full eclipse to a not-even-grazing non-eclipse.  In our initial tests, this made the eclipse depth of c completely unconstrained.   We therefore set a narrower prior on $e\sin\omega$ of $\mathcal{N}(0,0.002)$ based on tidal damping timescales (see \S \ref{subsec:tidal_heating}) to force $b_\mathrm{tr} \simeq b_\mathrm{ecl}$ and reweighted the joint prior accordingly.  We fit a multinormal distribution and added an option in \eureka{} to use this as a joint prior.   

We allow for negative eclipse depths and set a wide uniform prior between -1000 and 1000\,ppm for each planet.  For visit \texttt{DD09234\_1}, we fit both a transit of b and an eclipse of c and set a wide normal prior on the transit radius of b with a width of 1000\,ppm.   We use the quadratic limb darkening (LD) law and adopt as priors the coefficient values and uncertainties calculated using \texttt{exotic\_ld} \citep{Grant2024_exoticld} with the F1500W throughput function and the MPS-ATLAS library \citep{Kostogryz2023_mpsatlas} extrapolated to LTT~1445~A parameters ($T_\mathrm{eff}=3340\pm150$\,K; the library ends at 3500\,K).  We include light travel time.  We do not bin the data in our fits.  For systematics, we fit for an exponential ramp, a linear trend, and de-correlate against centroid drift.  For fits using GP, we adopt a Mat\'ern-3/2 kernel and set the timescale to be between 1-10 minutes for visits including c and 1-30 minutes for visit \texttt{DD12758}; because of the fast cadence of the observations, even 1 minute contains $\sim$30 integrations.   The priors are summarized in Table \ref{tab:priors}.  

\begin{figure}[ht]
    \centering
    \includegraphics[width=0.93\columnwidth]{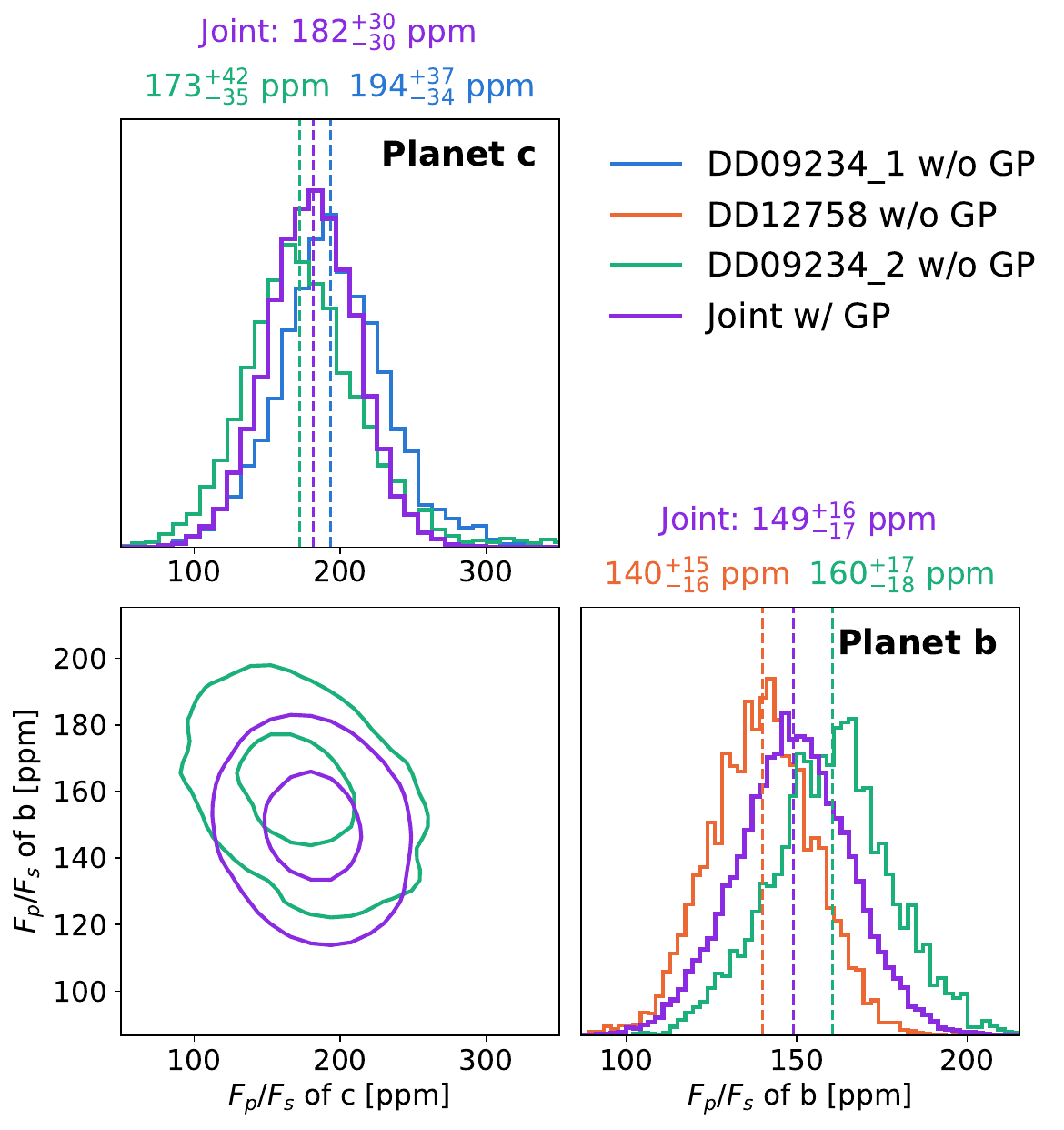}
    \caption{The posterior on eclipse depths from fits from the JI \eureka{} reduction, shown for each visit without GPs (blue, orange, and green; average in black) and for a joint fit including GP terms for visits where it is favored individually (purple).  Because the eclipses of the two planets overlap in visit \texttt{DD09234\_2} (green), the two eclipses are weakly anti-correlated, but this is partly mitigated by having two visits in the joint fit.}
    \label{fig:corner_fpfs}
\end{figure}

We show the light curves of the three visits and the best-fit models in Figure~\ref{fig:lightcurves}, for fits to separate visits.  We first show fits without GP to show the systematics.  Without GP, the separate fits give eclipse depths of $141^{+15}_{-16}$\,ppm (\texttt{DD12758}) and $160^{+17}_{-18}$\,ppm (\texttt{DD09234\_2}) for b, and $194^{+37}_{-34}$\,ppm (\texttt{DD09234\_1}) and $173^{+42}_{-35}$\,ppm (\texttt{DD09234\_2}) for c (Figure~\ref{fig:corner_fpfs}), showing repeatable eclipse depths to within $1\sigma$.  

Modeling the correlated noise with a GP leaves the median depths unchanged but roughly doubles the uncertainties in individual visit fits.  The inclusion of GP is strongly favored in \texttt{DD12758} ($\Delta\ln B=9.3$) and \texttt{DD09234\_2} ($\Delta\ln B=6.1$).  We perform a joint fit that includes GP terms for these two visits, which gives $(F_\mathrm{p}/F_\mathrm{s})_b = 149^{+16}_{-17}$\,ppm for planet b and $(F_\mathrm{p}/F_\mathrm{s})_c = 182 \pm 30$\,ppm for planet c, consistent with the weighted average of the separate fits.  The two depths are weakly anti-correlated where both eclipses fall in the same visit (correlation strength $\rho = -0.41$ in \texttt{DD09234\_2} individually and $-0.17$ in the joint fit).

\subsection{MF Reduction and Analysis}
\label{subsec:mf_red}

We performed an independent reduction following a similar approach to \citet{fortune25_lhs1140c}. We ran the first two stages of the standard \textsc{jwst} v3.0.0 pipeline after combining the \textsc{uncal} segments for each exposure. We excluded both the first and last groups, set a $7 \sigma$ jump detection threshold, and skipped the \textsc{photom} step. \textsc{photutils} was used to perform PSF centroiding, a background annulus with a narrow annulus of (12, 24) pixels was used, which includes some of the PSF wings but this is unavoidable given the small subarray. Aperture sum photometry was performed with a 5 pixel radius. The first 15 minutes of each observation was masked, in addition to the potential flare event in the light curve of \textsc{DD9234\_2}, clipping 1.5 hours between (2461303.85, 2461303.91) BJD TDB.

Transit light curve modeling was performed using \textsc{jaxoplanet} \citep{jaxoplanet} and accounting for light delay, while \textsc{tinygp} was used to implement an optimised Matérn-3/2 kernel without approximation \citep{Foreman-Mackey2024}. All three visits were joint-fit, allowing both planets to have non-zero eccentricities, parameterized to put uniform priors on $e$ and $\omega$. Specifically, we fit for $\sqrt{e_c}\cos{\omega_c}$, $\sqrt{e_c}\sin{\omega_c}$ for $c$, while we fit for $e_b\cos{\omega_b}$, $e_b\sin{\omega_b}$ and subtract $-\log{e_b}$ from our log-likelihood for $b$. Both yield the same priors \citep{eastman2013} and were chosen to improve sampling efficiency. Datapoints were binned by a factor of five when fitting, to reduce computational cost. Each visit was fit with a free offset, linear slope, and initial exponential ramp to model remaining settling. The eclipse depths were bounded within -200 to 400 ppm, and the planets were modeled as a uniform brightness disk. For the transit of planet $b$, quadratic limb darkening was assumed with the same limb darkening priors as from the JI \eureka{} analysis, but we also tested varying the parameters with uniform priors excluding non-physical ranges using the \citet{kipping_efficient_2013} parameterization (e.g. not allowing limb brightening). We used a multivariate normal prior using a covariance matrix calculated from the posterior of $c$'s NIRSpec and NIRISS/SOSS transits, constraining $t_0$, $a/R_*$, $R_p/R_*$, and its impact parameter. The period $P$ was fixed to the value in \citet{lavie23_planetary}, but the period uncertainty was propagated by adding to the prior variance of $t_0$. For planet $b$, independent Gaussian priors from the literature values in \citet{winters22_a} were used to constrain $t_0$, $a/R_*$, and the impact parameter $b$, while the transit depth $(R_p/R_*)^2$ was fitted with uniform priors. The period for $b$ was fixed, but with the uncertainty also propagated into $t_0$. MCMC inference was performed using the package \textsc{pymc}, using No U-Turn Sampling combined with Gibbs updates of slice sampling for some parameters close to prior bounds. 10 chains were run, where each chain had 2,000 tuning steps followed by 4,000 draws, with all parameters achieving $\hat{r}$-statistics $<1.01$ and with $>1000$ effective samples.

Eclipse depth results varying the inclusion of a GP, and whether planet $c$ was assumed to have a circular or eccentric orbit, are included in Table~\ref{tab:MF_depths}, although the inclusion of a GP and eccentric orbit for $c$ is preferred as the more conservative approach. Planet $b$ was always assumed eccentric as its eclipse was very tightly constrained with a constraint of $e_b \cos \omega_b = 0.00082\pm0.00021$, which corresponds to an eclipse that is $4\pm1$ minutes late compared to a circular orbit. 

\begin{table}
\centering
 \caption{Eclipse depth results from \textit{MF reduction} in ppm}
 \begin{tabular}{lrrrr}
   \toprule
    Planet & $c$ circ. & $c$ ecc. & $c$ circ.+GP & $c$ ecc.+GP\\
    \hline
    $b$ & $146\pm13$ & $147\pm13$ & $146\pm18$ & $148\pm 19$\\
    $c$ & $168_{-26}^{+27}$ & $163_{-33}^{+40}$ & $160_{-32}^{+31}$ & $154_{-35}^{+48}$
    \end{tabular}
    \label{tab:MF_depths}
\end{table}

When fitting for eccentricity of $c$ (with bounds of $\sqrt{e_c}\cos \omega_c \in (-0.04, 0.04)$ and $\sqrt{e_c}\sin \omega_c \in (-0.5, 0.5)$, we have a significantly larger upper uncertainty on the eclipse depth compared to a circular orbit, as the posterior becomes consistent with the maximum prior bound of $400$ ppm in a more eccentric posterior mode, regardless of the inclusion of a GP, which is also seen in the \textit{RV+LC Joint Fit}. Physically-motivated tidal arguments exclude this scenario from the JI \eureka{} analysis, but we see that without this assumption we may not constrain a meaningful upper bound on the eclipse depth. This posterior mode has low probability however, and the posterior eccentricity is consistent with a circular orbit.  We found that freely fitting limb darkening parameters with the Kipping parameterization can reduce the eclipse depth for $c$ appreciably, the case of a circular orbit for $c$ without a GP reduces the eclipse depth from $168_{-26}^{+27}$ ppm down to $146\pm29$ ppm, which is examined in more detail in Section~\ref{subsec:limbdarkening} for the JI \eureka{} reduction. Overall, our results for planets $b$ and $c$ match closely with the other reductions in this work.

\subsection{NA Reduction and Analysis}
We performed an independent reduction broadly following the approach detailed in \citet{allen25_hot}. However, we do not use optimal extraction as was done in the main reduction, as it was shown that this tends to increase the amplitude of the settling slope. Instead, we test aperture extraction using radii from 3 to 10 pixels, and find that using a radius of 7 pixels minimized the residual out of transit scatter. Background subtraction was done using an annulus between 15-22 pixels from the centroid. This does contain some of the outer PSF wings, but due to the small subarray this could not be avoided, and we tested changing the included pixels and saw no discernible difference as long as the annulus was sufficiently wide. 

Light curve fitting was done using \textsc{juliet} \citep{juliet}, with sampling done using the \textsc{dynesty} \citep{speagle20} dynamic nested sampler. We remove 30 minutes of integrations after the start of each new exposure to conservatively account for the exponential effect of the settling slope. In addition, we mask another 1.75 hours after the exposure break in DD12758, as this portion of the light curve has a much higher level of systematics that may be connected to flaring. A variety of systematics models were tested (multiple degree polynomials, exponential, multiple GP kernels including Matern 3/2, exponential, and exponential times Matern 3/2), but the model preferred through a comparison of $ln\,Z$ was a GP using the \textsc{celerite} \citep{celerite} (approximate) Matern with standardized time as a regressor, which has two free parameters per observation: the amplitude and length scale. This increases the uncertainty in measured eclipse depth, but we believe this is necessary given the amplitude of noise seen in the light curves. Each observation is also fit for a relative flux offset and an error inflation term. 

We use the results from the joint fit as priors in our analysis. For the transit of b, we allow the transit depth as a free parameter and center our quadratic limb darkening law to the values calculated in \S~\ref{sec:jegug_red}, but with 5x the uncertainty to allow for any ambiguity in the stellar models at 15 $\mu$m \citep[using $u1$, $u_2$ so as not to bias the small limb darkening values with a Lucy-Sweeney type bias,][]{Coulombe_2024}. We use a uniform prior on our eclipse depths from 0 to 1000 ppm, but note that fits extending the prior on eclipse to negative values gave consistent results. We assume a circular orbit for both planets, though small eccentricities can be tolerated through the uncertainty in the orbital parameters from the joint fit. We include the light travel delay, using the stellar radius derived in the joint fit. 

We perform fits on individual observations and joint observations, and found consistent results for the derived eclipse depths in both cases, though with significantly larger uncertainties in the individual fits. In particular, the eclipse of planet c is difficult to constrain in individual fits in certain parts of the parameter space, since in both observations it is overlapped by planet b. Therefore, we rely on our joint fit inference for comparison. We find eclipse depths of $165\pm32$ ppm and $134^{+16}_{-17}$ ppm for planets c and b, respectively. However, we note that while a GP model with systematics was preferred by the data, no one specific kernel was strongly preferred against the others, and that changing the specifics of the kernel can affect either eclipse depth at the $\sim15$ ppm level, though this is within the eclipse uncertainty. Overall, these values are broadly consistent with the results from the other reductions.

\subsection{Are the eclipses there?}  \label{subsubsec:eclipsesthere}

To test whether the eclipses have been detected, we perform (a) light curve fits that do not allow (non-physical) negative eclipse depths and (b) fits that do not include the eclipse of each planet in the JI \eureka{} reduction, per visit and jointly.  For visit \texttt{DD09234\_2}, we perform fits excluding b, c, and both.  From test (a), we find that forcing positive eclipse depths has no appreciable effect on the measured eclipse depths whatsoever.  From test (b), we show the resulting log-Bayes factor ($\Delta\ln{Z}$) of fits that exclude each eclipse in Table \ref{tab:eclipse_off}.   Without the use of GPs, all eclipses are favored with very strong evidence \citep{kass95}. 

\begin{table} 
\begin{tabular}{llcc}
\hline
 & & \multicolumn{2}{c}{$\Delta\ln{Z}$} \\
\cline{3-4}
Visit & Eclipse & w/o GP &  with GP \\
\hline
\texttt{DD09234\_1} & c & -- & -- \\
 & none & $-9.0$ & $ \hl{-3.8}$ \\
 \hline
\texttt{DD12758} & b & -- & -- \\
 & none & $-33$ & $-\hl{1.5}$ \\
\hline
\texttt{DD09234\_2} & b + c & -- & -- \\
 & no b & $-34 $ & $-5.5$ \\
 & no c & $-6.4$ & $ \hl{-0.2}$ \\
 & none & $-74$ & $-9.9$ \\
\hline
Joint & b + c & -- &  -- \\
& no c & -12 & -4.9 \\
\hline
\end{tabular}
\caption{Log Bayes factors for light curve fits from the JI \eureka{} analysis with and without each planet, defined as $\Delta\ln Z=\ln Z_{\mathrm{excluded}}-\ln Z_{\mathrm{full}}$, such that \textit{negative} values mean that the inclusion of the planet is favored.  All of the Bayes factors have sampling uncertainty of $\lesssim 0.3$.  The eclipses with less than very strong  evidence ($\Delta\ln{Z}<5$) are highlighted. }
\label{tab:eclipse_off}
\end{table}

With GP included, the detections are less strong.  The eclipse of b is favored with only positive evidence in \texttt{DD12758} ($\Delta\ln{Z}=-1.5$).  This is not so surprising given the particularly strong systematics in this visit.  However, the eclipse of b is independently favored with very strong evidence in \texttt{DD09234\_2}, and we do not test in our joint fit.  The eclipse of c is favored with only strong evidence in \texttt{DD09234\_1} ($\Delta\ln{Z}=-3.8$, 45:1 odds ratio assuming equal prior probability) and not detected in \texttt{DD09234\_2} ($\Delta\ln{Z}=-0.2\pm0.3$).  When the eclipse of c is excluded, the residual eclipse depth is absorbed by slightly different LD coefficients in visit 1 and deeper eclipse of b in visit 2, alongside larger GP amplitudes.  Regardless, in the joint fit of all visits, including the eclipse of planet c is favored with strong evidence ($\Delta\ln{Z}=-4.9$), even when GPs are included.  This supports the detection of its eclipse despite the different confounding signals in the individual visits.

\subsection{Does the eclipse depth of planet c depend on limb darkening priors?} \label{subsec:limbdarkening} 

Because the eclipse of planet c coincides with a transit of planet b in visit \texttt{DD09234\_1}, we find that the priors set on LD coefficients influence the inferred eclipse depth $(F_\mathrm{p}/F_\mathrm{s})_c$ via changing the orbital parameters of planet b and the shape of the baseline of c.  To quantify this effect, we test three prior assumptions on LD coefficients in the JI \eureka{} reduction: (a) widening the prior width by a factor of 10, (b) using a wide, completely uninformative prior, and (c) using LD coefficients calculated using a Kurucz stellar model.  The second test uses $u_1, u_2 \sim \mathcal{U}(-3,3)$; the third test uses $(u_1, u_2) \sim \mathcal{N}((0.014,0.101), (0.008, 0.011)^2)$.  

\begin{figure}
    \centering
    \includegraphics[width=0.99\linewidth]{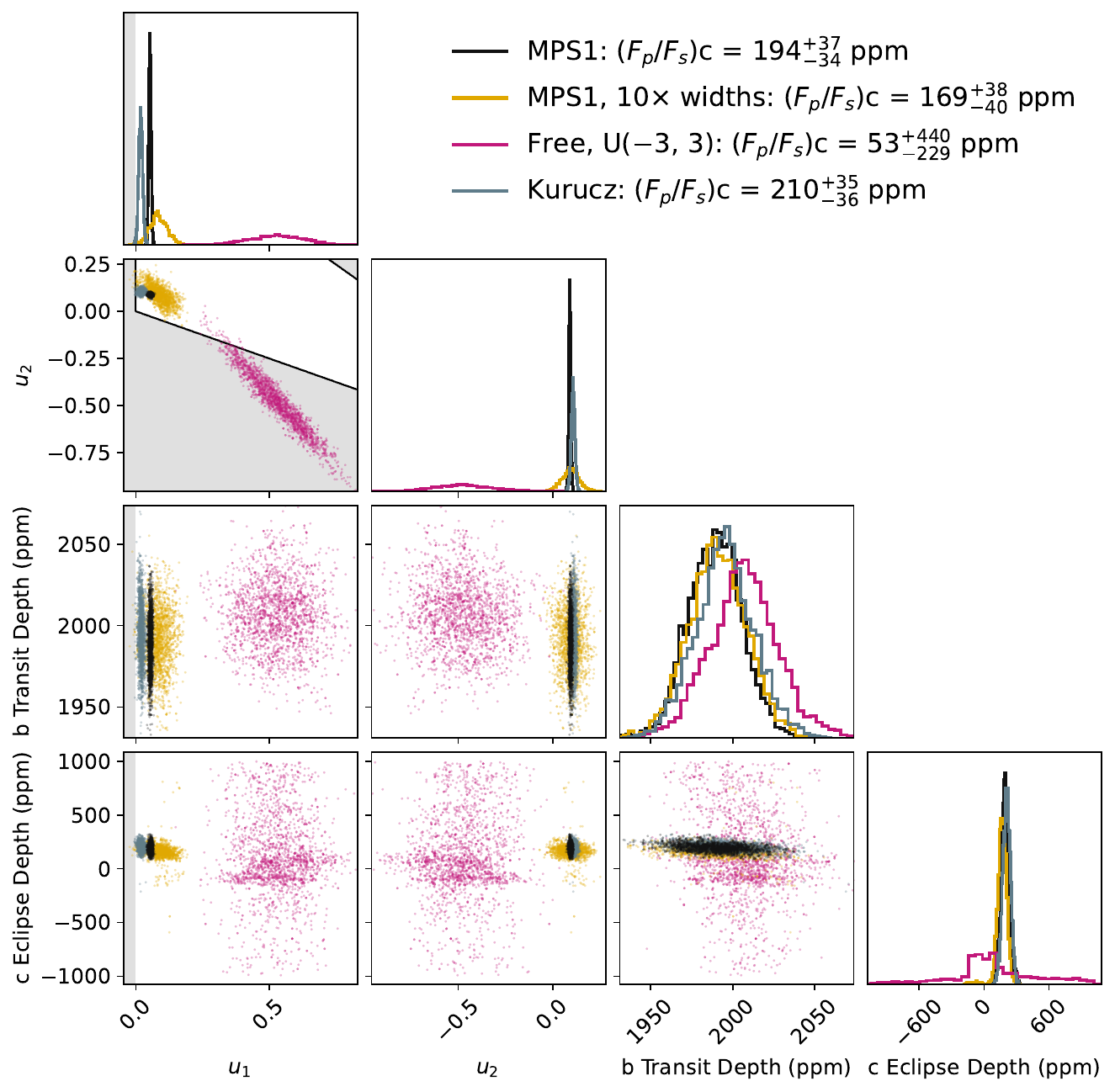}
    \caption{Eclipse depth of c and transit depth of b in visit \texttt{DD09234\_1} for various priors on LD coefficients $(u_1, u_2)$ from the JI \eureka{} fit.  The visit was fit individually.   MPS1 refers to the prior on $(u_1, u_2)$ derived from the MPS-ATLAS stellar model library and Kurucz refers to that from the Kurucz stellar model grid.  The unphysical region of $(u_1, u_2)$ (limb brightening) is shown in grey. }
    \label{fig:ld_tests}
\end{figure}

We show the resulting eclipse depth $(F_\mathrm{p}/F_\mathrm{s})_c$ in visit \texttt{DD09234\_1} in Figure~\ref{fig:ld_tests}.  Using ten times wider uncertainty (yellow) and using Kurucz stellar model (grey) move the median in either direction by $0.4-0.7\sigma$, while setting completely uninformative priors on LD coefficients yield a far less constrained eclipse depths that peak around $53$\,ppm and span basically the entire range of the prior space (fuchsia).  Most of this posterior is contained in the region of $(u_1, u_2)$ that lead to limb brightening and is unphysical.  

The degeneracy is not mediated by the transit depth of b, which is robust regardless of limb darkening priors.  Therefore it is small changes in the shape of the baseline that influence the eclipse depth of c, especially whether the bottom of the transit is round or flat.  The changes in eclipse depth of c due to LD coefficients alone is comparable to differences between data reductions (see Table \ref{tab:reductions}); while this could be partly mediated by the other eclipse of c in visit \texttt{DD09234\_2}, there the eclipse of planet c also overlaps with the eclipse of b.  We therefore deem it plausible that the dependence on LD coefficients is partly responsible for varying eclipse depths between reductions.

\section{Final RV+LC Joint Fit Including JWST eclipses} \label{sec:global_fit_results}

Having analyzed the eclipse data in \S \ref{sec:datared}, we now perform a final RV+LC joint fit including all eclipse observations to account for all parameter covariances and produce a final, self-consistent set of system parameters. 

\subsection{Updated Joint Fit Model Setup}

We update the joint fit from Section \ref{subsec:fit_model} by adding all eclipse observations from JWST (as well as the F1500W transit of b in \texttt{DD09234\_1}). We use the light curves from the fiducial JI \eureka{} reduction in \S\ref{sec:jegug_red} and the MIRI LRS light curves from \citet{wachiraphan25_the}, all binned to 60~s.  We trim the first 15, 30, and 60 minutes of each visit in \texttt{DD09234\_1}, \texttt{DD12758}, and \texttt{DD09234\_2}, respectively, but still include an exponential settling ramp.  We also trim 300 integrations at the start of the second exposure in \texttt{DD12758} and \texttt{DD09234\_2}, as also done in \S\ref{sec:jegug_red}. 

The eclipse impact parameter $b_\mathrm{ecl}$ is calculated from $b_\mathrm{tr}$ and $e\sin\omega$; this allows non-eclipsing (\textit{i.e.} not even grazing) scenarios.  Limb darkening coefficients at F1500W use the same \citet{kipping_efficient_2013} prior as for JWST near-IR instruments. We do not include light travel time (36~s for planet b across the orbit).   We include a GP term for visit \texttt{DD12758} only using the SHO GP kernel as used for the NIRISS SOSS data, with a log-uniform prior on the timescale between (10.3, 458)\,min, the latter corresponding to the visit length after trimming. 

The secondary-eclipse light curve function in \texttt{exoplanet v0.4.4} does not support more than one planet.  We therefore compute each planet's light curve directly, following the same formalism to estimate the surface brightness ratio, which is combined with the $(R_p/R_\star)^2$ to estimate the eclipse depth of the uniform disk for each planet in each bandpass that is used to fit the light curves.  

We also add a new GP term to the RV to de-trend a $\sim90$\,day signal discerned from the periodogram (discussed further in Appendix Section \ref{sec:rv_gp}).  This was also noticed in \citet{winters22_a} and attributed to stellar rotation.


\subsection{Final RV+LC Joint Fit Results}

\begin{figure*}[ht]
    \centering
    \includegraphics[width=0.93\textwidth]{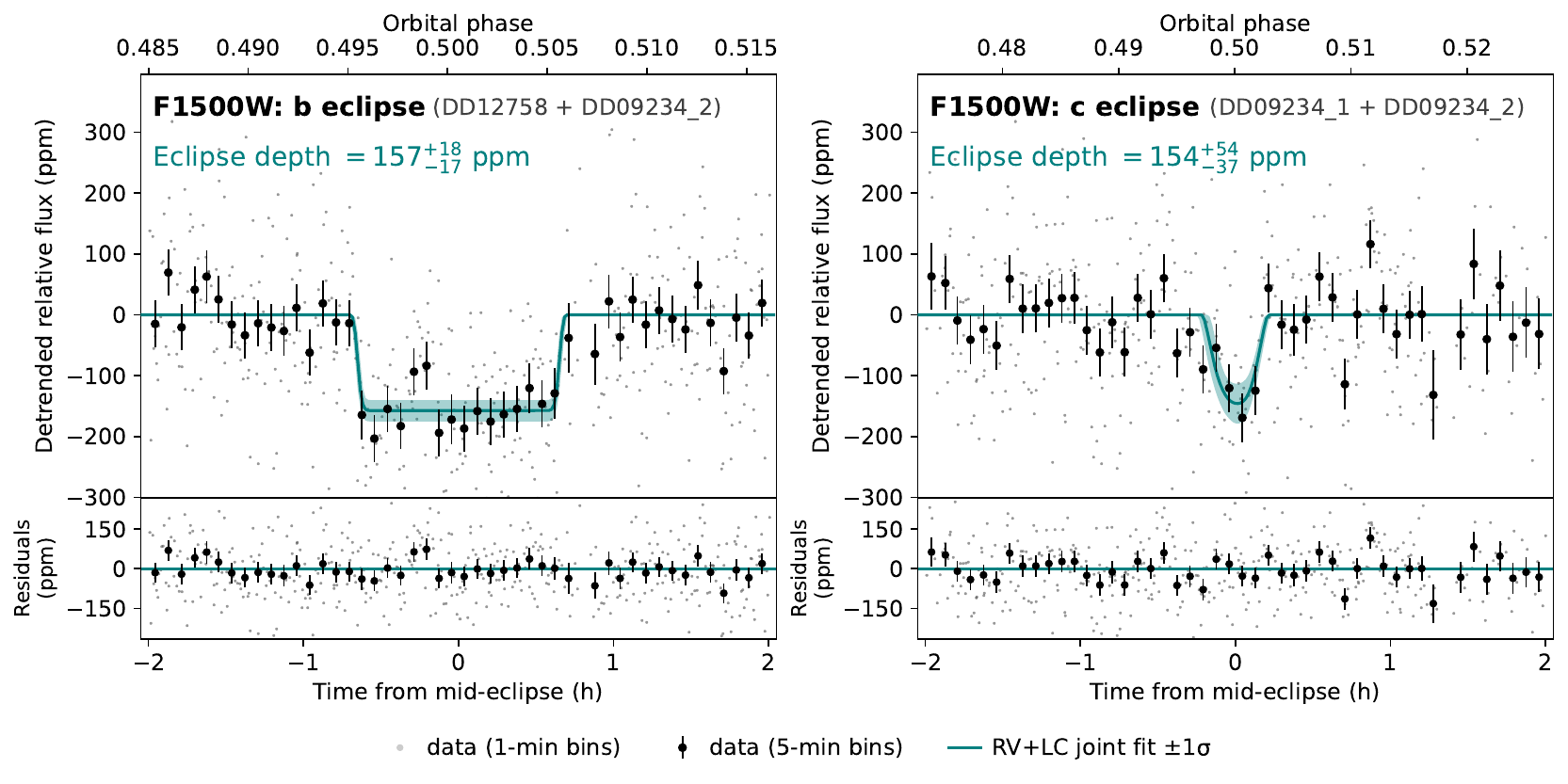}
    \caption{The phase-folded, de-trended light curve fits of LTT~1445~A~b (left) and LTT~1445~A~c (right), together with their residuals from the final RV+LC joint fit.   The data binned to 5-minute bins for visual clarity are shown as black points, while the 60\,s-binned data used for the joint fit are shown as grey dots.  The binned data include jitter.  The overlapping transit and eclipses are also removed for each de-trended light curve.  Because planet c has a grazing eclipse, the actual occultation in the light curve is slightly shallower than the best-fit eclipse depth.  The full light curves are shown in \autoref{fig:joint_lightcurves}. }
    \label{fig:joint_lightcurves_folded}
\end{figure*}

\begin{figure}
    \centering    \includegraphics[width=0.98\columnwidth]{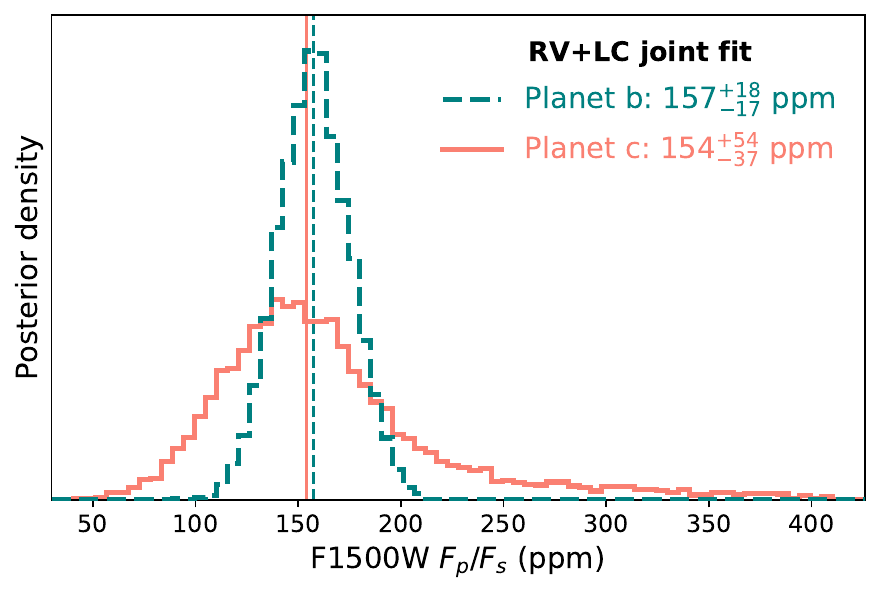}
    \caption{The posterior F1500W eclipse depths of the two planets from the final RV+LC joint fit, for planet b (dashed) and planet c (solid).  The medians are shown as vertical lines.  Planet c shows a long upper tail due to the grazing geometry. }
    \label{fig:joint_posterior_fpfs}
\end{figure}

We show the phase-folded, de-trended light curve fit from the final RV+LC joint fit for each planet in \autoref{fig:joint_lightcurves_folded}, wherein the overlapping transit and eclipse have also been removed.  Our final RV+LC joint fit gives $F_\mathrm{p}/F_\mathrm{s}= 157^{+18}_{-17}$\,ppm for planet b and $F_\mathrm{p}/F_\mathrm{s}= 154^{+54}_{-37}$\,ppm for planet c (\autoref{fig:joint_posterior_fpfs}).  The posterior for c in particular shows a long upper tail due to the grazing eclipse geometry.  

Each reduction and fit, including the final joint fit, makes a number of different choices, e.g. which portions of the light curves are masked, how priors on orbital parameters that govern grazing or on LD coefficients are chosen, or how systematics are handled.  We summarize these choices in Table \ref{tab:reductions} in Appendix \ref{app:compare_red}.   Regardless, the resulting eclipse depths are broadly consistent, close enough to ultimately not lead to meaningfully different model interpretations.   We adopt the joint fit values as our eclipse depths in subsequent analysis to use self-consistent system parameters in our models. 

\begin{table}[]
    \centering
    \caption{Summary table showing key parameters.  The full posterior distributions of fitted and derived parameters are shown in Table \ref{tab:jointfit} in Appendix \ref{app:global_posterior} and a comparison of all reductions and prior choices is summarized in Table \ref{tab:reductions} in Appendix \ref{app:compare_red}.  We adopt the values from the final RV+LC joint fit for subsequent analyses. }
    \begin{tabular}{@{}l@{\hspace{1.8em}}l|c|c}
    \hline
    & Parameter & b & c \\ \hline\hline
    \multicolumn{2}{@{}l|}{\textit{JI reduction}} & & \\
    & $F_\mathrm{p}/F_\mathrm{s}$ (ppm) & $149^{+16}_{-17}$ & $182\pm30$ \\
    \multicolumn{2}{@{}l|}{\textit{MF reduction}} & & \\
    & $F_\mathrm{p}/F_\mathrm{s}$ (ppm) & $148\pm19$ & $154^{+48}_{-35}$ \\
    \multicolumn{2}{@{}l|}{\textit{NA reduction}} & & \\
    & $F_\mathrm{p}/F_\mathrm{s}$ (ppm) & $134^{+16}_{-17}$ & $165\pm32$ \\ 
    \multicolumn{2}{@{}l|}{\textit{\textbf{RV+LC Joint fit}}} & & \\ 
    & $F_\mathrm{p}/F_\mathrm{s}$ (ppm) 
      & $157^{+18}_{-17}$ & $154^{+54}_{-37}$ \\ 
    & $R_\mathrm{p}$ ($R_\oplus$) & $1.296^{+0.027}_{-0.028}$ & $1.078\pm0.024$ \\
    & $M_\mathrm{p}$ ($M_\oplus$) & $2.38^{+0.17}_{-0.16}$ & $1.20^{+0.12}_{-0.11}$ \\
    & $T_\mathrm{eq}$ (K) & $425\pm19$ & $509\pm23$ \\
    & $e$ & $0.0026^{+0.0059}_{-0.0015}$ & $0.0028^{+0.0099}_{-0.0023}$ \\
    & $b_\mathrm{tr}$ & - & $0.9621^{+0.0012}_{-0.0011}$\\
    & $b_\mathrm{ecl}$ & - & $0.9630^{+0.0174}_{-0.0068}$ \\ \hline
    \end{tabular}
    \label{tab:global_fit_short}
\end{table}

We show the eclipse depths and key parameters from the joint fit in Table~\ref{tab:global_fit_short}.  To compare to previous literature values, planet b has radius $R_\mathrm{p}=1.296^{+0.027}_{-0.028}~ R_\oplus$, consistent with all previous literature values to within $2\,\sigma$ \citep{winters22_a, lavie23_planetary, oddo23_characterization}, while improving the precision by roughly a factor of 2.2.  Planet b has $e=0.0026^{+0.0059}_{-0.0015}$; this is consistent with the results of \citet{wachiraphan25_the} from LRS, who also found a nearly-circular orbit. 


Planet c has radius $R_\mathrm{p}=1.078\pm0.024 R_\oplus$, improving the precision by a factor of 2.3.  Its impact parameters $b_\mathrm{tr}=0.9621^{+0.0012}_{-0.0011}$ and $b_\mathrm{ecl}=0.9630^{+0.0174}_{-0.0068}$ suggest a grazing transit probability of 28\% and grazing eclipse probability of 52\%, but the actual extent to which it is grazing is relatively small; we discuss this further in \S\ref{subsec:grazing}.  

For masses, we find $M_\mathrm{p} = 2.38^{+0.17}_{-0.16} M_\oplus$ for planet b and $M_\mathrm{p}=1.20^{+0.12}_{-0.11} M_\oplus$ for planet c, improving the precision from previous literature values by a factor of 1.6-1.8.  Correctly accounting for the 90-day signal due to stellar rotation, robustly detected via leave-one-out analysis (see Appendix \ref{sec:rv_gp}), is critical for accurately constraining the masses; without it we find masses that are as much as $\sim1.8\sigma$ larger. 

The full posterior of fitted parameters and derived quantities are in Table \ref{tab:jointfit} in Appendix \ref{app:global_posterior}.

\section{Comparison to atmosphere and surface models} \label{sec:models}

To interpret the inferred eclipse depths, we first calculate the inferred brightness temperatures $T_\mathrm{B}$ and compare them to the expected maximum dayside temperature $T_\mathrm{max}$.  We then perform two types of modeling using radiative-convective equilibrium (RCE) atmosphere and surface models.  First, we perform a Bayesian climate-constrained grid-based retrieval (``grid-trieval'') on the F1500W and MIRI LRS eclipse depths of \lttab{} (\S \ref{subsec:retrievals}).  Then, we explore more atmosphere and surface possibilities with a suite of forward models for each planet (\S \ref{subsec:jegugmodels}). 

\subsection{Brightness and effective temperatures} \label{subsec:teff}

We convert each eclipse depth from the RV+LC joint fit to a brightness temperature, $T_\mathrm{b}$, by finding the uniform dayside blackbody temperature that matches it, using the F1500W throughput and a \texttt{SPHINX} stellar model \citep{iyer23}.  We use the \texttt{SPHINX} stellar spectrum grid interpolated to the median values of stellar parameters in \citet{winters22_a}: $T_\mathrm{eff}=3340\pm150$K, [M/H]$=-0.34\pm0.09$, $\log{g}=4.97^{+0.06}_{-0.08}$, and C/O$=0.5$.  We propagate the uncertainty in the stellar parameters to $T_\mathrm{b}$ via bootstrapping.  We compare $T_\mathrm{b}$ to the maximal blackbody ($f=2/3, A_\mathrm{B}=0$) temperature, $T_\mathrm{max}$, to calculate the brightness temperature ratio $\mathcal{R}$.  We do this per sample, such that both $T_\mathrm{B}$ and $T_\mathrm{max}$ have the propagated uncertainty from $T_\mathrm{eff}$ but the uncertainty is not double counted in $\mathcal{R}$.  We unpack the measured $\mathcal{R}$ value into constraints on Bond albedo and heat redistribution, using Eqn.\,3 of \citet{mansfield24}; we discuss the validity of this calculation for a single photometric measurement in \S \ref{subsec:lrs_15}. 

\subsubsection{\lttab{}}

For \lttab{}, we find $T_\mathrm{B,15}=528^{+39}_{-29}$\,K, consistent with the LRS value of $T_\mathrm{B,LRS}=524\pm19$\,K \citep{wachiraphan25_the}.  Combined with the LRS spectrum of 16 bins, we find $T_\mathrm{eff}=526 \pm 17$\,K.  The best-fit blackbody describes the panchromatic data well with $\chi^2=17.6$ at $\mathrm{dof}=16$.   

\begin{figure*}
    \centering
    \includegraphics[width=0.49\linewidth]{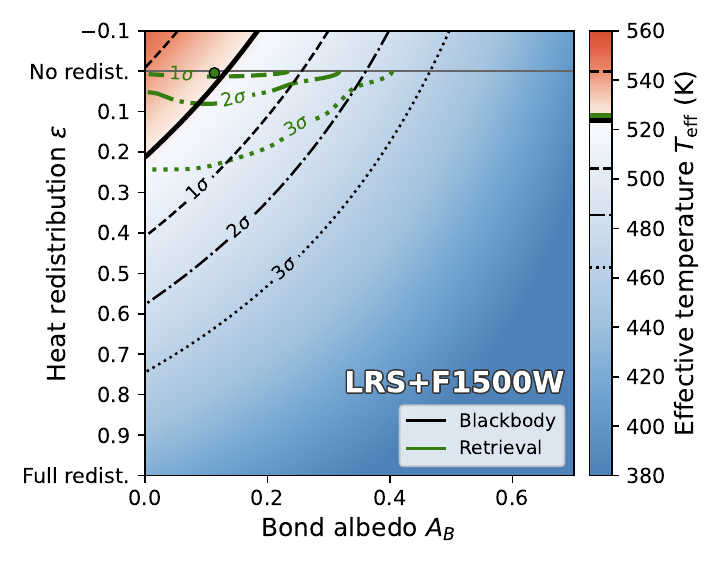}
    \includegraphics[width=0.49\linewidth]{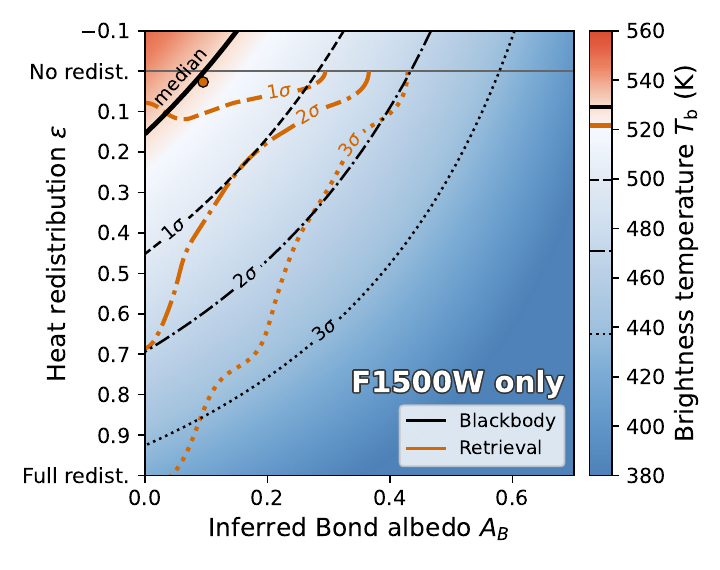}
    \caption{Constraints on heat redistribution $\epsilon$ and Bond albedo $A_\mathrm{B}$ of \lttab{}, using the brightness temperature jointly constrained by combined LRS+F1500W (left) and F1500W only (right).  The brightness temperatures and its 1, 2, and $3\sigma$ uncertainties are shown in the contour and colorbar.  The upper axis limit $\epsilon=0$ corresponds to no heat redistribution ($f=2/3$), and the lower limit $\epsilon=1$ corresponds to full heat redistribution ($f=1/4$); \textit{inferred} $\epsilon<0$ corresponds to thermal beaming.  The Bayesian 1-, 2-, and 3-$\sigma$ constraints on $\epsilon$ and $A_\mathrm{B}$ from the retrievals (see \autoref{fig:retrieval}), which considers the spectral effects of gas absorption and constrains both $\epsilon$ and $A_\mathrm{B}$, are shown as color contours with the medians as markers.}
    \label{fig:fab}
    
    \vspace{3em}
    
    \includegraphics[width=0.99\textwidth]{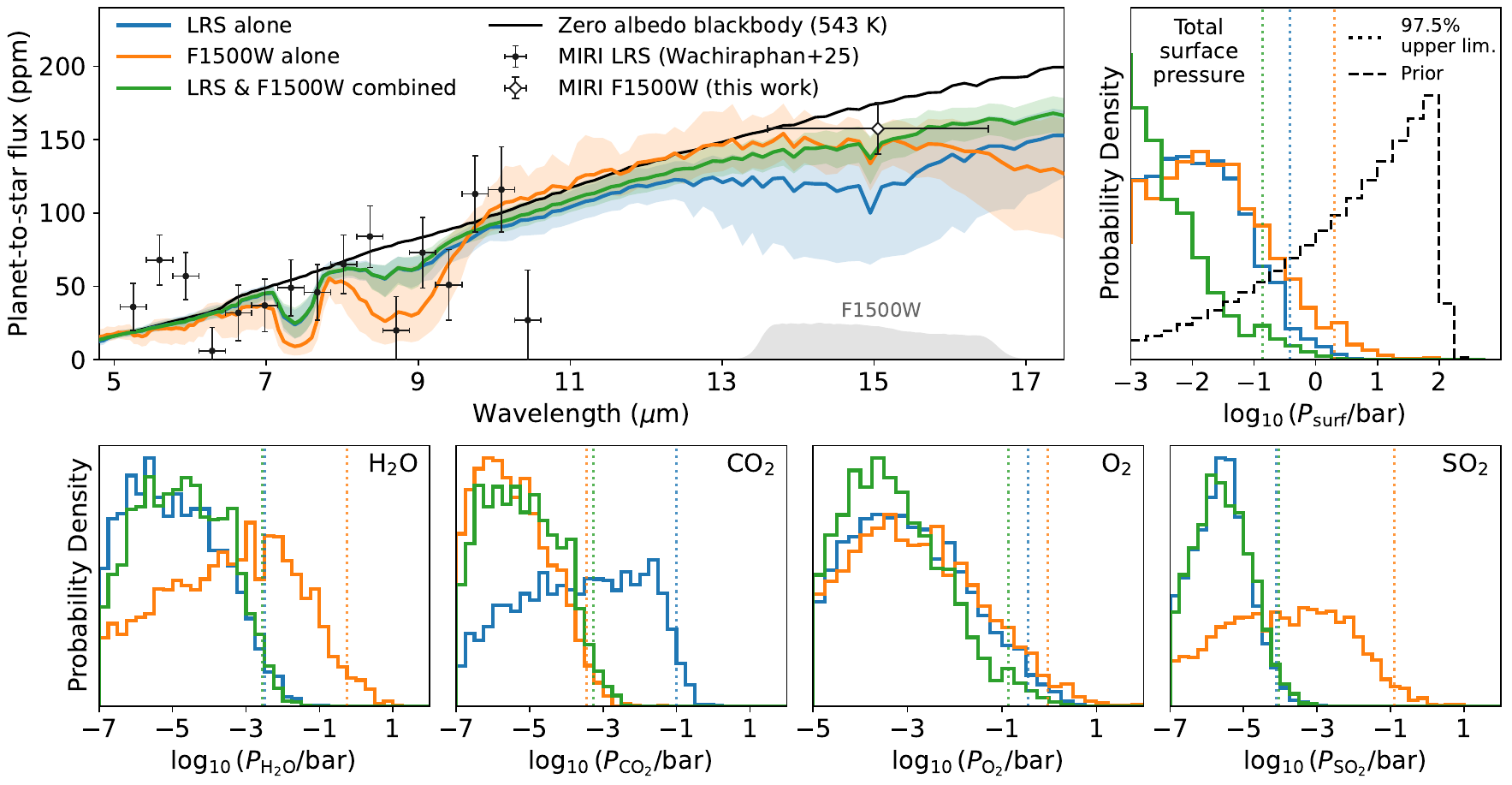}
    \caption{Climate-constrained Bayesian fits to the existing MIRI LRS spectrum \citep[blue;][]{wachiraphan25_the}, the new F1500W photometry (orange; this work), and to both of the datasets combined (green). Model spectra in the top left panel are the median and 1$\sigma$ fits to the data. Vertical dotted lines are 97.5\% upper limits for the total pressure and the partial pressure of each gas. The total pressure is derived from the sum of the partial pressures and hence has an implicit non-uniform prior. When considered in isolation, the MIRI LRS observations alone are compatible with up to $\sim0.1$ bar CO$_2$. Adding the F1500W photometry decreases the CO$_2$ upper limit by about two orders of magnitude to $\lesssim6\times10^{-4}$ bar at 97.5\% posterior probability.}
    \label{fig:retrieval}
\end{figure*}

The maximal blackbody temperature with propagated uncertainty from orbital parameters and stellar effective temperature is $T_\mathrm{max}=544\pm24$\,K; F1500W gives a brightness temperature ratio ($T_\mathrm{B}/T_\mathrm{max}$) at 15 micron of $\mathcal{R}_{15}=0.96\pm0.05$ and the combined data gives an effective temperature ratio of $\mathcal{R}_\mathrm{eff} = 0.97\pm0.04$, consistent with the dark, bare rock limit.   We show the Bond albedo $A_\mathrm{B}$ and heat redistribution factor $f$ (reparameterized as $\epsilon=(12/5)(2/3-f)$, as per \citet{cowan11}) contours in Figure~\ref{fig:fab}, following previous LRS observations in e.g. \citet[][]{mansfield24}.  

While brightness temperature inferred from eclipse observations generally depend on the stellar spectrum used \citep{parkcoy25_population}, the stellar spectrum interpolated using the \texttt{BT-Settl} grid \citep{allard_models_2012} has a 3\% brighter flux in the F1500W bandpass and leads to negligible differences from using the \texttt{SPHINX} model.  This is unsurprising as the discrepancies are the largest for late M stars \citep{fauchez25}, while LTT~1445~A is an M2.5V star.

\subsubsection{\lttac{}}

For \lttac{}, we infer a 15-micron brightness temperature of $T_\mathrm{B,15} = 625^{+109}_{-79}$\,K.  The grazing geometry leads to a long upper tail on the eclipse depth and hence $T_\mathrm{B,15}$; the 95.5\% uncertainties on $T_\mathrm{B,15}$ are $^{+352}_{-154}$\,K and 99.7\% widths are $^{+495}_{-232}$\,K.  The maximal blackbody temperature is $T_\mathrm{max}=650\pm29$\,K, resulting in a brightness temperature ratio $\mathcal{R} = 0.96^{+0.17}_{-0.12}$.   While this is consistent with the dark, bare rock limit to within $<0.3\sigma$, the uncertainties are large enough to preclude drawing any conclusion about the planet's true nature from this alone.

\subsection{Bayesian grid-trieval models} \label{subsec:retrievals}

\subsubsection{Model description}

For \lttab{}, we perform a climate-constrained Bayesian retrieval, using the model and methods described in \citet{wogan26_a}.  The approach fits the data by linearly interpolating a grid of pre-computed spectra for atmospheres at radiative-convective equilibrium.  Each temperature-pressure profile was computed with the climate module of the \texttt{Photochem} software package \citep{wogan25_photochem}, then corresponding emission spectra were computed with \texttt{PICASO} \citep{batalha19_picaso}.  The grid spans a mixture of partial pressures for CO$_2$, H$_2$O, SO$_2$ and O$_2$ as well as a range of T$_\mathrm{eq}$, surface albedo, and an uncertain heat re-distribution parameter \citep{koll22}.  The model grid was computed using the planet radius and mass for planet b from \citet{pass23_hst} ($1.34 R_\oplus$ and $2.73 M_\oplus$). We find that adopting the updated radius and mass values presented in this paper results in changes to the predicted F1500W eclipse depths by less than 1 ppm for the atmospheres tested, which is negligible compared with the observational uncertainty.

Interpolating the spectra grid yields the planet emission flux (i.e., $F_{p,\lambda}$) as a function of the seven parameters described above. That flux is then combined with a stellar spectrum (i.e., $F_{\star,\lambda}$) and planet to star radius ratio (i.e., $R_p/R_\star$) to compute the planet to star flux ratio: $\delta_\lambda = (F_{p,\lambda}/F_{\star,\lambda})(R_p/R_\star)^2$. Model F1500W eclipse depths are computed as the ratio of the planet and stellar fluxes integrated over the instrumental throughput with photon-counting weights, scaled by $(R_p/R_\star)^2$.  We use the same \texttt{SPHINX} stellar models adopted for the brightness-temperature calculations in Section~\ref{subsec:teff}. Among the stellar parameters in \texttt{SPHINX}, only the uncertainty in $T_\mathrm{eff}$ meaningfully affects the predicted eclipse depths. We therefore sample $T_\mathrm{eff}$ while holding the other stellar parameters fixed.

We use \texttt{PyMultiNest} \citep{buchner_x-ray_2014}, with 1000 live points, to fit the nine-parameter model to the LRS spectrum, F1500W photometry, and their combination, assuming Gaussian LRS errors and a two-piece Gaussian approximation to the slightly asymmetric F1500W uncertainty in the RV+LC joint fit ($157^{+18}_{-17}$ ppm). Priors follow \citet{wogan26_a}, except that we construct a correlated prior for $T_\mathrm{eq}$, $T_\mathrm{eff}$, and $R_p/R_\star$ from the final RV+LC joint system-fit posterior (Table \ref{tab:jointfit}).

\subsubsection{Retrieval results}

We perform a grid-retrieval fit to the combined LRS and F1500W spectrum, as well as each instrument independently. Figure~\ref{fig:retrieval} shows each fit compared to the data (top left), as well as the marginalized posterior distributions for the total pressure (top right), and each gas (bottom). No molecule is detected. All three datasets are consistent with an airless surface, while the combined observations constrain any atmosphere to be thin.

When considered in isolation, the LRS observations rule out H$_2$O and SO$_2$ pressures above $\sim3\times10^{-3}$ and $\sim8\times10^{-5}$ bar, respectively, at 97.5\% posterior probability, while permitting CO$_2$ and O$_2$ pressures as high as $\sim0.1$ and $0.4$ bar. Here, O$_2$ is meant to represent the sum of all optically thin, yet plausible gases like N$_2$, CO, or O$_2$ \citep{wogan26_a}. These LRS-only constraints are broadly consistent with \citet{wogan26_a}, which analyzed the same dataset, but with different system parameters.

The new F1500W photometry rules out the possibility of a $\sim0.1$ bar CO$_2$ atmosphere permitted by the LRS observations. Combining both datasets constrains the CO$_2$ partial pressure to $<6\times10^{-4}$ bar at 97.5\% posterior probability, tightening the LRS-only upper limit by more than two orders of magnitude. The combined observations also constrain the total surface pressure to $\lesssim0.1$ bar at the same probability level. If the planet had a thin atmosphere, this analysis suggests it would be optically thin and composed of a gas like N$_2$, CO, or O$_2$. An airless rock is also a reasonable explanation of the spectra. Figure~\ref{fig:retrieval} uses the 16-bin LRS spectrum; repeating the combined LRS + F1500W analysis with the 8-bin spectrum from \citet{wachiraphan25_the} yields very similar conclusions, with a total surface-pressure upper limit of $\sim0.1$ bar in both cases at 97.5\% posterior probability.

We show the derived posteriors on heat redistribution $f$ and Bond albedo $A_\mathrm{B}$ from the retrieval on \autoref{fig:fab}.  Because the retrieval uses the spectral information to rule out atmospheres, with $f$ and $A_\mathrm{B}$ calculated per sample, it is more constraining than simply unpacking the inferred brightness temperature.  This is true even of the F1500W-only retrieval as well, albeit to a lesser extent; the retrieval is able to rule out e.g. thin CO$_2$ atmospheres that do not strongly redistribute heat but cause spectral absorption, which leads to a tighter bound on $f$.

\subsection{Suite of forward models} \label{subsec:jegugmodels}

\subsubsection{Model description}

In addition to the Bayesian retrievals performed in \S \ref{subsec:retrievals}, we also model various atmospheres and surfaces for planets b and c.  We use the RCE code \helios{} \citep{malik17_helios, malik19_self, whittaker22}, which has been widely used for the interpretation of numerous emission observations of rocky exoplanets \citep[e.g.,][]{whittaker22, ih23_constraining, mansfield24, xue24, parkcoy25_population, august25_hot, meier2025_toi1468b, fortune25_lhs1140c, allen25_hot, holmberg26_hot, xue25_the, zhang24}. 

To calculate the day-to-night heat redistribution self-consistently, \helios{} uses the scaling approximation for heat redistribution from \citet{koll22} that depends on the longwave optical depth $\tau_\mathrm{LW}$, surface pressure, and temperature.  We noticed that \helios{} had in fact used the scaling in the preprint \texttt{v1} version \citep{Koll2019} (also presented in a companion paper \citep{koll19_identifying}), which had a slightly different dependence on the longwave optical depth $\tau_\mathrm{LW}$.  Given that numerous works have used this code, we quantify the relative error incurred in Appendix \ref{app:redistribution} as an all-inclusive erratum.  The models in this work uses the updated scaling.  

For atmospheres, we consider gas absorption and scattering from CO$_2$, H$_2$O, CH$_4$, SO$_2$, O$_2$, and N$_2$.  We now consider scattering from CH$_4$ and SO$_2$ as well, previously not included in \helios{}, using data and methods from \citet{he21_scattering} and \citet{ranjan17}.  We  include collision-induced absorption (CIA) between included pairs using the data from the HITRAN database \citep{terragni25}.  We note in particular that CO$_2$-CO$_2$ CIA features are present in the mid-infrared range at 7-8 micron, which both create spectral features in the LRS bandpass and can alter the temperature-pressure (T-P) profile by of order $\sim 60$ K at deeper pressures (in LTT~1445~A planets).  We generated correlated-$k$ tables at the same wavelength grid as the \texttt{SPHINX} models ($R=250$ between 0.1--20 $\mu$m) to avoid re-binning the pre-binned stellar model grid and used coarser bins of $R=50$ out to 500 $\mu$m, resulting in 1488 spectral bins with 20 Gaussian points for iterative runs.  We assume $T_\mathrm{int}=0$\,K (see \S \ref{subsec:tidal_heating}).   We used opacity sampling of $R=10^5$ for post-processing a converged T-P profile to generate the emergent eclipse spectrum, which we then bin down by using the instrument throughput weighted by the \texttt{SPHINX} spectrum.

We also explore the effects of aerosols. Optically thick aerosol layers are one source of degeneracy for \textit{transmission} observations, as they can have the same signature on the spectrum as a bare surface.  For this reason, we examine whether the eclipse observations can rule out these scenarios.   We consider the following species: a sulfuric acid (H$_2$SO$_4$) cloud, soot and Tholin haze, and dust decks using Earth's ash and Mars' dust using the MRO data \citep{palmer75, khare84_optical, deguine20_complex, wolff09}.  While clouds and hazes have been generally understood to lead to shallow eclipse depths due to their reflectivity, they can be potentially more absorptive around M stars whose emission peak in the IR \citep{steinrueck25, maurel25_constraints}.  Here, we also found and corrected a bug in \helios{} in which the asymmetry parameter ($g_0$) was not read in correctly, leading to overestimating the reflectivity for larger particles.  Except for sulfuric acid cloud, we simply prescribe a range of plausible vertical distributions of each aerosols, based on the optical depth; we test a range of aerosol thickness from $\tau=0.1$ to $\tau=10$ at a reference wavelength of 10 $\mu$m.

For surfaces, we model bare rock emission using the powder data from \citet{paragas25} in the sub-surface temperature gradient model from \citet{lyu26_subsurface} accounting for surface roughness. 
Our sub-surface model closely follows \citet{lyu26_subsurface} but is different in two ways.  First, we calculate the wavelength-dependent skin depth $d_\lambda$ from the extinction cross sections \citep{henderson97_nearsurface2} rather than from imaginary refractive indices only \citep{henderson94_nearsurface1}, which means we also consider scattering in the $d_\lambda$ rather than pure absorption.  This generally has the effect of reducing $d_\lambda$ contrast with wavelength.  Secondly, we account for how surface roughness affects the distribution of incident angles, rather than assuming a smooth sphere.  Surface roughness creates a preferential visibility of hot facets that direct more emission back towards the star (and hence the eclipse observer), leading to thermal beaming \citep{spencer90_rough, lagerros98_thermal, rozitis11}.  Surface roughness is described by the RMS slope of a Gaussian slope distribution, defined as $\tan(\mathrm{RMS})=\langle\tan^2(\theta)\rangle^{1/2}$ for facets of tilt $\theta$ \citep{bandfield15}.  The RMS sets how much of the surface lies at each tilt, which determines each facet's illumination; RMS=0 is equivalent to a smooth sphere, while thermal-infrared observations of the Moon indicate RMS slopes of 20-35\,deg \citep{rozitis11, bandfield15}.

\subsubsection{\lttab}

\begin{figure*}
    \centering
    \includegraphics[width=0.97\textwidth]{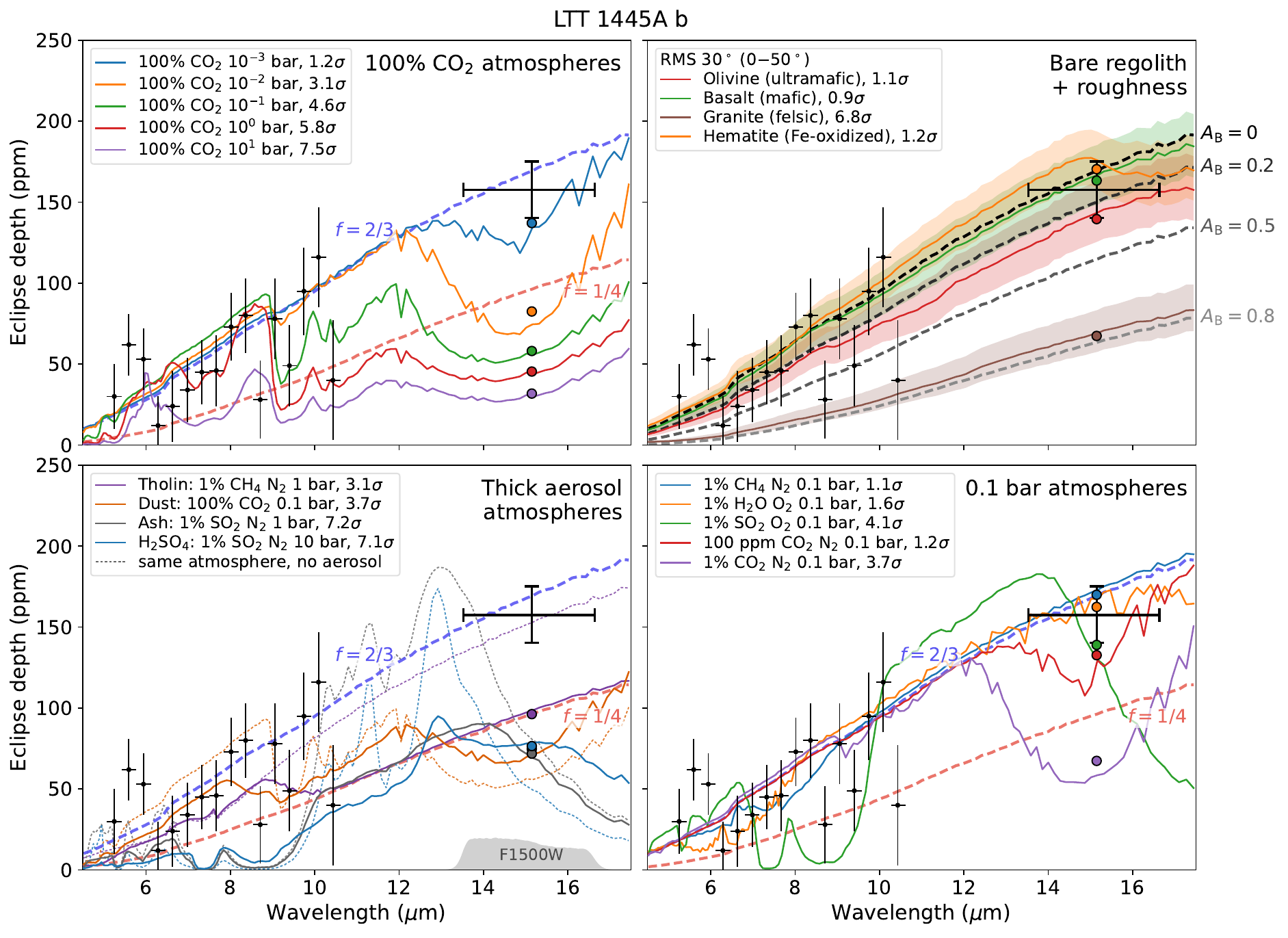}
    \caption{The observed eclipse spectrum of \lttab{} against various \helios{} forward models of atmospheres and surfaces.  Each panel focuses on: bare surfaces (\textit{top right}), pure CO$_2$ atmospheres of varying pressures (\textit{top left}), models with different types of aerosols (\textit{bottom left}), and 0.1 bar atmospheres with single absorbers (\textit{bottom right}).  The $\sigma$ values are calculated from a two-sided $\chi^2$ distribution with $\mathrm{dof}=17$.  The surface models show a range of surface roughness from RMS 0--50 deg with the center at 30 deg, wherein RMS=0 corresponds to a smooth sphere.  We show the F1500W bandpass function weighted by the stellar spectrum.  The blackbody emission at the predicted temperature is shown for a grid of Bond albedo values for bare surfaces and no redistribution ($f=2/3$) and full redistribution ($f=1/4$) for atmospheres, which do not consider any spectral effects; the small wiggles in these curves are from the stellar spectrum.  We do not show this plot in brightness temperature as the 1$\sigma$ uncertainties in the LRS eclipse spectrum are too large.}
    \label{fig:helios_b}
\end{figure*}

For planet b, we focus on scenarios not yet considered in \S~\ref{subsec:retrievals}: various bare surfaces, reducing atmospheres, and atmospheres with aerosols.  
We show a representative set of surface and atmosphere models in \autoref{fig:helios_b}.  We calculate the goodness-of-fit ($\chi^2$) for each model and convert it to equivalent standard errors using a two-sided $\chi^2$ distribution of $\mathrm{dof}=17$.  We find that, a range of bare surface models are consistent, and atmospheres need be thin or transparent to remain consistent.  

Among the bare surface models, the best-fit surface models are basalt with $\mathrm{RMS}=20$\,deg ($0.8\sigma$) and olivine with $\mathrm{RMS}=40$\,deg ($0.8\sigma$), while (a priori unlikely) granite surface is clearly ruled out ($>6\sigma$).  The best-fit inferred Bond albedo is $A_\mathrm{B}=0.12\pm0.09$ ($\chi^2=17.6$, $\mathrm{dof}=16$).  

Among atmospheres models, only atmospheres of thin surface pressure $\leq 0.1$\,bar with absorbers $<1$\% remain consistent with the full LRS-F1500W spectrum.  We show a representative set of 0.1 bar atmospheres, which are near the upper limit of how thick an atmosphere the retrieval analysis allows, though we now also consider CH$_4$.  Notably, the handful of points in the LRS spectrum between 6-9 $\mu$m that appear $\sim1\sigma$ below the $f=2/3$ blackbody could be explained by either CH$_4$ or H$_2$O absorption.  A 0.1 bar 1\% CH$_4$ + N$_2$ atmosphere ($1.1\sigma$) and a 0.1 bar 1\% H$_2$O + O$_2$ atmosphere ($1.6\sigma$) remain consistent with the data.  On the other hand, a 0.1 bar 1\% SO$_2$+O$_2$ atmosphere is inconsistent at $4.1\sigma$, as it predicts a shallow eclipse in the LRS wavelength range.  Among 0.1 bar atmospheres with CO$_2$, models are only consistent when CO$_2$ is less than $1\%$: a 0.1 bar 100\,ppm CO$_2$+N$_2$ atmosphere is consistent at $1.2\sigma$, while a 0.1 bar 1\% CO$_2$+N$_2$ atmosphere is clearly ruled out at $3.7\sigma$. 

For atmosphere with aerosols, we show models that contain tholin, ash, Martian dust, and sulfuric acid, in matching background atmospheres of 1\%  CH$_4$, 1\% SO$_2$, 100\% CO$_2$, and 1\% SO$_2$ to match relevant Solar System conditions.  For tholin and ash, we show a thick deck of $\tau=10$ near the surface; for Martian dust, we show a thin layer $\tau=1$ lofted at 0.05 bar.  We also show sulfuric acid cloud, whose vertical loading follows that from the clear atmosphere T-P profile crossing the condensation curve.  All of these aerosols have a bolometric cooling effect and do not completely mute the spectral feature of the absorber, which is then ruled out by the 15 $\mu$m observation.  Martian dust, in particular, actually raises the 15-$\mu$m eclipse depth by $\sim30$\,ppm while bringing the LRS bandpass down; it is still ruled out by the F1500W observation nonetheless due to the strong CO$_2$ absorption.  As the eclipse data rules out these aerosol scenarios, a featureless transmission spectrum, if observed, can be interpreted as a true bare rock rather than a thick aerosol layer. 

\subsubsection{\lttac}


\begin{figure*}
    \centering
    \includegraphics[width=0.85\linewidth]{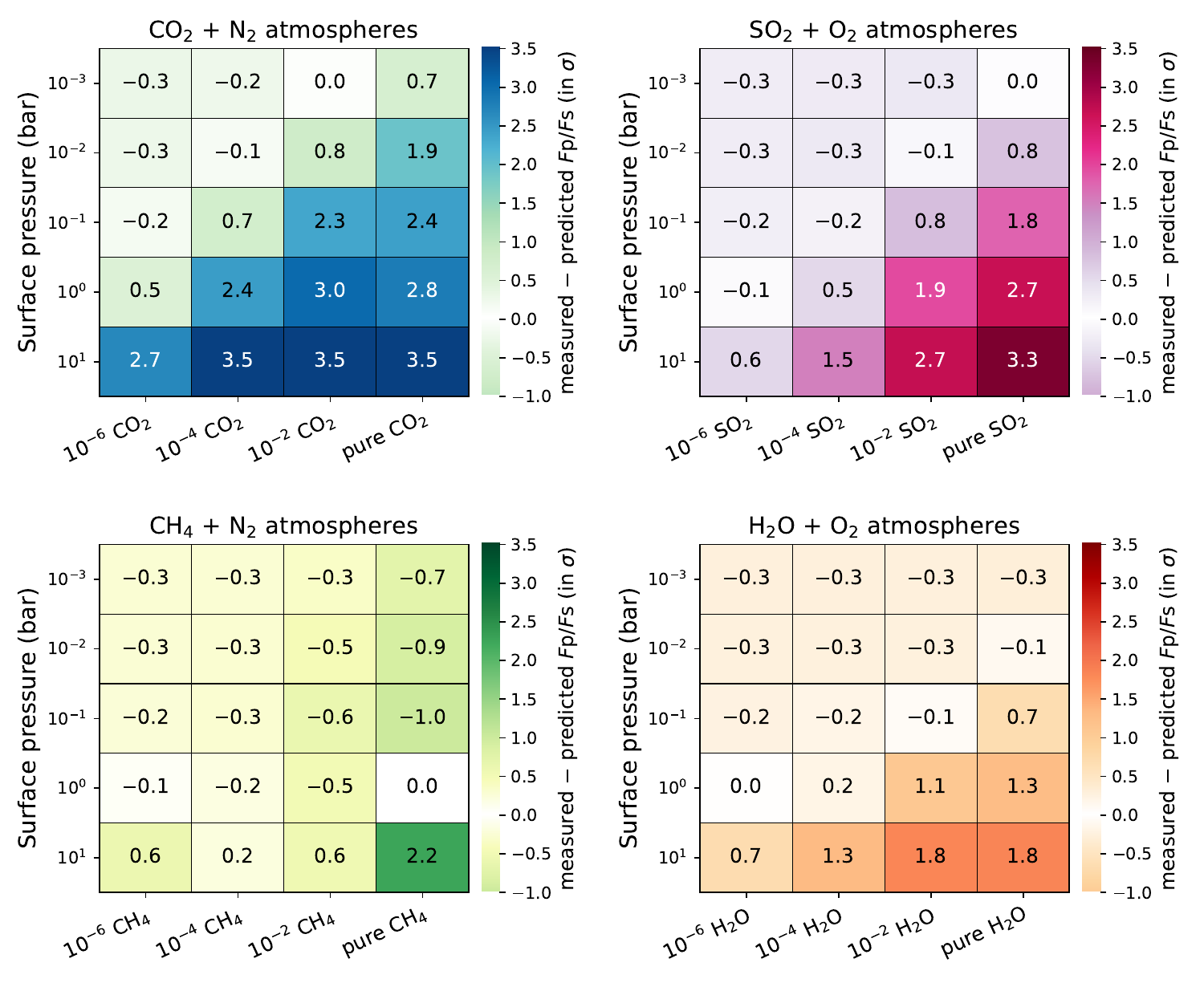}
    \caption{Agreement between the observed eclipse depth of \lttac{} and predicted eclipse depths of atmospheres containing a single absorber, one of CO$_2$, SO$_2$, CH$_4$, and H$_2$O, in units of the observed uncertainty.  In each panel, a grid of volume mixing ratio of the absorber and total surface pressure $P_\mathrm{surf}$ is shown.  A \textit{positive} value means that the observed eclipse is \textit{deeper} than the predicted model eclipse depth.  Only the thickest CO$_2$- and SO$_2$-bearing atmospheres considered are ruled out by the F1500W observations. }
    \label{fig:c_psurfco2}
\end{figure*}

F1500W observations are most effective in constraining the partial pressure of CO$_2$ and (to a lesser extent) SO$_2$ \citep{ih23_constraining, zieba23, maurel25_constraints}.  In Figure~\ref{fig:c_psurfco2}, we show how well models that contain a single gas as the sole absorber match the data in a grid of the absorber volume mixing ratio and total surface pressure $P_\mathrm{surf}$, assuming a N$_2$ or O$_2$ background. 

For CO$_2$ atmospheres, our primary interest, the data can only rule out combinations of 10 bar atmosphere of $\geq 100$\,ppm CO$_2$ to $>3\sigma$, while 1, 0.1, and 0.1 bar atmospheres of 100\,ppm, 1\%, and 100\% CO$_2$, respectively, are inconsistent at $>2\sigma$.  Models that use O$_2$ rather than N$_2$ atmospheres show minor differences due to CIA, but does not change the conclusion.   Atmospheres that have partial pressure of CO$_2$ less than $P_\mathrm{CO_2} <0.01$\,bar are all allowed.  A Venus-like 100 bar CO$_2$ atmosphere (not modeled) is clearly ruled out, while a Mars-like 0.01 bar CO$_2$ atmosphere is consistent at 1.9$\sigma$.

\begin{figure*}
    \centering
    \includegraphics[width=0.97\textwidth]{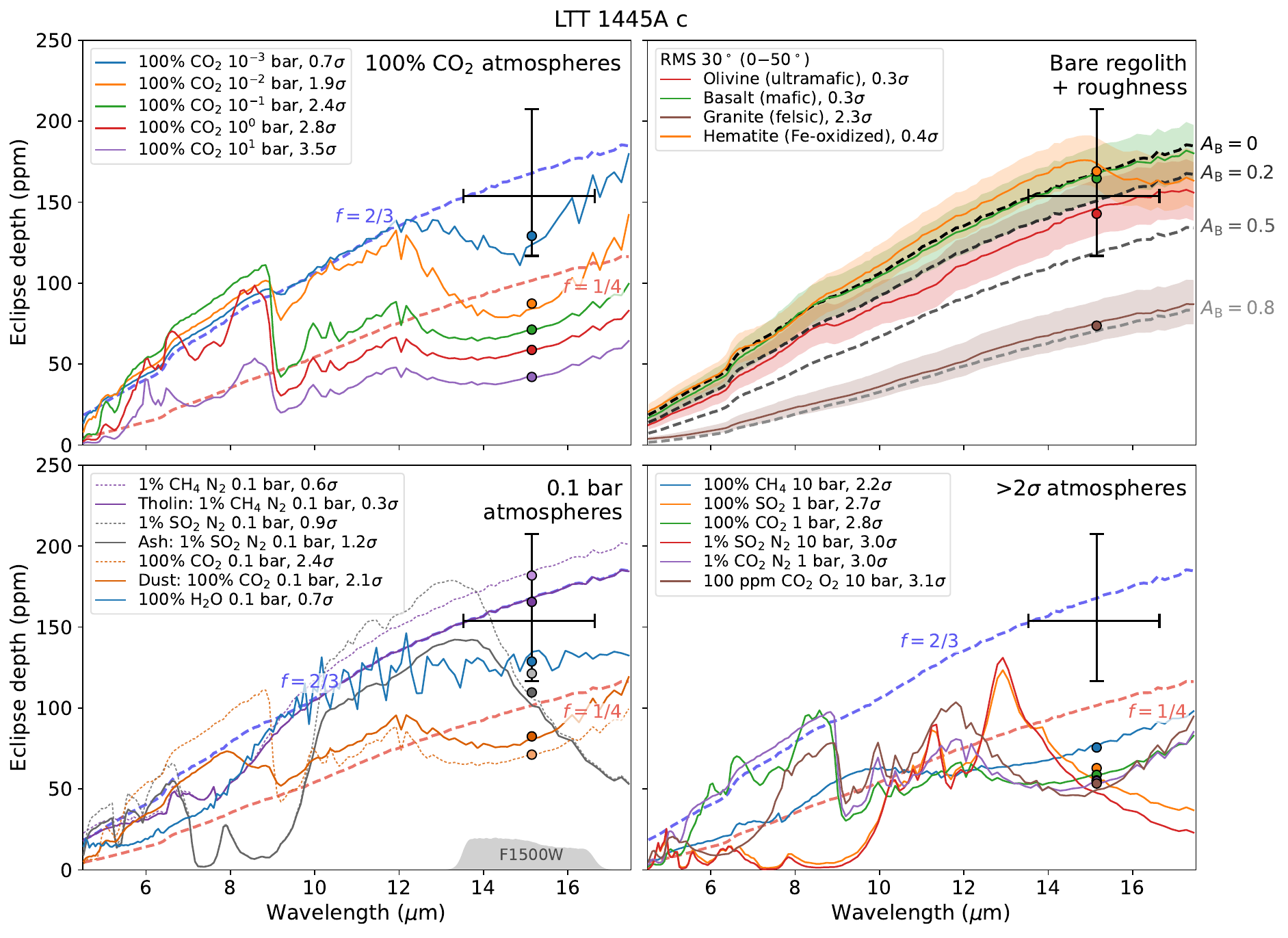}
    \caption{Same as \autoref{fig:helios_b} but for \lttac{}.  The large observed uncertainty allows for a wide range of atmosphere and surface scenarios; we show the models that disagree at $>2\sigma$ in the bottom right panel, which are generally thick atmospheres.  Note that the measured eclipse depth is slightly non-Gaussian (see \autoref{fig:joint_posterior_fpfs}); the $\sigma$ values are calculated from the equivalent percentiles to a Gaussian distribution.}
    \label{fig:helios_c}
\end{figure*}

We show a suite of models including aerosols and bare surfaces in \autoref{fig:helios_c}.  Because of the large uncertainty, a wide range of scenarios are consistent.  Among bare surface models, all regolith types considered including granite are consistent with the observation to within $2.3\sigma$ (top right panel).  Likewise, a number of 0.1 bar atmospheres are consistent, especially atmospheres that contain CH$_4$ as the main absorber, as CH$_4$ lacks absorption at these wavelengths.  The presence of aerosols of moderate optical depth ($\tau=1$) only move the eclipse depth by $<0.2\sigma$ of the observed uncertainty (bottom left).  The only atmosphere models that are ruled out by the observation generally have a large partial pressure of the absorber with spectral feature at 15 $\mu$m or a large enough surface pressure ($P_\mathrm{surf}>1$\,bar) to sustain heat redistribution (bottom right).  

We therefore deem it difficult to conclude whether planet c has a substantial atmosphere.  A bare rock ranging from dark to bright surfaces describes the observation well, but also any moderate atmosphere ($P_\mathrm{surf}\sim1$\,bar) that is not dominated by CO$_2$ or SO$_2$ does too.  The most definitive statement about planet c we can make is that it does not have a Venus-like thick CO$_2$ atmosphere.

\section{Discussion} \label{sec:discuss}

\subsection{Does planet c have grazing transit and eclipse?} \label{subsec:grazing}

Whether planet c has a grazing transit or eclipse has been uncertain in past works.  To find a definitive answer with our refined system parameters, we use the RV+LC joint fit posterior from \S \ref{sec:global_fit_results} to calculate the probability of planet c having a grazing eclipse, $P(\mathrm{grazing})$, and the occulted fractional area of the planet disk, $f_\mathrm{ecl}$.    

A planet with radius $r=R_\mathrm{p}/R_\mathrm{s}$, semi-major axis $a$, and inclination $i$ has an impact parameter
\begin{equation*}
    b = \left( \frac{a}{R_\mathrm{s}} \right) \cos i \left( \frac{1-e^2}{1\pm e\sin\omega} \right),
\end{equation*}
\noindent where the plus sign in the denominator denotes transit ($b_\mathrm{tr}$) and the minus sign eclipse ($b_\mathrm{ecl}$).  This planet has a grazing transit/eclipse if $1 - r < b < 1 + r$ \citep{winn10}.  Using the eclipse impact parameter $b_\mathrm{ecl}$, we calculate the occulting fraction at mid-eclipse, $f_\mathrm{ecl}$, defined as:
\begin{equation*}
    f_\mathrm{ecl}(b_\mathrm{ecl}) =\left( \frac{\text{Overlapping area of planet and star}}{\text{Area of the planet's disk}} \right),
\end{equation*}
at mid-eclipse, wherein ``area'' should be construed as the projected area as observed by us.   In other words, $f_\mathrm{ecl}=1$ corresponds to a full, non-grazing eclipse, and $f_\mathrm{ecl}=0$ corresponds to a \textit{not-even-grazing} non-eclipse.  We calculate $f_\mathrm{ecl}$ following the methods of \citet{mandel02}.   We convert each sample in the posterior of the RV+LC joint and JI \eureka{} fit and calculate the fraction of posterior samples that meet the grazing condition, $P(\mathrm{grazing})$, and the distribution of $f_\mathrm{ecl}$.  We also similarly define $f_\mathrm{tr}$ for mid-transit and find its distribution.  

\begin{figure*}
    \centering
    \includegraphics[width=0.95\linewidth]{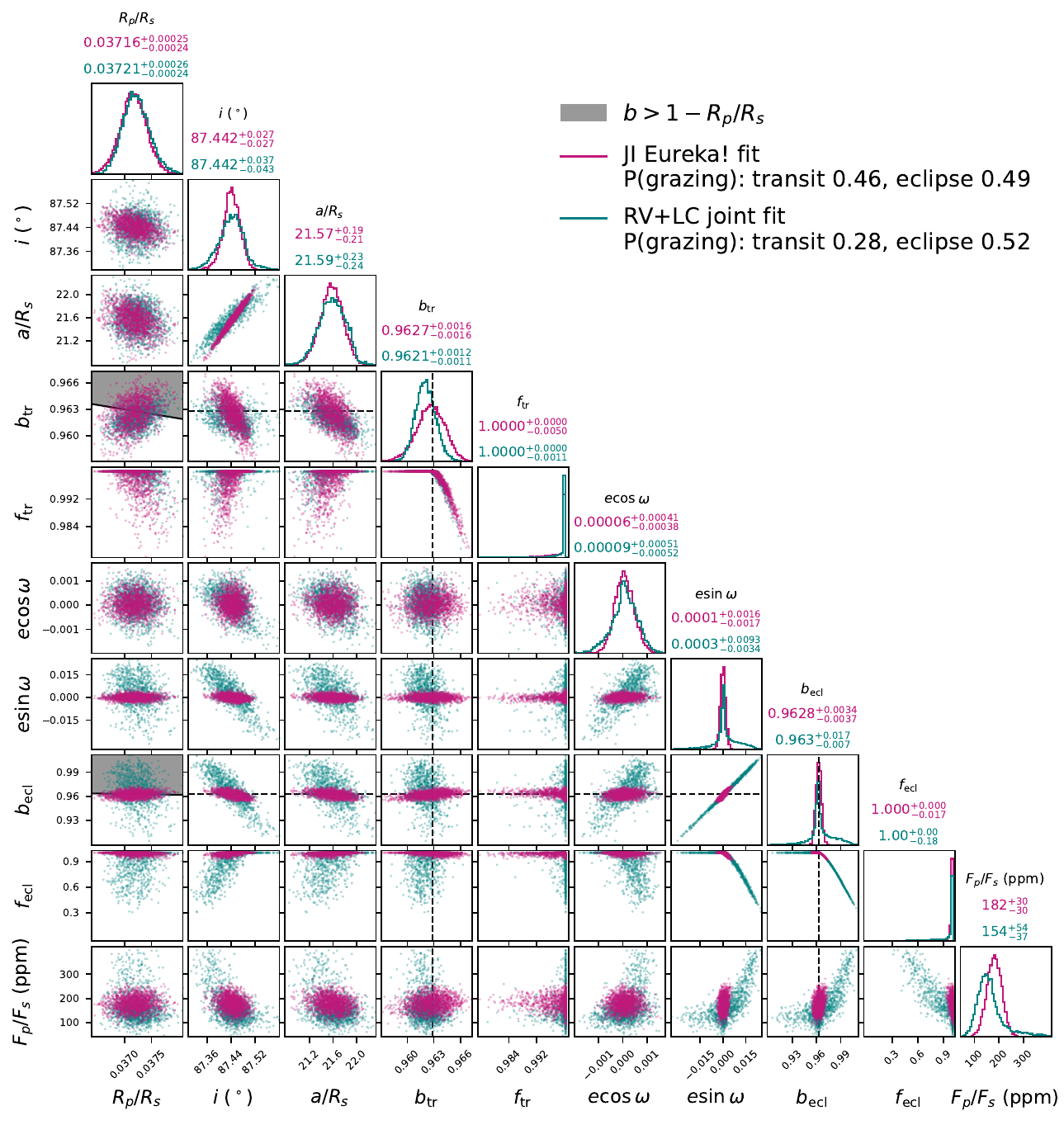}
    \caption{Posterior distributions of orbital parameters, occulted area fractions $f_\mathrm{tr}$ and $f_\mathrm{ecl}$, and the eclipse depth of \lttac{}.  The distribution from the final RV+LC joint fit (\S \ref{sec:global_fit_results}) is shown in teal, and the JI \eureka{} fit (\S \ref{sec:jegug_red}) to all three visits is plotted in purple.   The grey regions in $R_\mathrm{p}/R_\mathrm{s}$-$b$ panels show where grazing transit and eclipse occur.  The grazing probabilities for transit and eclipse are calculated from comparing $b_\mathrm{tr}$ and $b_\mathrm{ecl}$ with $1-R_\mathrm{p}/R_\mathrm{s}$, while the occulted area fractions take the geometry of overlapping circles into account.  A full, non-grazing eclipse corresponds to $f_\mathrm{ecl}=1$.   Planet c has a 52\% chance of having a grazing eclipse and the 97.5\% lower limit on $f_\mathrm{ecl}$ is 53\%.  The \eureka{} fit set a narrower prior on $e_c\sin\omega_c$ and forced $b_\mathrm{tr} \simeq b_\mathrm{ecl}$ based on tidal damping timescales (\S \ref{subsec:tidal_heating}); this leads to a tighter lower limit on $f_\mathrm{ecl}$ of 96\% instead. }
    \label{fig:c_graze}
\end{figure*}

We show the dependence of $f_\mathrm{ecl}$ on the orbital parameters and its effect on the inferred eclipse depth of planet c in Figure~\ref{fig:c_graze}.  In the RV+LC joint fit, we find that planet c has a grazing transit probability of $P(\mathrm{grazing})=0.28$ but a 97.5\% (one-sided $2\sigma$) lower limit on  the grazing fraction $f_\mathrm{tr}$ of $>0.993$.  So while planet c possibly has a grazing transit, it is \textit{very} close to a full transit.  

For eclipse, planet c has a grazing eclipse probability of $P(\mathrm{grazing})=0.52$, and none of the 10,000 posterior samples has a not-even-grazing non-eclipse.  Including the samples with total eclipse ($f_\mathrm{ecl}=1$), the occulted fraction $f_\mathrm{ecl}$ has a 97.5\%  lower limit of 0.53, a much weaker constraint than that of the transit's, due to the fact that $b_\mathrm{ecl}$ also inherits the uncertainty of $e\sin\omega$.  

The comparison to the JI \eureka{} fit demonstrates how strongly this limit depends on $e\sin\omega$.  Its prior on $e\sin\omega$ was narrowed based on the expected tidal damping, yielding a posterior that is also narrower: $e\sin\omega=0.0001^{+0.0016}_{-0.0017}$ against $0.0003^{+0.0093}_{-0.0034}$ in the final RV+LC joint fit.   This narrower prior on $e\sin\omega$ raises the lower limit on $f_\mathrm{ecl}$ to 0.96 from 0.53, closer to the limit on $f_\mathrm{tr}$.  The tighter limit therefore rests on the assumption that tides have damped the eccentricity of planet c.  The \eureka{} fit finds a similar grazing eclipse probability of $P(\mathrm{grazing})=0.49$, but this agreement is partly coincidental: $b_\mathrm{tr}$ lies almost exactly at the grazing limit $1-R_p/R_\star$, so $P(\mathrm{grazing})$ is sensitive to small shifts in $b_\mathrm{tr}$.  Applying the same tidal prior directly to the RV+LC samples gives $P(\mathrm{grazing})=0.38$; $f_\mathrm{ecl}$ is therefore the more robust measure of the grazing eclipse geometry.  


\subsection{Dynamical modeling of the LTT~1445~A system}

\subsubsection{Are the LTT~1445~A siblings in orbital resonance?} \label{subsec:resonance}

The LTT~1445~A siblings have an orbital period ratio close to several mean-motion resonances (MMR). The pair is 0.07\% from the 12:7 commensurability, 1.98\% from the 7:4 commensurability, and 2.92\% from the 5:3 commensurability.  

For a planet pair to be in true orbital resonance, there must be a critical resonant angle that librates around a fixed point over time rather than circulating from 0 to 2$\pi$ \citep[see e.g. Fig. 4 of][]{yee26}. Two-body resonance angles have the form $\phi = p\lambda_\mathrm{b} - q\lambda_\mathrm{c} + j_\mathrm{c}\varpi_\mathrm{c} + j_\mathrm{b}\varpi_\mathrm{b}$ \citep{murray_dermott99}, where $\lambda_\mathrm{c}$ and $\lambda_\mathrm{b}$ are the mean longitudes of the inner (c) and outer (b) planets, $\varpi_\mathrm{c}$ and $\varpi_\mathrm{b}$ are their longitudes of periastron, $p$ and $q$ are the integers describing the MMR (e.g., $p=7$ and $q=4$ for the 7:4 MMR), and $j_\mathrm{c}$ and $j_\mathrm{b}$ are non-positive integers $\leq 0$ that sum to $-5$ for fifth order (12:7), $-3$ for third order (7:4), and $-2$ for a second-order resonance (5:3).  Therefore there are six possible resonant angles for the 12:7 resonance, four for the 7:4, and three for the 5:3.  

We perform \texttt{rebound} $N$-body simulations \citep{Tamayo2020} to examine the evolution of all possible critical angles for the 12:7, 7:4, and 5:3 resonances.  We draw 1000 random initial system parameters ($M_i$, $e_i$, $\omega_i$, $a_i$, and $i_i$) from the final joint fit results in Table \ref{tab:jointfit} and evolve the system for 200 years in 1000 independent simulations. 

We find circulation from 0 to 2$\pi$ in all resonant angles for all simulations, confirming that the LTT~1445~A siblings are \textit{not} in any of the 12:7, 7:4, or 5:3 resonances.  This is not unexpected despite the pair being very close to the 12:7 commensurability.  The strength of a 5th-order resonance scales as $\propto e^5$, so the 12:7 resonance is very weak at the small eccentricities of the siblings, and the periapses can easily drift out of (anti-)alignment.  We are not aware of an exoplanet pair confirmed to be in a 12:7 resonance; such configurations are known mainly among minor bodies in mean motion resonance with Neptune \citep{ragozzine07}. 

\subsubsection{Predictions for TTV} \label{subsec:ttv}

Planet pairs that are near but not in resonance can still exhibit transit timing variations (TTV) \citep{lithwick12, Agol2016}.  We compute the expected TTV amplitudes for the LTT~1445~A planet pair using the \texttt{TTVFaster} package \citep{Agol2016}.
We initialize $5\times10^5$ synthetic system parameters ($M_i$, $P_i$, $e_i$, $\omega_i$, and $T_{0,i}$) drawn from the joint fit distribution.
We consider up to 10th order resonances and simulated the system for 10 years, then computed the maximum TTV amplitude from the \texttt{TTVFaster} model output for each configuration. 

We find that LTT~1445~A~c has a TTV amplitude of 0.45 $\pm$ 0.04 minutes and LTT~1445~A~b has a TTV amplitude of 0.20 $\pm$ 0.02 minutes, which are negligible for the purposes of observation scheduling but potentially detectable by combining extreme-precision timing measurements from existing and planned JWST observations in future work.  A potential exterior third planet with an orbital period of $\sim$24.3 days \citep{lavie23_planetary} would not affect the TTVs of the inner pair given that this candidate planet is nowhere near any resonance and has a period ratio $> 4$ \citep{lithwick12}.  The other LTT~1445~BC stellar components should also not affect the TTVs given the proximity of LTT~1445~A~c and b to their host star while the other stellar components are tens of AU away \citep[e.g.,][]{Zuniga-Fernandez2025}. 

\subsubsection{Are the LTT~1445~A siblings tidally heated?} \label{subsec:tidal_heating}

Despite not being in resonance, the LTT~1445~A siblings have a very compact orbital configuration within 0.04 AU. The expected tidal orbital circularization timescales \citep[][Equation 1]{Puranam2018} are $<$ 20 Myr assuming that the planets have Earth-like tidal dissipation efficiencies and that planet-planet interactions are negligible.  However, gravitational interactions between planets in compact systems can excite low-level forced eccentricities that are stable over Gyr timescales in the face of tidal circularization \citep[e.g.,][]{Peterson2023}.  In this case, the system reaches an equilibrium state where the planet-planet eccentricity excitation is balanced with tidal circularization.  The timescale to settle to this equilibrium forced eccentricity depends on the planet’s reduced tidal quality factor $Q’_p$ \citep[e.g.,][]{Goldreich1966,Peterson2023,Greklek-McKeon2025a,Greklek-McKeon2025b}.  These small but persistent eccentricities can lead to tidal heating and volcanic activity that can alter the structure and atmosphere of the planet, similar to Jupiter’s moon Io \citep{Spencer2000, Veeder2004, DeKleer2019, Peterson2023}. 

To estimate the forced eccentricities in the LTT~1445~A system, we perform long-term dynamical simulations incorporating a constant time-lag tidal model \citep{Eggleton1998} using the tides-spin implementation of \citep{Lu2023} in \texttt{REBOUNDx 1.6.1} \citep{Tamayo2020} with the same modeling framework as \cite{Greklek-McKeon2025a}.  This $N$-body modeling includes the effects of tidal forces between the star and planets and planet-planet gravitational interactions to track orbital evolution.  We initialize the system with planet masses and orbital parameters drawn from Table \ref{tab:jointfit}, and then run the \texttt{REBOUNDx} simulations for $2\times10^6$/$Q'_p$ years, so that the damping timescale can easily be mapped onto various possible values of $Q’_p$.  Assuming Earth-like $Q'_p$ = 100 values and that the system is older than $\sim$1 Gyr \citep{lavie23_planetary}, we find steady-state forced eccentricities of $e_c\lesssim 3\times10^{-4}$ for planet c and  $e_b\lesssim 1\times10^{-4}$ for planet b. These dynamical modeling results are consistent with the observational constraints on $e_i$, though both constraints have 2-$\sigma$ upper limits of $>0.01$ and cannot \textit{confirm} our predicted $e_i$ from dynamical modeling. 

Using the forced eccentricities, we can estimate the tidal heat fluxes that the planets should have assuming Earth-like $Q'_p$ = 100.  Using Equation 4 of \cite{jackson08} to calculate the tidal luminosity and then converting this to a tidal heat flux in W/m$^2$, we find that LTT~1445~A~c has a tidal heat flux $\sim$100x smaller than that of Io \citep[$\sim$2 W/m$^2$,][]{Veeder2012} and $\sim2\cdot10^{-5}\times$ its incident stellar flux, while LTT~1445~A~b has a tidal heat flux $\sim3\times10^{-4}$ less than Io and $\sim8\cdot10^{-6}\times$ its incident stellar flux. These tidal heat fluxes are far below the values expected to support volcanic activity or vigorous atmospheric outgassing \citep{jackson08,Seligman2024}, indicating that tidal heating is likely not an important present-day process for either planet.

\subsection{MIRI settling behaviors} \label{subsec:systematics}

\subsubsection{Visit-start Settling behavior} \label{subsubsec:settling}

\citet{fortune25_lhs1140c} noted a potential correlation between the previous MIRI filter in position before each observation, and the initial strength and sign of the settling ramp. This trend was identified by comparing all \texttt{SUB256} observations (18 visits) within the Hot Rocks Survey (GO 3730), where it was found to yield initial upward slopes when shorter wavelength filters (F560W, F770W F1000W) were in place before the F1500W science observations, and downward slopes when longer wavelength filters were in place beforehand. The proposed explanation is that JWST will slew towards the target and begin fine guidance tracking before switching to the F1500W filter, often pointing at the target for $>10$ minutes with whichever filter was in place last. As there is no shutter on MIRI, the detector pixels will be exposed to flux from the target as viewed through a different filter, while being clocked out and reset in idle mode. The filter swap typically only occurs 1-2 mins before the beginning of the observation, which is much shorter than the observed persistence timescales, and so the detector may still be affected by the previous filter. Sometimes, the previous filter was the same as the science filter.

\begin{figure*}
    \centering
    \includegraphics[width=\textwidth]{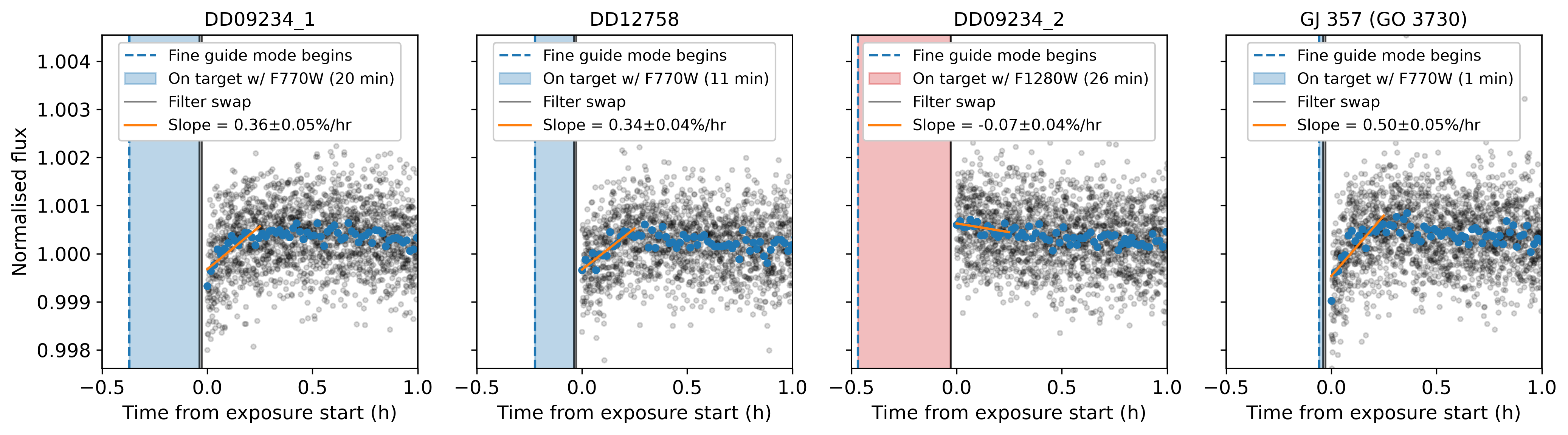}
    \caption{Initial settling comparison for the three visits of LTT~1445~A, and a GJ 357 observation analyzed in \citet{zgraggen26_hot}, which all used the SUB64 subarray, and were all uniformly reduced using the \textit{MF reduction} procedure. We include normalised flux values in black, and flux in 1-minute time bins in blue. Visits which previously had the F770W filter in position show similar upward ramps, while the \textsc{DD09234\_2} visit previously had the F1280W filter in place and shows a downward ramp. Slopes are fit to the first 15 minutes of observations as a rough measure of initial settling, similar to \citet{fortune25_lhs1140c}. The GJ 357 visit spends a significantly shorter time on target before observations, yet shows a similarly strong slope to the first two visits of LTT~1445~A.}
    \label{fig:initial_settling}
\end{figure*}

The settling timescale has been noted to be strongly dependent on the target brightness \citep{connors2025}. Therefore, a direct comparison between the strength of settling in the LTT~1445~A observations and the significantly dimmer targets observed with the \texttt{SUB256} subarray in \citet{fortune25_lhs1140c} could be misleading. We instead compare the three visits between each other, as well as to the only other comparable \texttt{SUB64} observation published to-date, which is an observation of GJ 357 b as analyzed in \citet{zgraggen26_hot}. Coincidentally, three of these four visits had the F770W filter in position prior to switching filter, while the \texttt{DD09234\_2} visit had the F1280W filter in position. Therefore, we may expect the \texttt{DD09234\_2} visit to have a different slope to the other three visits if this trend holds. Figure~\ref{fig:initial_settling} appears to confirm this, with the three F770W visits producing similar, strong upward slopes, while only the F1280W visit shows a downward slope. The observations were reduced uniformly following the \textit{MF reduction} in Section~\ref{subsec:mf_red}, including a new reduction of the GJ~357 observation to ensure consistency. We give the best-fit slope to the first 15 minutes of observations as a rough measure of the strength of settling, reported in units of percentage flux change per hour, excluding the first integration of each visit because they are affected by the `first exposure' effect \citep{morrison23}. This matches the \citet{fortune25_lhs1140c} procedure, except we chose to use 15 minutes rather than 30 minutes for the settling slope fit, due to the shorter settling timescale for these bright targets.

We measure time on target with the previous filter by calculating the difference between when the telescope enters `fine guide' mode, to when a filter move is detected (using the \textsc{IMIR\_HK\_FW\_CUR\_POS} engineering mnemonic). The Fine Guidance Sensor will remain in fine guide mode throughout the observation, with the MIRI PSF locked in its target position with high precision from the time this mode is entered. While this approach may slightly underestimate the true time on target, as earlier modes of guide star acquisition may happen to be close to the target position, we note slews of $>1$ arcsec ($>9$ MIRI pixels) for all of these observations during guide star acquisition or later, which implies the time pointed at the approximate target location with the previous filter can only add another $1-2$ minutes at most to the times given in Figure~\ref{fig:initial_settling}.

Intriguingly, the GJ 357 b observation is only briefly exposed ($\sim 1$ minute) to the F770W filter while on target, as highlighted in Figure~\ref{fig:initial_settling}, whereas the other observations enter fine guide mode $>10$ minutes prior to switching filter. Previous observations have shown that even if the previous filter is a strong predictor of initial settling, there have been clear outliers to this trend when the telescope spends a brief period of time on target prior to observations \citep{fortune25_lhs1140c, xue25_the, connors26_gj}, which could naturally be explained by not having enough time for persistence effects to build up. So this similar, yet stronger upward slope observed for the GJ 357 observation may simply be coincidental. Regardless, the visits of LTT~1445~A appear to show further evidence that the settling slope can be much more consistent between observations which share the same previous filter. This has already been tested as an experiment for the RWDDT observations of GJ 3929 b, by intentionally switching to the MIRI LRS P750L filter position before slewing toward the target\footnote{https://rockyworlds.stsci.edu/rw-website-schedule.html\#strategy}, although this had mixed results possibly due to different amounts of time on target \citep{xue25_the, connors26_gj}.

\subsubsection{New-exposure settling behavior} \label{subsubsec:newexposure}


As we have shown in \S\ref{sec:datared}, our light curves also show \textit{mini-settling} behaviors at the start of each new exposure. For time series observations that exceed the on-board storage limit of 196,608 frames, it is recommended to increase the number of Exposures/Dither, where Dither = None. Though the telescope does not re-point, the new exposure introduces a mini-settling ramp, which we mitigated by masking 10-30\,minutes of data.  As a cautionary tale, the egress of planet b in visit \texttt{DD12758} overlaps with an exposure break, which makes its eclipse duration and timing uncertain, as shown in the orange contours in \autoref{fig:lightcurves}.  The uncertainty this introduces is partly mitigated by having multiple visits, though it is worth noting that the egress of planet b is also very close to an exposure break in visit \texttt{DD09234\_2} (within 4 minutes). 
Here we discuss how prevalent this mini-settling behavior is, and try to make recommendations for mitigation in future observations.


During a MIRI exposure break, there is no re-pointing, target acquisition, or filter swap, and the MIRI subarray is clocked out at a constant rate regardless of whether an exposure is running \citep{morrison23}.  However, pixels are held in reset continuously instead of integrating, and the first integration of a new exposure then follows a brief set of continuous idle resets with no accumulated signal, which leads to a different behavior from following the single reset between consecutive integrations  \citep{morrison23, carter23, dyrek24}.  An exposure break lasts tens of seconds of sustained reset rather than hours (typical between visits), so we would expect a smaller and faster time-dependent ramp than at the visit-start, as seen in this work (\autoref{fig:lightcurves}). 

To see if this new-exposure ramp is a common behavior for MIRI time series observations (TSOs), we examine a set of MIRI observations with multiple exposures.  We look at LRS observations of GJ~1214~b (K=8.8) \citep[GO 1803,][]{kempton_reflective_2023} and GJ~367~b (K=5.8) \citep[GO 2508,][]{zhang24}, and two F1500W observations of LTT~1445~A (K=6.5) from this work.  Some exposure breaks were attributed to high-gain antenna (HGA) moves in \citet{kempton_reflective_2023}, while \citet{zhang24} mitigated a binned $\sim1000$\,ppm new-exposure ramp that decayed within $\sim 100$\,s via masking 170 s in their analysis.  

Analyzing these four TSOs without any sigma clipping in the time series, we find that a similar mini-settling behavior is indeed present in all of the exposure breaks.  We also find that the first integration after an exposure break is lower by 1-2\% and does not fit the ramp (which can be either positive or negative), likely a \textit{mini-first exposure effect}.  In GO~2508, first integrations of each new exposure are highly irregular: $-0.6\%$, $-7.7\%$ and $+0.2\%$ from the median after the first exposure break, and $-32\%$, $-0.05\%$ and $-4.3\%$ after the second.  In GO~1803, after the last (fourth) exposure break, there are three integrations that are lower by 2000$\times$ the rms scatter, resembling a HGA move. In Cycle 1, HGA moves were much more common, and several are seen in this dataset \citep[See Extended Data Fig. 3 of][wherein 3 of the 4 exposure breaks have been identified as HGA moves]{kempton_reflective_2023}. 

If an exposure break can also cause a more long-term shift in the baseline, then simply masking the mini-settling is not a complete mitigation, as the eclipse depth will be degenerate with the offset amplitude.  This is true even if the eclipse timing and duration were precisely known a priori.  For this reason, we also test whether an exposure break can also cause a long-term shift in the baseline.  In each TSO, we first mask the first hour and any eclipses or transits and remove $>20\sigma$ outliers from a running median.  We then simultaneously fit a cubic trend as well as a sinusoidal curve for phase curves and a single exponential ramp that allows a permanent offset at the start of each new exposure and a free offset for the very first integration of each new exposure (effectively masking it).  This ramp has the form $A_k\exp[-(t-t_k)/\tau_k]+b_k$ when $t\geq t_k$, where $t_k$ denotes the start of a new exposure.  To compare how anomalous the new-exposure offset is compared to typical red noise, we pretend that there is an exposure break every 10 minutes in each light curve and fit for a ramp with a free amplitude and offset but the same decay timescale as the median timescale of real new-exposure ramps in the TSO.  We then compare the amplitude and offset of the real new-exposure ramp to the rms values of the pretend ramps. 

\begin{deluxetable*}{lllcccrrrrr}
\tablewidth{0pt}
\tabletypesize{\small}
\tablecaption{
Summary of \textit{mini-settling} and \textit{mini-first exposure}
behavior in MIRI TSOs (see text) and permanent offsets at the exposure breaks, which are numbered $k=1,2,\ldots$ in each visit. The rms values after $A_k$ and $b_k$ indicate the rms of the amplitudes and offsets of the pretend ramps fit every 10~minutes, which are a result of systematics (\textit{not} the uncertainty of the real new-exposure ramp). The larger $A_k$ and $b_k$ relative to the rms value, the more significant the mini-settling and long-term offset relative to those induced by pure red noise.
\label{tab:exposure_breaks}
}
\tablehead{
\colhead{Star} & \colhead{PID} & \colhead{Mode} & \colhead{Cadence} & \colhead{Break ($k$)} & \colhead{1st int.} &
\colhead{$\tau_k$} & \colhead{$A_k$} & \colhead{rms $A$} & \colhead{$b_k$} & \colhead{rms $b$} \\
\colhead{} & \colhead{} & \colhead{} & \colhead{(s)} & \colhead{} & \colhead{} &
\colhead{(s)} & \colhead{(ppm)} & \colhead{(ppm)} & \colhead{(ppm)} & \colhead{(ppm)}
}
\startdata
LTT~1445~A & DD12758 & F1500W & 2.21 & 1 & $-1.2\%$ & $74 \pm 52$ & $+1300 \pm 1600$ & $170$ & ---\tablenotemark{a} & --- \\
 & DD 9234\_2 & F1500W & 1.96 & 1 & $-1.1\%$ & $64 \pm 34$ & $+620 \pm 240$ & $187$ & $+8$ & $82$ \\
 & & & & 2 & $-1.1\%$ & $139 \pm 63$ & $+490 \pm 160$ & $187$ & $-36$ & $82$ \\\hline
GJ 1214 & GO 1803 & LRS & 6.84 & 1 & $-1.6\%$ & $22 \pm 5$ & $-2900 \pm 600$ & $590$ & $+6$ & $44$ \\
 & & & & 2 & $-1.6\%$ & $14 \pm 4$ & $-4100 \pm 1200$ & $590$ & $+2$ & $44$ \\
 & & & & 3 & $-1.6\%$ & $14 \pm 4$ & $-4100 \pm 1200$ & $590$ & $-81$ & $44$ \\
 & & & & 4 & HGA? & --- & --- & & $-85$ & $44$ \\\hline
GJ 367 & GO 2508 & LRS & 0.95 & 1 & $-0.6\%$ & $27.1 \pm 1.3$ & $+3930 \pm 140$ & $257$ & $+72$ & $100$ \\
 & & & & 2 & $-32\%$\tablenotemark{b} & $35.7 \pm 2.2$ & $+2550 \pm 120$ & $257$ & $+128$ & $100$
\enddata
\tablenotetext{a}{Planet b's eclipse is masked until 85\,s after the break.}
\tablenotetext{b}{A single outlier; the third integration is also 4\% low.}
\end{deluxetable*}

We show the results in Table \ref{tab:exposure_breaks}.  The amplitude of the real new-exposure ramp $A_k$ is 3-15$\times$ the rms of the pretend ramps, indicating that the mini-settling ramp is a real effect distinguishable from red noise.  In visit \texttt{DD12758}, the visit that led us to investigate the new-exposure behavior, the masked eclipse of planet b leaves no data until 85\,s after the break, so $A_k$ and $b_k$ cannot be precisely measured without modeling the eclipse.  The real new-exposure offsets $b_k$ are within 1.9$\times$ the rms of the pretend offsets caused purely by red noise, indicating that an exposure break does not lead to an anomalous long-term offset.  We therefore recommend first fitting a ramp to measure the decay timescale and masking a sufficient amount of time accordingly (including the first integration), as done in \citet{zhang24} and in this work.  


\subsection{The MIRI LRS and F1500W synergy} \label{subsec:lrs_15}

Among JWST eclipse observations of rocky planets, \lttab{} is one of only two targets that have been observed with both MIRI LRS and F1500W alongside LHS~1478~b \citep{august25_hot}.  It is one of only five planets with $T_\mathrm{irr} \leq 1800$K that have been observed in emission in more than one bandpass with JWST alongside TRAPPIST-1~b,~c \citep{ducrot25_combined}, and LHS~3844~b \citep{zieba26_lhs3844}, which also have phase curve observations.  As F1500W has been used in isolation for the lion's share of rocky planet observations, it is worth discussing the complementary strengths of LRS and F1500W. 

An oft-mentioned yet rather underplayed strength of F1500W is that CO$_2$ absorption (or emission in a thermal inversion) has an extremely strong effect in its bandpass.  This point is worth belaboring.  F1500W does not primarily look for cooling via heat redistribution while CO$_2$ absorption is a mere bonus; CO$_2$ absorption has an effect on the 15-$\mu$m flux comparable to or up to a few orders of magnitude stronger than that of heat redistribution, depending on the surface pressure.  The inferred absence of thick CO$_2$ atmospheres on many rocky planets would not change if a thick atmosphere did not redistribute heat \citep{fortune25_lhs1140c, wirth26_analytic, wogan26_a}.  This means that inferring heat redistribution factor $f$ and an effective albedo from F1500W measurements without considering the composition is poorly posed unless the inferred constraints are very tightly confined to one corner of the $f$-$A$ parameter space, as the 15-micron flux does not necessarily represent the dayside energy budget since the dominant effect is spectral.  
The natural axes for 15-micron observations are instead $P_\mathrm{surf}$ and  $\mathrm{VMR}_\mathrm{CO_2}$ \citep{zieba23, maurel25_constraints}, and the leap from inferring the absence of a CO$_2$ atmosphere to a dark, bare rock requires assuming that CO$_2$ truly is ubiquitous in secondary atmospheres \citep[e.g.][]{ih23_constraining}.  

LRS observations provide complementary information by accurately constraining the dayside energy budget, as most of the planetary flux is captured.  LRS alone captures 50\% of \lttab{}'s thermal emission were it a blackbody at the measured brightness temperature ($524\pm19$\,K), and LRS and F1500W together captures 57\% ($526\pm17$\,K).  For illustration, a 20\% change in the remaining 43\% of the planetary flux would only change the effective temperature by $\pm11$\,K.  The combined LRS-F1500W wavelength coverage is broad enough to therefore claim that the best-fit brightness temperature approximates the true effective temperature of the planet.  This makes the inferred $f$-$A_\mathrm{B}$ an actual characterization of the dayside (Figure~\ref{fig:fab}).   

Moreover, beyond constraining the energy budget with the broadband measurement, LRS observations are also spectroscopically sensitive to the rest of the expected IR absorbers that F1500W cannot easily constrain, such as H$_2$O and SO$_2$, as seen in Figure~\ref{fig:retrieval}.  If CO$_2$ was not ubiquitous, an atmosphere that only contains these absorbers could pose false negatives for observations using F1500W in isolation.  A steam atmosphere is particularly interesting, as around an M star H$_2$O can cause greenhouse warming near the photospheric pressures and bring the brightness temperature closer to the dark, bare rock limit \citep{malik19}.  

We therefore caution against interpreting F1500W observations as a definitive test for atmospheres beyond other possibilities, as their primary role is to efficiently rule out Venus-like thick CO$_2$ atmospheres.  Complementary efforts via follow-up observations using LRS eclipses or different geometries (transits and phase curves) are necessary to conclusively characterize a planet's atmosphere.





\section{Conclusions} \label{sec:conclusion}

How do the RWDDT eclipse observations of LTT~1445\,A\,b and c change our understanding of their atmospheres and the M~dwarf Cosmic Shoreline?

\begin{enumerate}
    \item \textbf{LTT~1445~A~b.} The prior constraint based on the published HST and JWST transit and eclipse observations was that planet b either lacks an atmosphere or has a CO$_2$ atmosphere of at most moderate pressure ($\sim$1\,bar).  The additional eclipse observations at 15~$\mu$m now set an upper limit on CO$_2$ partial pressure of $\leq0.6$\,mbar and limit on atmospheres of $P_\mathrm{surf}\lesssim0.1$\,bar.  A dark bare-rock model with $A_\mathrm{B} = 0.1$ explains the data well.  We conclude that planet b has no substantial atmosphere.

    \item \textbf{LTT~1445~A~c.} No published results had yet constrained the atmosphere of planet c.\footnote{Our analysis is independent of its transmission \textit{spectrum} in GO~7073 (Brande \& Wiser et al., in prep.).}  We nonetheless expected it to be airless, given its position relative to the nominal Shoreline and its smaller escape velocity and greater instellation than planet b. The new observations neither confirm nor overturn this expectation, as the grazing geometry leaves its eclipse depth imprecisely constrained.  The observations rule out atmospheres of $\gtrsim 1$\,bar with $\geq 1\%$ CO$_2$, i.e. Venus-like atmospheres, but allow many atmosphere scenarios of various compositions and surface pressures as well as bare surfaces of a wide range of albedos.  Given that planet b appears to lack a thick atmosphere, we expect the same for planet c.

    \item \textbf{M~dwarf Cosmic Shoreline.}  \lttab{} joins the sample of observed rocky exoplanets on the ``dry'' side of the nominal Shoreline that lack thick atmospheres (\autoref{fig:cosho}).  \lttac{} is too weakly constrained to test the Shoreline on its own, and our expectation that it lacks a thick atmosphere presupposes the hypothesis being tested.  The M~dwarf Cosmic Shoreline in its current canonical form therefore remains consistent with the RWDDT observations of this system. 
    Because \lttab{} lies closer to the nominal Shoreline than any other target observed in eclipse so far, its  lack of a substantial atmosphere indicates that the true separating line cannot lie much farther toward the dry side than the nominal line, even allowing for its intrinsic stochasticity.  The decisive test will come from the targets on the ``wet'' side of the Shoreline, where atmospheres are \textit{predicted} to survive; that is where upcoming RWDDT observations will be the most informative.
\end{enumerate}

If planet c is revealed to in fact have an atmosphere, it would directly refute the core logic of the XUV-driven Cosmic Shoreline, especially as one of a pair within a single system.  However, the grazing geometry (as much as the overlaps with b, which should be avoidable in follow-up observations) leads to a larger uncertainty on its eclipse depth than expected.  For this reason, it becomes a more costly target to precisely characterize in secondary eclipse than its short eclipse duration and the bright star would suggest.  We therefore argue that observing time for follow-ups on planet c is probably better spent observing other targets that extend the observed sample. 

Our RWDDT eclipse results narrow down what the concurrent and upcoming transmission programs need to test.  For planet b, the eclipses already exclude thick atmospheres and CO$_2$ partial pressures above 0.6\,mbar.  The NIRISS and NIRSpec observations of \lttab{} in the Cosmic Shoreline program (GO~7073) and NIRCam observations in GO~7251 will therefore test the remaining scenarios: tenuous or transparent atmospheres ($P_\mathrm{surf}\lesssim0.1$\,bar) whose heat transport and spectral features are too weak to be measurable in eclipse.  Transmission can potentially be more sensitive to such atmospheres than emission, as the slanted optical path through the limb raises the optical depth relative to the vertical.  Moreover, repeated NIRCam visits will help separate any planetary signal from epoch-to-epoch stellar contamination.  A flat transmission spectrum would support the bare-rock interpretation, as the eclipse observations also rule out thick aerosol layers; any detected features would be reconciled with the eclipse observations in a joint analysis of emission and transmission.

With 500 hours of observing time, the RWDDT is a large undertaking, and its scientific return will depend on analyses beyond those of any single team.  As our work demonstrates, accurately interpreting eclipse observations also requires extensive archival analysis, data reduction, and modeling. To coordinate these efforts, the RWDDT Community Involvement Initiative\footnote{\url{https://rockyworlds.stsci.edu/rw-website-community.html}} launched in 2026, of which this work is a part.  We invite the community interested in e.g. MIRI data reduction, atmospheric and surface modeling, or population-level analysis to join the Initiative to strengthen the science of RWDDT.




\begin{acknowledgments}

We thank the Rocky Worlds DDT Core Implementation Team (CIT) for planning and executing the observations in the current study.  We also thank P. Wachiraphan for publicly sharing the light curves from JWST GO 2708. 

JI is not affiliated with the RWDDT in any capacity other than through the Community Involvement Initiative.  
JI, MF, NHA, and MH ran all data reductions starting from publicly available raw data or data products available in MAST; no data products produced by the Rocky Worlds DDT CIT were used in this work. 
HDL compiled published RV and Magellan data for use in the joint fits, helped investigate MIRI settling behavior in exposure breaks (starting from \texttt{uncal} data and processing it with \texttt{Eureka!}), and supported and provided feedback on the manuscript.

JI acknowledges support from the Alfred P. Sloan Foundation under grant G-2025-25284.  JI thanks D. Koll, N. Wallack, A. Savel, and P. Gao for useful discussions.  JI acknowledges using Claude by Anthropic for improving visualization, formatting tables and data, performing literature search on and constructing analysis of MIRI exposure break behavior, and optimizing and validating bug fixes in \helios{}.  All scientific ideas and writing originate from the authors, who are humans and take full responsibility. 

SK acknowledges using Anthropic Claude AI Opus 5.5 for (1) debugging convergence and sampling issues within \texttt{pymc3}, (2) incorporating eclipse modeling into existing joint RV + transit models for multi-planet systems, (3) modifying the leave-one-out calculation within \texttt{arviz} to include the correlated likelihood from the GP. SK acknowledges responsibility for manually verifying the accuracy of all these outputs. 

NW, MWM, and JLY acknowledge support for JWST Program GO 7073, provided by NASA through a grant from STScI under NASA contract NAS5-03127. NW acknowledges the use of Codex GPT-6.1 Sol by OpenAI for help with formatting figures and double checking the correlated system prior used in the climate-constrained retrieval analyses (Section \ref{subsec:retrievals}).

This research has made use of the Exoplanet Follow-up Observation Program (ExoFOP; DOI: 10.26134/ExoFOP5) website, which is operated by the California Institute of Technology, under contract with the National Aeronautics and Space Administration under the Exoplanet Exploration Program.

Resources supporting this work were provided by the Resnick High Performance Computing Center, a facility supported by Resnick Sustainability Institute at the California Institute of Technology. This content is solely the responsibility of the authors and does not necessarily represent the views of Caltech.

NHA thanks the Trottier Family Foundation for their generous support through the Trottier Postdoctoral Fellowship.

MF acknowledges using Opus 5.5 for debugging convergence issues within \textsc{pymc}.

CIC acknowledges support by (i) NASA Headquarters through award number 80GSFC24M0006 and (ii) the GSFC Sellers Exoplanet Environments Collaboration (SEEC), which is supported by NASA's Planetary, Astrophysics and Heliophysics Science Divisions' Research Program.

\end{acknowledgments}



%
\facilities{JWST, Magellan II, ARC (ARCTIC), ESPRESSO, HARPS, HIRES, MAROON-X, PFS}


\software{
\texttt{ArviZ v0.11.1} \citep{kumar_arviz_2019},
\texttt{AstroImageJ} \citep{collins_astroimagej_2017},
\texttt{astropy} \citep{robitaille_astropy:_2013, astropy_collaboration_astropy_2018},
\texttt{celerite} \citep{celerite},
\texttt{celerite2 v0.1.0} \citep{foreman-mackey_fast_2017, foreman-mackey_scalable_2018},
\texttt{corner} \citep{foreman-mackey_cornerpy_2016},
\texttt{dynesty} \citep{speagle20},
\texttt{Eureka!} \citep{bell22_eureka},
\texttt{exoplanet v0.4.4} \citep{foreman-mackey_exoplanet-devexoplanet_2021, foreman-mackey_exoplanet_2021},
\texttt{ExoTiC-LD} \citep{Grant2024_exoticld},
\texttt{HELIOS} \citep{malik17_helios, malik19_self, whittaker22},
\texttt{ipython} \citep{perez_ipython:_2007},
\texttt{jaxoplanet} \citep{jaxoplanet},
\texttt{juliet} \citep{juliet},
\texttt{jwst} \citep{bushouse_jwst},                
\texttt{matplotlib} \citep{hunter_matplotlib:_2007},
\texttt{numpy} \citep{oliphant_numpy:_2006},
\texttt{pandas} \citep{mckinney-proc-scipy-2010},
\texttt{Photochem} \citep{wogan25_photochem},
\texttt{PICASO} \citep{batalha19_picaso},
\texttt{PyMC} \citep{pymc2023},                 
\texttt{PyMultiNest} \citep{buchner_x-ray_2014},
\texttt{MultiNest} \citep{feroz_multinest_2009},           
\texttt{REBOUNDx v1.6.1} \citep{Tamayo2020, Lu2023},
\texttt{scipy} \citep{oliphant_python_2007, virtanen_scipy_2020},
\texttt{starry} \citep{luger_starry_2019, agol_analytic_2020},
\texttt{Theano v1.0.5} \citep{the_theano_development_team_theano_2016},
\texttt{tinygp} \citep{Foreman-Mackey2024},
\texttt{TTVFaster} \citep{Agol2016}
}

\appendix
\counterwithin{figure}{section}                          
\renewcommand{\thefigure}{\thesection\arabic{figure}}

\section{ARC~3.5~m/ARCTIC observations of \lttac{}} \label{app:arctic}


We observed two transits of LTT~1445~A~c on UT 2022 November 19 and 2023 September 24 with the Astrophysical Research Consortium (ARC) Telescope Imaging Camera \citep[ARCTIC;][]{huehnerhoff_astrophysical_2016} on the ARC 3.5 m Telescope at Apache Point Observatory. The LTT~1445~BC stellar components at $\sim$7\arcsec\ separation prevented the use of ARCTIC's Engineered Diffuser \citep{stefansson_toward_2017, Stefansson2018_diffuserSPIE}, which is normally used to shape the PSF into a stable top-hat for high-precision photometry, since the diffuser-assisted PSF has FWHM $\sim$8.7\arcsec.

We instead used the narrowband Semrock 857/30~nm filter (842--872~nm), which is specifically designed to avoid wavelength regions affected by telluric water absorption (see transmission curve in Figure 1 of \citealt{stefansson_toward_2017}), thus mitigating the effects of changing airmass and observing conditions. Despite the narrow bandpass, we used 1$\times$1 binning (0.114\arcsec/pix) to prevent saturation, as the LTT~1445 stars are very bright for ARCTIC.

The night of 2022 November 19 was plagued by variable seeing and patchy clouds, in addition to the observations occurring at higher airmass (ranging from 3.24 to 1.63); despite efforts to de-trend against these parameters, we are still left with residual photometric variations comparable to the transit depth, so we exclude this night from our joint fit. The night of 2023 September 24 had clear skies, low humidity, and stable seeing. We observed in quad-amplifier, fast-readout mode (9s readout time in 1$\times$1 binning) and defocused to FWHM $\sim$2.5\arcsec\ with 10s exposures, with the target rising from airmass of 1.74 to 1.53 before setting again to 1.91.

We performed differential aperture photometry, including standard bias and flat correction\footnote{ARCTIC dark current is negligible for exposure times shorter than $\sim$1 minute, and dark subtraction is not recommended in these cases (\url{https://www.apo.nmsu.edu/mainpage/arctic/arcticguide/##4p3}).}, on the data using \texttt{AstroImageJ} \citep{collins_astroimagej_2017}. Due to the brightness of LTT~1445~A, the only available reference star within the 7.8\arcmin\ ARCTIC FOV is LTT~1445~BC. The B and C components are separated by $\sim$1.3\arcsec\ and are effectively completely blended by our mild defocus.

\begin{figure*}
    \centering
    \includegraphics[width=0.98\textwidth]{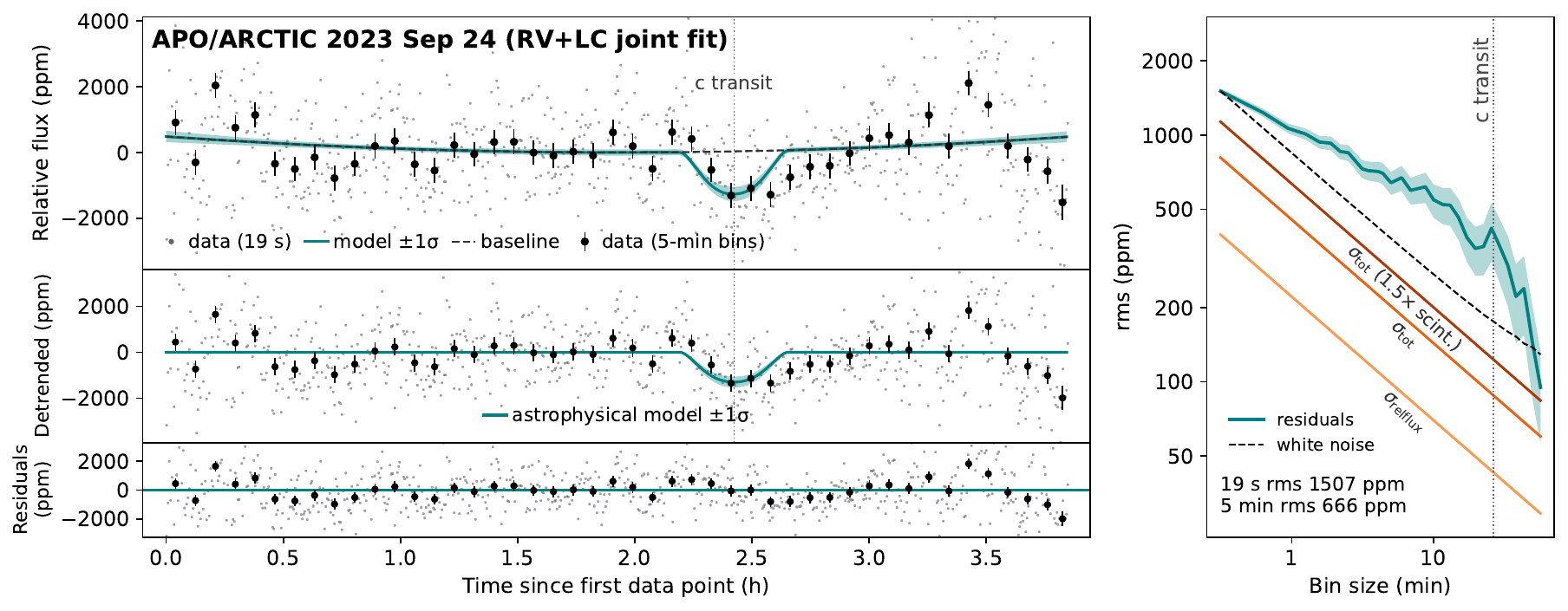}
    \caption{ARCTIC light curve fit of a transit of planet c from the final RV+LC joint fit, which also includes 4 JWST transits of the planet.  The dashed line (``white noise'') shows 19-s scatter of the residuals scaled down by $\sqrt{N}$, i.e. the rms if every residual point were independent, measured from the data.   Colored lines show the predicted noise contribution from differential photometry ($\sigma_{\rm{relflux}}$), the total expected noise ($\sigma_{\rm{tot}}$) including scintillation noise as given by Young's approximation \citep{Young1967_scintillation}, and $\sigma_{\rm{tot}}$ with the empirically-derived factor of 1.5x scintillation noise \citep{Osborn2015_scintillation}.}
    \label{fig:arctic_transit}
\end{figure*}

We calculate the theoretical noise expected from differential photometry ($\sigma_{\rm{relflux}}$), scintillation ($\sigma_{\rm{scint}}$), and the total added in quadrature ($\sigma_{\rm{tot}}$) as a function of bin size following \citet{stefansson_toward_2017}. As expected for a bright star like LTT~1445~A, scintillation dominates the expected noise (705 vs.\ 394~ppm at the 19~s cadence). \autoref{fig:arctic_transit} shows that the observed scatter exceeds even the empirical estimate for $\sigma_{\rm{tot}}$ from \citet{Osborn2015_scintillation} with 1.5$\times$ the theoretical scintillation noise, by a factor of 1.3 unbinned and 2.4--2.7 in 5--30-minute bins, suggesting additional correlated noise, likely from varying observing conditions. The joint-fit residuals have an rms of 1510~ppm at the unbinned 19~s cadence, binning down to $670^{+80}_{-60}$~ppm in 5-minute bins.

\section{Identifying a 90-day stellar rotation RV signal}\label{sec:rv_gp}

After the initial joint fit (Section \ref{subsec:fit_model}) we perform generalised Lomb Scargle (GLS) periodogram \citep{lomb_least-squares_1976,scargle_studies_1982, zechmeister_generalised_2009} on the residuals and find a peak at $\sim$ 90 days that exceeds $<$ 0.1\% False Alarm Probability (computed using the bootstrap method), similar to \citet{winters22_a}. We therefore also added an SHO GP kernel to model the RVs with quality factor $Q=1/3$ and free amplitude $\sigma_{\rm GP}$ and undamped period $\rho_{\rm GP}$.

We compared three versions of the joint model that are otherwise identical. First, we compare the eccentric model with and without an RV GP. For the RVs we used Pareto-smoothed importance-sampling leave-one-out cross-validation \citep[PSIS-LOO;][]{vehtari_practical_2017,2015arXiv150702646V} to estimate the expected log point-wise predictive density from the posteriors ($\rm{elpd}_{LOO}$).  The photometric parameters of the three fits are largely consistent, and thus we compare only the RVs. Models without the GPs can directly be compared using the LOO method implemented in\texttt{arviz\_stats.loo{}} \citep{kumar_arviz_2019}, however this cannot account for the correlated log-likelihood from the GP kernel and hence we replace the pointwise likelihoods in the importance sampling with conditional densities that include the covariance from the GP \citep{burkner_efficient_2021}.

The model including an RV GP is favoured with a $\Delta{\rm elpd_{LOO}}=67.0\pm12.0$, i.e., having a significance of 5.6. Since the LOO method already penalizes the more flexible model (with GP hyperparameters), the improvement seen here is significant \citep{sivula_uncertainty_2025}. Importantly, a GLS periodogram on the RV residuals after including a GP no longer shows the peak at $\sim$ 90 days. The SHO kernel with $Q=1/3$ is non-periodic and its power is roughly constant at timescales longer than $\sim$~80 days and falls off steeply at short periods, and has $<0.1\%$ power at the orbital periods of b and c. We compare the results with and without the GPs in \autoref{tab:jointfit}.


\section{Joint fit results and posteriors} \label{app:global_posterior}

\begin{figure*}
    \centering
    \includegraphics[width=0.98\textwidth]{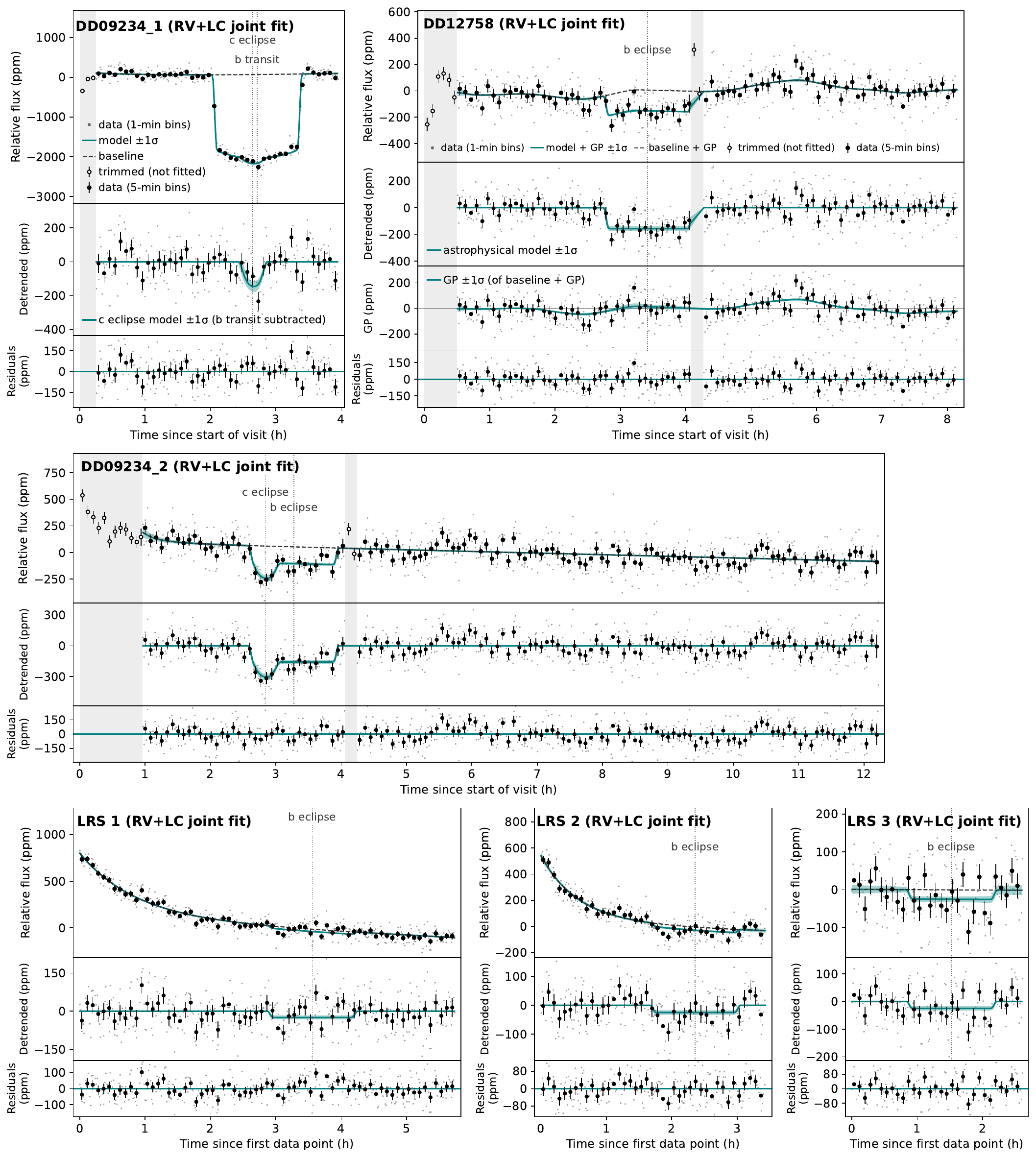}
    \caption{Same as \autoref{fig:lightcurves}, but using the final RV+LC joint fit and also showing de-trended light curves.  Fits to the MIRI/LRS observations (GO 2708) in \citet{wachiraphan25_the} are also included.  A GP was used on visit \texttt{DD12758} only.  For visit \texttt{DD09234\_1}, the transit of planet b is also removed in the de-trended light curve.  The data binned to 5 minute bins for visual clarity are shown as black points, while the 60\,s-binned data used for the joint fit are shown as grey dots.}
    \label{fig:joint_lightcurves}
\end{figure*}

\begin{figure*}[ht!]
    \centering
    \includegraphics[width=0.99\textwidth]{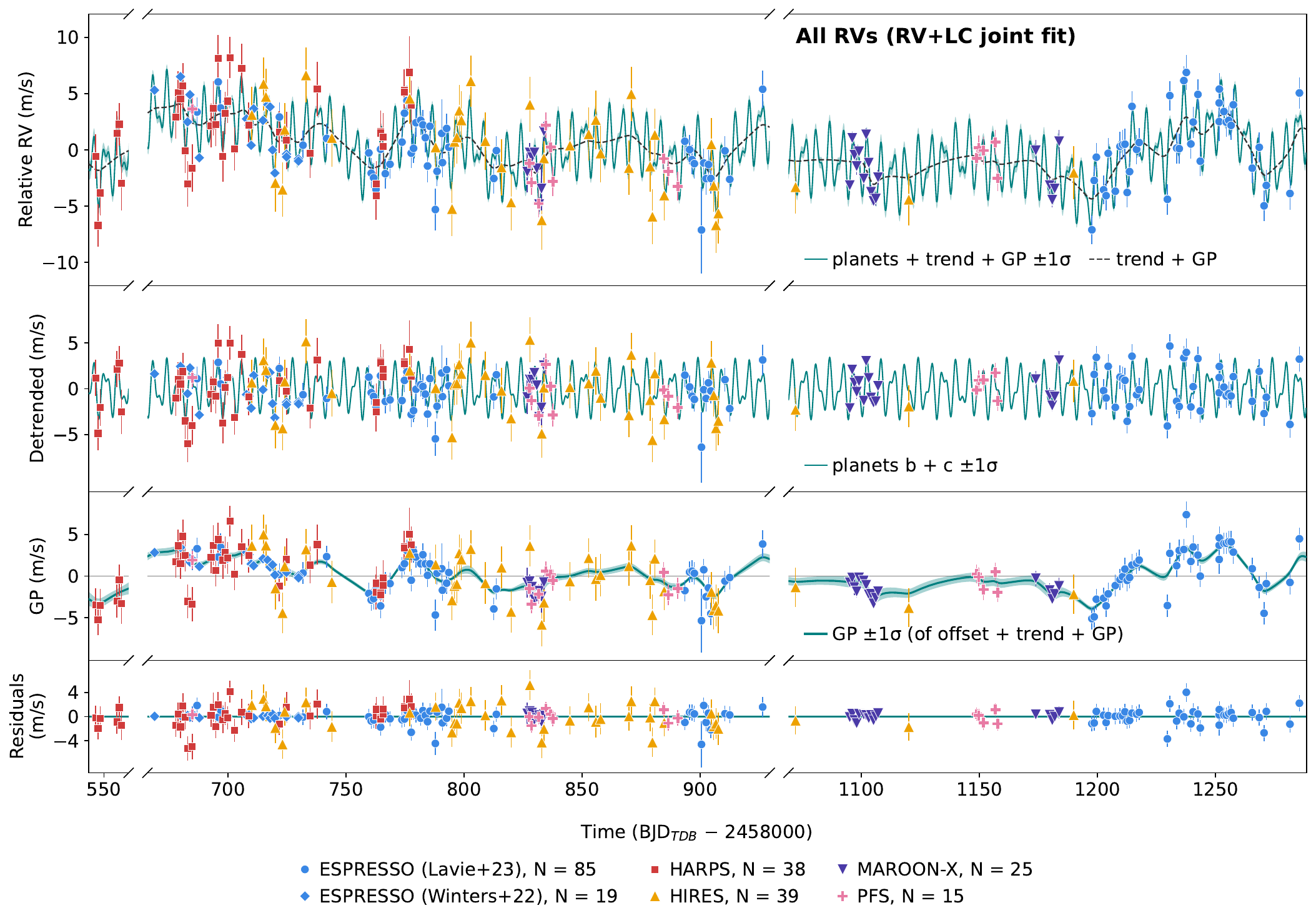}
    \includegraphics[width=0.9\textwidth]{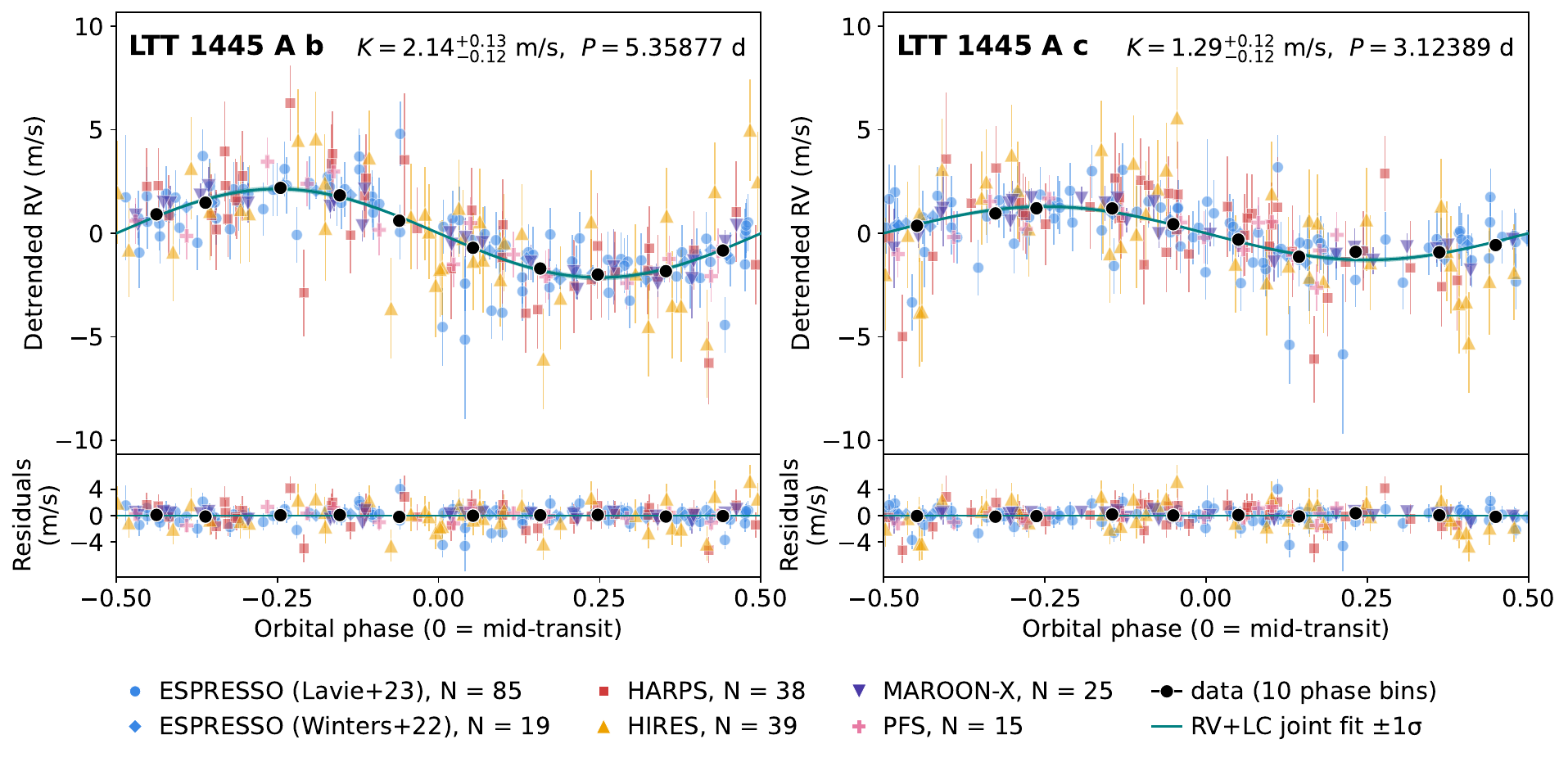}
    \caption{\textit{Top}: Radial velocities of LTT~1445~A \citep{winters22_a, lavie23_planetary} and the RV+LC joint fit (Table~\ref{tab:jointfit}).  From the top: RVs relative to each instrument's median offset, with the full model (planets b and c, quadratic trend and GP; teal) and the trend and GP alone (dashed); RVs with the offset, trend and GP subtracted, with the model of the two planets; RVs with the offset, planets and trend subtracted, with the GP; and residuals from the full model.  Error bars are the reported errors and the instrument's median jitter added in quadrature. 
    \textit{Bottom}: Phase-folded RVs of LTT~1445~A~b (left) and c (right), with the offsets, trend, GP and the other planet subtracted, on one RV scale.}
    \label{fig:rv_fits}
\end{figure*}

We show the priors and posteriors for the system parameters and derived quantities from the joint fits here in Table \ref{tab:jointfit}, for the RV+transit-only joint fit in \S \ref{sec:newglobal} and the full joint fit in \S \ref{sec:global_fit_results}.  We also show the joint fit results to the RV data in \autoref{fig:rv_fits} and to the RWDDT light curves in \autoref{fig:joint_lightcurves}.

\begin{deluxetable*}{llcc}
\tabletypesize{\scriptsize}
\tablewidth{0pt}
\tablecolumns{4}
\renewcommand{\arraystretch}{0.85}
\tablecaption{Priors and posteriors of the joint fit of the transits, eclipses and RVs.\label{tab:jointfit}  The RV and transit-only joint fit is also shown, which also excludes the F1500W transit in visit \texttt{DD09234\_1}.  Note that the change in planet masses between the fits arise from the inclusion of a 90-day GP term in the RV, rather than the inclusion of eclipse data. }
\tablehead{\colhead{Parameter} & \colhead{Prior} & \colhead{RV + transit only} & \colhead{RV + transit + eclipse + RV GP}}
\startdata
\sidehead{\textbf{Fitted parameters}}
$M_\star$ ($M_\odot$) & $\mathcal{N}_{[0,\,1.5]}(0.257, 0.014)$ & $0.259^{+0.013}_{-0.014}$ & $0.259 \pm 0.014$ \\
$R_\star$ ($R_\odot$) & $\mathcal{N}_{[0,\,1.5]}(0.268, 0.027)$ & $0.2673^{+0.0087}_{-0.0076}$ & $0.2655^{+0.0055}_{-0.0056}$ \\
$T_\mathrm{eff}$ (K) & $\mathcal{N}_{[2000,\,7000]}(3340, 150)$ & $3340 \pm 150$ & $3340 \pm 150$ \\
\hline
\sidehead{LTT~1445~A~b}
$P$ (d) & $\mathcal{N}(5.3587657, 0.0000043)$ & $5.35876803 \pm 0.00000011$ & $5.35876807 \pm 0.00000011$ \\
$T_0$ (BJD$_\mathrm{TDB} - 2461000$) & $\mathcal{N}(285.008153, 0.0004)$ & $285.007907 \pm 0.000017$ & $285.007917^{+0.000016}_{-0.000017}$ \\
$R_p/R_\star$ & $\log\mathcal{N}(\ln 0.0451, 1)$ & $0.044751^{+0.000071}_{-0.000066}$ & $0.044740^{+0.000068}_{-0.000066}$ \\
$b$ & $\mathcal{U}(0, 1 + R_p/R_\star)$ & $0.185^{+0.060}_{-0.092}$ & $0.183^{+0.052}_{-0.077}$ \\
$M_p$ ($M_\oplus$) & $\mathcal{U}(0.1, 3\times10^6)$ & $2.69 \pm 0.22$ & $2.38^{+0.17}_{-0.16}$ \\
$e$ & $\mathcal{B}(1.12, 3.09)$ & $0.030^{+0.030}_{-0.020}$ & $0.0026^{+0.0059}_{-0.0015}$ \\
$\omega$ ($^\circ$) & $\mathcal{U}(-180, 180)$ & $60^{+90}_{-206}$ & $-25^{+101}_{-57}$ \\
$F_p/F_\star$, F1500W (ppm) & $(R_p/R_\star)^2\,\mathcal{U}(0, 0.3)$ & \nodata & $157^{+18}_{-17}$ \\
$F_p/F_\star$, MIRI LRS (ppm) & $(R_p/R_\star)^2\,\mathcal{U}(0, 0.3)$ & \nodata & $24.9^{+7.6}_{-7.7}$ \\
\hline
\sidehead{LTT~1445~A~c}
$P$ (d) & $\mathcal{N}(3.1239035, 0.0000036)$ & $3.12389314^{+0.00000061}_{-0.00000059}$ & $3.12389311 \pm 0.00000060$ \\
$T_0$ (BJD$_\mathrm{TDB} - 2461000$) & $\mathcal{N}(258.451978, 0.0006)$ & $258.452095^{+0.000060}_{-0.000058}$ & $258.452072^{+0.000044}_{-0.000045}$ \\
$R_p/R_\star$ & $\log\mathcal{N}(\ln 0.0396, 1)$ & $0.03712^{+0.00029}_{-0.00025}$ & $0.03721^{+0.00026}_{-0.00024}$ \\
$b$ & $\mathcal{U}(0, 1 + R_p/R_\star)$ & $0.9624^{+0.0013}_{-0.0012}$ & $0.9621^{+0.0012}_{-0.0011}$ \\
$M_p$ ($M_\oplus$) & $\mathcal{U}(0.1, 3\times10^6)$ & $1.38 \pm 0.17$ & $1.20^{+0.12}_{-0.11}$ \\
$e$ & $\mathcal{B}(1.12, 3.09)$ & $0.036^{+0.039}_{-0.025}$ & $0.0028^{+0.0099}_{-0.0023}$ \\
$\omega$ ($^\circ$) & $\mathcal{U}(-180, 180)$ & $14^{+82}_{-78}$ & $70^{+20}_{-164}$ \\
$F_p/F_\star$, F1500W (ppm) & $(R_p/R_\star)^2\,\mathcal{U}(0, 0.3)$ & \nodata & $154^{+54}_{-37}$ \\
\hline\hline
\sidehead{\textbf{Derived parameters}}
\multicolumn{3}{l}{\rlap{\makebox[0.615\textwidth][l]{$\rho_\star$ (g cm$^{-3}$) \dotfill}}} & $19.50 \pm 0.64$ \\
\hline
\sidehead{LTT~1445~A~b}
\multicolumn{3}{l}{\rlap{\makebox[0.615\textwidth][l]{$R_p$ ($R_\oplus$) \dotfill}}} & $1.296^{+0.027}_{-0.028}$ \\
\multicolumn{3}{l}{\rlap{\makebox[0.615\textwidth][l]{$\rho_p$ (g cm$^{-3}$) \dotfill}}} & $6.04^{+0.42}_{-0.41}$ \\
\multicolumn{3}{l}{\rlap{\makebox[0.615\textwidth][l]{$K$ (m s$^{-1}$) \dotfill}}} & $2.14^{+0.13}_{-0.12}$ \\
\multicolumn{3}{l}{\rlap{\makebox[0.615\textwidth][l]{$a/R_\star$ \dotfill}}} & $30.94^{+0.33}_{-0.34}$ \\
\multicolumn{3}{l}{\rlap{\makebox[0.615\textwidth][l]{$a$ (AU) \dotfill}}} & $0.03819^{+0.00068}_{-0.00069}$ \\
\multicolumn{3}{l}{\rlap{\makebox[0.615\textwidth][l]{$i$ ($^\circ$) \dotfill}}} & $89.66^{+0.14}_{-0.10}$ \\
\multicolumn{3}{l}{\rlap{\makebox[0.615\textwidth][l]{$T_{14}$ (h) \dotfill}}} & $1.3626^{+0.0014}_{-0.0013}$ \\
\multicolumn{3}{l}{\rlap{\makebox[0.615\textwidth][l]{$10^4\, e\cos\omega$ \dotfill}}} & $8.6^{+1.8}_{-1.6}$ \\
\multicolumn{3}{l}{\rlap{\makebox[0.615\textwidth][l]{$10^3\, e\sin\omega$ \dotfill}}} & $-0.4^{+3.7}_{-5.3}$ \\
\multicolumn{3}{l}{\rlap{\makebox[0.615\textwidth][l]{eclipse delay (min) \dotfill}}} & $4.22^{+0.89}_{-0.81}$ \\
\multicolumn{3}{l}{\rlap{\makebox[0.615\textwidth][l]{$S$ ($S_\oplus$) \dotfill}}} & $5.43^{+1.03}_{-0.92}$ \\
\multicolumn{3}{l}{\rlap{\makebox[0.615\textwidth][l]{$T_\mathrm{eq}$ (K) \dotfill}}} & $425 \pm 19$ \\
\hline
\sidehead{LTT~1445~A~c}
\multicolumn{3}{l}{\rlap{\makebox[0.615\textwidth][l]{$R_p$ ($R_\oplus$) \dotfill}}} & $1.078 \pm 0.024$ \\
\multicolumn{3}{l}{\rlap{\makebox[0.615\textwidth][l]{$\rho_p$ (g cm$^{-3}$) \dotfill}}} & $5.28^{+0.53}_{-0.52}$ \\
\multicolumn{3}{l}{\rlap{\makebox[0.615\textwidth][l]{$K$ (m s$^{-1}$) \dotfill}}} & $1.29 \pm 0.12$ \\
\multicolumn{3}{l}{\rlap{\makebox[0.615\textwidth][l]{$a/R_\star$ \dotfill}}} & $21.59^{+0.23}_{-0.24}$ \\
\multicolumn{3}{l}{\rlap{\makebox[0.615\textwidth][l]{$a$ (AU) \dotfill}}} & $0.02665^{+0.00047}_{-0.00048}$ \\
\multicolumn{3}{l}{\rlap{\makebox[0.615\textwidth][l]{$i$ ($^\circ$) \dotfill}}} & $87.442^{+0.037}_{-0.043}$ \\
\multicolumn{3}{l}{\rlap{\makebox[0.615\textwidth][l]{$T_{14}$ (h) \dotfill}}} & $0.4283 \pm 0.0023$ \\
\multicolumn{3}{l}{\rlap{\makebox[0.615\textwidth][l]{$b_\mathrm{ecl}$ \dotfill}}} & $0.9630^{+0.0174}_{-0.0068}$ \\
\multicolumn{3}{l}{\rlap{\makebox[0.615\textwidth][l]{$10^4\, e\cos\omega$ \dotfill}}} & $0.9^{+5.1}_{-5.2}$ \\
\multicolumn{3}{l}{\rlap{\makebox[0.615\textwidth][l]{$10^3\, e\sin\omega$ \dotfill}}} & $0.3^{+9.3}_{-3.4}$ \\
\multicolumn{3}{l}{\rlap{\makebox[0.615\textwidth][l]{eclipse delay (min) \dotfill}}} & $0.3 \pm 1.5$ \\
\multicolumn{3}{l}{\rlap{\makebox[0.615\textwidth][l]{$S$ ($S_\oplus$) \dotfill}}} & $11.1^{+2.1}_{-1.9}$ \\
\multicolumn{3}{l}{\rlap{\makebox[0.615\textwidth][l]{$T_\mathrm{eq}$ (K) \dotfill}}} & $509 \pm 23$
\enddata
\tablecomments{$\mathcal{N}_{[a,\,b]}$: normal truncated to $[a, b]$. $\mathcal{B}$: beta. $e$ and $\omega$ are sampled as $(\sqrt{e}\cos\omega, \sqrt{e}\sin\omega)$ on the unit disk with the \citet{kipping_parametrizing_2013} prior on $e$. $F_p/F_\star$: full-eclipse depth of a uniform disk, sampled as the surface brightness ratio. Eclipse delay: mid-eclipse minus half an orbit after mid-transit. $T_\mathrm{eq}$: zero Bond albedo, full redistribution.}
\end{deluxetable*}
\onecolumngrid
\clearpage

\section{Comparison of reductions} \label{app:compare_red}


\newcommand{\redtabstep}[1]{\parbox[t]{1.00in}{\raggedright\strut#1\strut}}
\newcommand{\redtabcell}[1]{\parbox[t]{1.35in}{\raggedright\strut#1\strut}}
\startlongtable
\begin{deluxetable*}{lllll}
\tabletypesize{\scriptsize}
\tablewidth{0pt}
\tablecolumns{5}
\renewcommand{\arraystretch}{0.95}
\setlength{\tabcolsep}{3pt}
\tablecaption{Data reduction, light curve fitting, priors, and sampler choices of the three independent analyses in \S \ref{sec:datared} and the RV+LC joint fit in \S \ref{sec:global_fit_results}.  We highlight the choices deemed most consequential in \textbf{bold}, discussed throughout the main text.  The rms values of JI are from joint fit to all visits with GPs; the depths of MF assume an eccentric orbit for c and include a GP. \label{tab:reductions}}
\tablehead{\colhead{Step or parameter} & \colhead{JI} & \colhead{MF} & \colhead{NA} & \colhead{Final RV+LC joint fit}}
\startdata
\sidehead{\textit{Stage 1-2}}
\redtabstep{Pipeline} & \redtabcell{Default \texttt{jwst} pipeline} & \redtabcell{\texttt{jwst} v3.0.0, first two stages} & \redtabcell{Broadly follows A25, without optimal extraction} & \redtabcell{F1500W: JI reduction. LRS: W25 binned light curves} \\
\redtabstep{Starting product} & \redtabcell{Stage 2 \texttt{calints}} & \redtabcell{\texttt{uncal}, segments combined per exposure} & \redtabcell{\texttt{rateints}} & \redtabcell{As JI} \\
\redtabstep{Groups rejected} & \redtabcell{Last only; no RCSD} & \redtabcell{First and last} & \redtabcell{All frames included (negligible effect when removed in A25)} & \redtabcell{As JI} \\
\redtabstep{Jump threshold} & \redtabcell{Pipeline default} & \redtabcell{$7\sigma$} & \redtabcell{Pipeline default} & \redtabcell{As JI} \\\hline
\sidehead{\textit{Stage 3-4}}
\redtabstep{Centroiding} & \redtabcell{PSF centroiding with \texttt{photutils}}  & \redtabcell{PSF centroiding with \texttt{photutils}} & \redtabcell{PSF centroiding with \texttt{photutils}} &  \redtabcell{As JI} \\
\redtabstep{Aperture radius} & \redtabcell{7 px} & \redtabcell{5 px} & \redtabcell{7 px} & \redtabcell{As JI} \\
\redtabstep{Aperture selection} & \redtabcell{Grid of 4--9 px in 0.5 px steps; best tradeoff of point-to-point and binned (10--60 min) scatter} &  \redtabcell{Minimised mean white noise jitter between 4, 5, 6 px} & \redtabcell{Grid of 3--10 px; minimizes out-of-transit scatter of residuals} & \redtabcell{As JI} \\
\redtabstep{Partial-pixel treatment} & \redtabcell{Weighted by exact area}  & \redtabcell{Weighted by exact area} & \redtabcell{Weighted by exact area} &  \redtabcell{As JI}\\
\redtabstep{Background annulus} & \redtabcell{24--36 px} & \redtabcell{12--24 px} & \redtabcell{15--22 px} & \redtabcell{As JI} \\
\redtabstep{Outlier rejection} & \redtabcell{Iterative $5\sigma$ clip against an 80-integration moving mean (3, 11, 14 integrations removed)} &  \redtabcell{5$\sigma$ clip and replace with a 100-integration moving mean filter before aperture extraction} & \redtabcell{3$\sigma$ rejection against 21-integration median filter} & \redtabcell{As JI} \\\\ \hline\hline
\sidehead{\textit{Masking and binning of TSO data}}
\redtabstep{\textbf{Binning of data}} & \redtabcell{None} & \redtabcell{Every 5 integrations} & \redtabcell{None} & \redtabcell{60 s} \\
\redtabstep{Visit-start mask} & \redtabcell{300 integrations (9.8 min in DD9234\_1,2; 11.1 min in DD12758)} & \redtabcell{15 min} & \redtabcell{30 min} & \redtabcell{15, 30, 60 min in DD9234\_1, DD12758, DD9234\_2} \\
\redtabstep{New-exposure masks} & \redtabcell{300 integrations after each new exposure} &  \redtabcell{15 min after each new exposure} & \redtabcell{30 min after each new exposure} & \redtabcell{300 integrations after the second exposure starts in DD12758 and DD9234\_2} \\
\redtabstep{\textbf{Other masks}} & None  & \redtabcell{1.5 h in DD9234\_2, BJD$_{\rm TDB}$ 2461303.85--2461303.91 (possible flare)} & \redtabcell{Further 1.75 h after the exposure break in DD12758 (stronger systematics, possibly flaring)} & None \\\hline
\sidehead{\textit{Light curve model}}
\redtabstep{Modeling code} & \redtabcell{\texttt{Eureka!}} & \redtabcell{\texttt{jaxoplanet}, \texttt{tinygp}} & \redtabcell{\texttt{juliet}} & \redtabcell{\texttt{exoplanet} v0.4.4 with direct multi-planet eclipse} \\
\redtabstep{Data fit} & \redtabcell{Each DD visit separately and all three jointly (adopted)} & \redtabcell{All three DD visits jointly} & \redtabcell{Each DD visit separately and all three jointly (adopted)} & \redtabcell{All DD visits, MIRI/LRS eclipses, JWST, Magellan, and ARCTIC transits, 220 RVs (Table~1)} \\
\redtabstep{Systematics model} & \redtabcell{Offset, linear trend, exponential ramp, de-correlation against $x$, $y$ centroid position} & \redtabcell{Offset, linear trend, exponential ramp per visit} & \redtabcell{Flux offset and GP per visit; no explicit ramp} & \redtabcell{Offset, linear trend, exponential ramp per visit} \\
\redtabstep{\textbf{GP kernel}} & \redtabcell{Mat\'ern-3/2 (\texttt{celerite2})} & \redtabcell{Mat\'ern-3/2 (\texttt{tinygp})} & \redtabcell{Approximate Mat\'ern (\texttt{celerite}) in standardized time} & \redtabcell{SHO, $Q = 1/\sqrt{2}$ (\texttt{celerite2})} \\
\redtabstep{GP visits} & \redtabcell{DD12758, DD9234\_2} & \redtabcell{All three} & \redtabcell{All three} & \redtabcell{DD12758 only} \\
\redtabstep{White noise} & \redtabcell{Jitter in quadrature per visit} &  \redtabcell{Fit as a free parameter for each visit} & \redtabcell{Error inflation per visit} & \redtabcell{Jitter in quadrature per light curve} \\
\redtabstep{Light travel time} & \redtabcell{Included} & \redtabcell{Included} & \redtabcell{Included} & \redtabcell{Not included} \\\hline
\sidehead{\textit{Priors}}
\redtabstep{Source of system priors} & \redtabcell{RV+transit joint fit (\S3) + Tidal arguments on $e\sin\omega$} & \redtabcell{c: posterior of its JWST transits; b: W22} & \redtabcell{RV+transit joint fit (\S3)} & \redtabcell{Fit simultaneously (Table~7)} \\
\redtabstep{Geometry} & \redtabcell{Multivariate normal over $R_{\rm p}/R_\star$, $i$, $a/R_\star$, $e\cos\omega$, $e\sin\omega$ of b and c (Table~2)} & \redtabcell{c: multivariate normal over $t_0$, $a/R_\star$, $R_{\rm p}/R_\star$, $b$. b: independent normals on $t_0$, $a/R_\star$, $b$} & \redtabcell{Normal on $a/R_\star$, $b$ for both planets, $\mathcal{U}(0, 0.01)$ for $(R_{\rm p,b}/R_\star)^2$} & \redtabcell{$b_{\rm tr} \sim \mathcal{U}(0, 1 + R_{\rm p}/R_\star)$; $R_{\rm p}/R_\star \sim \log\mathcal{N}(\ln m, 1)$, $m = 0.0451$ (b), 0.0396 (c)} \\
\redtabstep{Ephemeris} & \redtabcell{Normal on $t_0$, $P$ (Table~2) from the initial joint fit} & \redtabcell{$P$ fixed (c: L23), uncertainty added to $t_0$ prior variance} & \redtabcell{Normal on $t_0$, $P$} & \redtabcell{Normal on $T_0$, $P$ (Table~7) (W22)} \\
\redtabstep{Eccentricity} & \redtabcell{Free: $e\cos\omega$, $e\sin\omega$ in multivariate normal, with $e\sin\omega_c \sim \mathcal{N}(0, 0.002)$ from tidal damping; $e\cos\omega_b = 0$ in DD9234\_1} & \redtabcell{Free: uniform in $e$, $\omega$. For c, $\sqrt{e}\cos\omega \sim \mathcal{U}(-0.04, 0.04)$, $\sqrt{e}\sin\omega \sim \mathcal{U}(-0.5, 0.5)$. For b, $e\cos\omega \sim \mathcal{U}(-0.0016, 0.0016)$, $e\sin\omega \sim \mathcal{U}(-0.25, 0.25)$} & \redtabcell{Circular for both} & \redtabcell{Free: $e \sim \mathcal{B}(1.12, 3.09)$ (K13a), $\omega \sim \mathcal{U}(-180^\circ, 180^\circ)$, sampled as $(\sqrt{e}\cos\omega, \sqrt{e}\sin\omega)$} \\
\redtabstep{\textbf{Eclipse impact parameter of c}} & \redtabcell{Derived from $b_{\rm tr}$ and $e\sin\omega$ ($b_{\rm ecl} \simeq b_{\rm tr}$ forced by the $e\sin\omega_c$ prior)} & \redtabcell{Derived from its eccentricity, the transit impact parameter, and the system scale $a/R_*$} & $b_{\rm tr}$ = $b_{\rm ecl}$ & \redtabcell{From $b_{\rm tr}$ and $e\sin\omega$; non-eclipsing orbits allowed} \\
\redtabstep{Eclipse depth} & \redtabcell{$\mathcal{U}(-1000, 1000)$ ppm} & \redtabcell{$\mathcal{U}(-200, 400)$ ppm} & \redtabcell{$\mathcal{U}(0, 1000)$ ppm (consistent with $\mathcal{U}(-1000, 1000)$ ppm)} & \redtabcell{$(R_{\rm p}/R_\star)^2\,\mathcal{U}(0, 0.3)$: up to $\approx 600$ ppm (b), $\approx 415$ ppm (c)} \\
\redtabstep{Transit of b in DD9234\_1} & \redtabcell{$R_{\rm p}/R_\star \sim \mathcal{N}(0.045, 0.001)$} & \redtabcell{$(R_{\rm p}/R_\star)^2 \sim \mathcal{U}(0, 1)$ } & \redtabcell{$R_{\rm p}/R_\star \sim \mathcal{U}(0, 0.01)$} & \redtabcell{As geometry above} \\
\redtabstep{\textbf{Limb darkening law}} & \redtabcell{Quadratic; \texttt{exotic\_ld} with MPS-ATLAS} & \redtabcell{Quadratic; values of JI} & \redtabcell{Quadratic in $u_1$, $u_2$; values of JI} & \redtabcell{Quadratic in $q_1$, $q_2$ (K13b)} \\
\redtabstep{Limb darkening prior} & \redtabcell{$u_1 \sim \mathcal{N}(0.053, 0.005)$, $u_2 \sim \mathcal{N}(0.089, 0.006)$} & \redtabcell{As JI} & \redtabcell{Means of JI, $5\times$ widths} & \redtabcell{$q_1, q_2 \sim \mathcal{U}(0, 1)$} \\
\redtabstep{Stellar parameters} & \redtabcell{$R_\star = 0.2673\,R_\odot$ fixed} & \redtabcell{$R_\star = 0.265\,R_\odot$ fixed only for light delay} & \redtabcell{$R_\star = 0.276\,R_\odot$ for light travel time} & \redtabcell{$M_\star \sim \mathcal{N}_{[0, 1.5]}(0.257, 0.014)$, $R_\star \sim \mathcal{N}_{[0, 1.5]}(0.268, 0.027)$, $T_{\rm eff} \sim \mathcal{N}_{[2000, 7000]}(3340, 150)$} \\
\redtabstep{Baseline} & \redtabcell{$c_0 \sim \mathcal{N}(1, 0.005)$, $c_1 \sim \mathcal{N}(0, 0.01)$, $x_{\rm pos}, y_{\rm pos} \sim \mathcal{U}(-0.1, 0.1)$} & $c_0$, $c_1$ uniform & \redtabcell{baseline offset up to $\pm0.1$ per visit} & \redtabcell{Offset (ppm), trend (ppm d$^{-1}$) $\sim \mathcal{N}(0, 10^7)$} \\
\redtabstep{Exponential ramp} & \redtabcell{$r_0 \sim \mathcal{U}(-0.05, 0.01)$, $r_1 \sim \mathcal{LU}(5, 1500)$} & \redtabcell{$r_0 \sim \mathcal{U}(-0.1, 0.1)$, $r_1 \sim \mathcal{LU}(5, 75)$ min} &  & \redtabcell{Amplitude $\sim \mathcal{U}(-5000, 5000)$ ppm, timescale $\sim \mathcal{LU}(5, 240)$ min} \\
\redtabstep{Jitter (ppm)} & \redtabcell{$\sigma_{\rm white} \sim \mathcal{U}(0, 2000)$} &  $\sigma_{\rm white} \sim \mathcal{LU}(10^{-8}, 10^{0})$ & \redtabcell{$\sigma_{\rm white}\sim \mathcal{LU}(0.1, 1000)$} & \redtabcell{Jitter $\sim \mathcal{LU}(10^{-4}, 10^{6})$} \\
\redtabstep{\textbf{GP amplitude}} & \redtabcell{$\ln\sigma^2_{\rm GP} \sim \mathcal{U}(-24, -14)$} & \redtabcell{$\log_{10}\sigma_{\rm GP} \sim \mathcal{N}(-4.5, 0.5)$} & \redtabcell{$\sigma_{\rm GP}\sim \mathcal{LU}(10^{-8}, 10^{2})$} & \redtabcell{$\sigma \sim \mathcal{LU}(1, 1000)$ ppm} \\
\redtabstep{\textbf{GP timescale}} & \redtabcell{$\ln\rho\,({\rm d}) \sim \mathcal{U}(-7.27, -4.97)$, i.e., 1--10 min} & \redtabcell{$\ln\rho\,({\rm d})\sim\mathcal{U}(-7.27, -3.18)$, i.e., 1-60 min} & \redtabcell{$\sigma_{\rm \rho}\sim \mathcal{LU}(10^{-8}, 10^{2})$} & \redtabcell{$\rho \sim \mathcal{LU}(10.3, 458)$ min} \\\hline
\sidehead{\textit{Sampling}}
\redtabstep{Sampler} & \redtabcell{\texttt{dynesty} static nested sampling} & \redtabcell{NUTS (\texttt{pymc}), with Gibbs slice sampling for parameters near prior bounds} & \redtabcell{\texttt{dynesty} dynamic nested sampling} & \redtabcell{HMC (\texttt{PyMC3}), after staged MAP optimization} \\
\redtabstep{Sampler settings} & \redtabcell{1500 live points} & \redtabcell{10 chains, 2000 tuning steps, 4000 draws} & \redtabcell{500 live points, $rwalk$ sampling method} & \redtabcell{4 chains, 5000 tuning steps, 2500 draws, target acceptance 0.95} \\
\redtabstep{Convergence} & \redtabcell{Estimated $\Delta\ln{Z}<0.001$} & \redtabcell{$\hat{r} < 1.01$, $> 1000$ effective samples} & \redtabcell{Estimated $\Delta\ln{Z}<0.001$} & \redtabcell{$\hat{R} \leq 1.01$, large bulk and tail effective sample size} \\\hline\hline\hline
\sidehead{\textit{Results}}
\redtabstep{Unbinned rms (ppm)} & \redtabcell{626, 579, 628} & --  & \redtabcell{621, 583, 640} &  -- \\
\redtabstep{5$\times$-bin rms (ppm)} & \redtabcell{284, 264, 285} & \redtabcell{295, 276, 298} & \redtabcell{293, 273, 304} &  -- \\
\redtabstep{60 s rms (ppm)} & \redtabcell{116, 112, 111}  & \redtabcell{122, 125, 129} & \redtabcell{121, 115, 133} & \redtabcell{122, 111, 130} \\
\redtabstep{5 min rms (ppm)} & \redtabcell{63, 40, 41} & \redtabcell{58, 56, 60} & \redtabcell{63, 58, 65} & \redtabcell{62, 47, 66} \\
\redtabstep{$F_{\rm p}/F_{\rm s}$ of b (ppm)} & \redtabcell{$149^{+16}_{-17}$} & \redtabcell{$148 \pm 19$} & \redtabcell{$134^{+16}_{-17}$} & \redtabcell{$157^{+18}_{-17}$} \\
\redtabstep{$F_{\rm p}/F_{\rm s}$ of c (ppm)} & \redtabcell{$182 \pm 30$} & \redtabcell{$154^{+48}_{-35}$} & \redtabcell{$165 \pm 32$} & \redtabcell{$154^{+54}_{-37}$} \\
\enddata
\tablecomments{$\mathcal{N}(\mu, \sigma)$: normal; $\mathcal{N}_{[a,\,b]}$: normal truncated to $[a, b]$; $\mathcal{U}(a, b)$: uniform; $\mathcal{LU}(a, b)$: log-uniform; $\log\mathcal{N}(\ln m, \sigma)$: lognormal with median $m$; $\mathcal{B}$: beta. Integrations removed and rms values are in the order DD9234\_1, DD12758, DD9234\_2. A25: \citet{allen25_hot}; K13a: \citet{kipping_parametrizing_2013}; K13b: \citet{kipping_efficient_2013}; L23: \citet{lavie23_planetary}; W22: \citet{winters22_a}; W25: \citet{wachiraphan25_the}.}
\end{deluxetable*}
\clearpage
\section{Incorrect heat redistribution scaling in \helios{}} \label{app:redistribution}

The widely-used RCE code \helios{} adopts the scaling for atmospheric heat redistribution on tidally locked planets from \citet{koll22}.   There was a minor update in this expression for the scaling factor $f$ between the initial submission \citep{Koll2019} (also used in the companion paper \citep{koll19_identifying}) and the accepted manuscript \citep{koll22}.  This was not a typographical error but a result of now connecting horizontal wind speed $U$ directly to the heat absorbed, $U^3 \propto Q \propto \tau_\mathrm{LW}$, and considering the nightside in radiative equilibrium; the updated scaling is indeed more physically accurate (D. Koll, \textit{private comm.}).

\begin{figure*}
    \centering
    \includegraphics[width=0.9\linewidth]{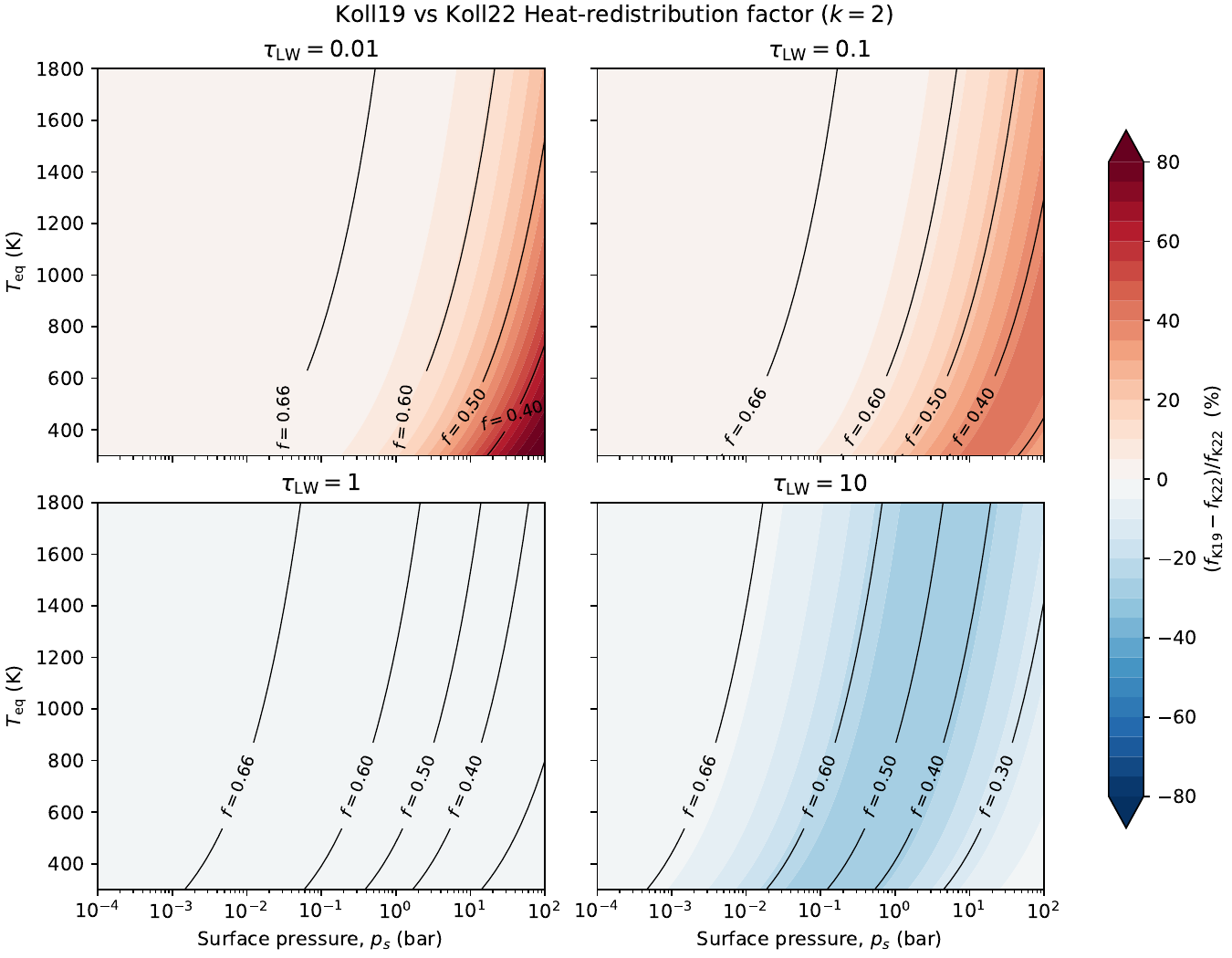}
    \caption{Relative difference in heat redistribution factor $f$ at various values of longwave optical depths $\tau_\mathrm{LW}$, as a function of surface pressure ($P_\mathrm{s}$) and equilibrium temperature ($T_\mathrm{eq}$).  The black lines denote the \textit{updated} redistribution factor $f$, and the color contours denotes how much the \textit{original} scaling in \helios{} over- or underestimated the dayside energy budget.  They are equal when $\tau_\mathrm{LW}=1$. }
    \label{fig:koll_comp}
\end{figure*}

We became aware that \helios{} had used the older, incorrect expression for the redistribution factor, $f$:

\begin{equation}
f = \frac{2}{3} - \frac{5}{12}\,\frac{X}{k + X},
\qquad
X \equiv \tau_\mathrm{LW} \left(\frac{p_\mathrm{s}}{1\,\mathrm{bar}}\right)^{2/3} \left(\frac{T_\mathrm{eq}}{600\,k}\right)^{-4/3},
\label{eq:koll22}
\end{equation}

\noindent wherein the updated scaling has a $1/3$ exponentiation in $\tau_\mathrm{LW}$, such that $X \propto \tau_\mathrm{LW}^{1/3}$, meaning the old expression either over- or underestimated the dayside energy budget, as shown in Figure~\ref{fig:koll_comp}.  The discrepancy is most relevant when $\tau_\mathrm{LW} \ll 1$ or $\tau_\mathrm{LW} \gg 1$; note that the assumptions of the scaling approximation already break down when $\tau_\mathrm{LW} \gg 1$, and its validity in this regime is more empirical and based on benchmarks against global circulation models \citep{koll22}. 

We quantified $f$ across the full grid of atmosphere models of TRAPPIST-1~b in \citet{ih23_constraining}; $\tau_\mathrm{LW}$ spanned $10^{-6}$-44, and the dayside energy budget was overestimated by as much as 19\% and underestimated by as much as 27\%.  For TRAPPIST-1 b ($T_\mathrm{eq}=398$ K), this error was most significant at surface pressures of 1 and 10 bar and less than 2\% at lower pressures.  Given that other rocky planets whose observations were analyzed with \helios{} have hotter $T_\mathrm{eq}$, this error is not as critical to those analyses.  The only planet cooler is TRAPPIST-1~c, and its observation was not interpreted using \helios{} \citep{zieba23}, though the work cites the older scaling in the companion paper \citet{koll19_identifying}.   We note that $k$, a catch-all term that absorbs a number of physical unknowns and varies from planet to planet (fixed to $k=2$ in \helios{}), also has a nominal factor of two uncertainty; this alone produces an uncertainty of up to $\sim17$\% in $f$ near intermediate values of $f$.  

\bibliography{ltt1445a_refs, MyLibrary}{}

@article{agol_analytic_2020,
  title = {Analytic {{Planetary Transit Light Curves}} and {{Derivatives}} for {{Stars}} with {{Polynomial Limb Darkening}}},
  author = {Agol, Eric and Luger, Rodrigo and {Foreman-Mackey}, Daniel},
  year = 2020,
  month = mar,
  journal = {Astron. J.},
  volume = {159},
  pages = {123},
  issn = {0004-6256},
  doi = {10.3847/1538-3881/ab4fee},
  urldate = {2021-07-28}
}

@article{allard_models_2012,
  title = {Models of Very-Low-Mass Stars, Brown Dwarfs and Exoplanets},
  author = {Allard, F. and Homeier, D. and Freytag, B.},
  year = 2012,
  month = jun,
  journal = {Philos. Trans. R. Soc. Lond. Ser. A},
  volume = {370},
  pages = {2765--2777},
  issn = {0962-8436},
  doi = {10.1098/rsta.2011.0269},
  urldate = {2021-03-29}
}

@article{astropy_collaboration_astropy_2018,
  title = {The {{Astropy Project}}: {{Building}} an {{Open}}-Science {{Project}} and {{Status}} of the v2.0 {{Core Package}}},
  shorttitle = {The {{Astropy Project}}},
  author = {{Astropy Collaboration} and {Price-Whelan}, A. M. and Sip{\H o}cz, B. M. and G{\"u}nther, H. M. and Lim, P. L. and Crawford, S. M. and Conseil, S. and Shupe, D. L. and Craig, M. W. and Dencheva, N. and Ginsburg, A. and VanderPlas, J. T. and Bradley, L. D. and {P{\'e}rez-Su{\'a}rez}, D. and {de Val-Borro}, M. and Aldcroft, T. L. and Cruz, K. L. and Robitaille, T. P. and Tollerud, E. J. and Ardelean, C. and Babej, T. and Bach, Y. P. and Bachetti, M. and Bakanov, A. V. and Bamford, S. P. and Barentsen, G. and Barmby, P. and Baumbach, A. and Berry, K. L. and Biscani, F. and Boquien, M. and Bostroem, K. A. and Bouma, L. G. and Brammer, G. B. and Bray, E. M. and Breytenbach, H. and Buddelmeijer, H. and Burke, D. J. and Calderone, G. and Cano Rodr{\'i}guez, J. L. and Cara, M. and Cardoso, J. V. M. and Cheedella, S. and Copin, Y. and Corrales, L. and Crichton, D. and D'Avella, D. and Deil, C. and Depagne, {\'E}. and Dietrich, J. P. and Donath, A. and Droettboom, M. and Earl, N. and Erben, T. and Fabbro, S. and Ferreira, L. A. and Finethy, T. and Fox, R. T. and Garrison, L. H. and Gibbons, S. L. J. and Goldstein, D. A. and Gommers, R. and Greco, J. P. and Greenfield, P. and Groener, A. M. and Grollier, F. and Hagen, A. and Hirst, P. and Homeier, D. and Horton, A. J. and Hosseinzadeh, G. and Hu, L. and Hunkeler, J. S. and Ivezi{\'c}, {\v Z}. and Jain, A. and Jenness, T. and Kanarek, G. and Kendrew, S. and Kern, N. S. and Kerzendorf, W. E. and Khvalko, A. and King, J. and Kirkby, D. and Kulkarni, A. M. and Kumar, A. and Lee, A. and Lenz, D. and Littlefair, S. P. and Ma, Z. and Macleod, D. M. and Mastropietro, M. and McCully, C. and Montagnac, S. and Morris, B. M. and Mueller, M. and Mumford, S. J. and Muna, D. and Murphy, N. A. and Nelson, S. and Nguyen, G. H. and Ninan, J. P. and N{\"o}the, M. and Ogaz, S. and Oh, S. and Parejko, J. K. and Parley, N. and Pascual, S. and Patil, R. and Patil, A. A. and Plunkett, A. L. and Prochaska, J. X. and Rastogi, T. and Reddy Janga, V. and Sabater, J. and Sakurikar, P. and Seifert, M. and Sherbert, L. E. and {Sherwood-Taylor}, H. and Shih, A. Y. and Sick, J. and Silbiger, M. T. and Singanamalla, S. and Singer, L. P. and Sladen, P. H. and Sooley, K. A. and Sornarajah, S. and Streicher, O. and Teuben, P. and Thomas, S. W. and Tremblay, G. R. and Turner, J. E. H. and Terr{\'o}n, V. and {van Kerkwijk}, M. H. and {de la Vega}, A. and Watkins, L. L. and Weaver, B. A. and Whitmore, J. B. and Woillez, J. and Zabalza, V. and {Astropy Contributors}},
  year = 2018,
  month = sep,
  journal = {Astron. J.},
  volume = {156},
  pages = {123},
  issn = {0004-6256},
  doi = {10.3847/1538-3881/aabc4f},
  urldate = {2021-07-28}
}

@article{bennett_additional_2025,
  title = {Additional {{JWST}}/{{NIRSpec Transits}} of the {{Rocky M Dwarf Exoplanet GJ}} 1132 b {{Reveal}} a {{Featureless Spectrum}}},
  author = {Bennett, Katherine A. and MacDonald, Ryan J. and Peacock, Sarah and Perez, Junellie and May, E. M. and Moran, Sarah E. and Alderson, Lili and {Lustig-Yaeger}, Jacob and Wakeford, Hannah R. and Sing, David K. and Stevenson, Kevin B. and Batalha, Natasha E. and {L{\'o}pez-Morales}, Mercedes and Alam, Munazza K. and Lothringer, Joshua D. and Fu, Guangwei and Kirk, James and Valenti, Jeff A. and Mayorga, L. C. and Sotzen, Kristin S.},
  year = 2025,
  month = oct,
  journal = {Astron. J.},
  volume = {170},
  pages = {205},
  publisher = {IOP},
  issn = {0004-6256},
  doi = {10.3847/1538-3881/adf198},
  urldate = {2026-09-29}
}

@misc{berta-thompson_3d_2025,
  title = {The {{3D Cosmic Shoreline}} for {{Nurturing Planetary Atmospheres}}},
  author = {{Berta-Thompson}, Zach K. and Wachiraphan, Patcharapol and Murray, Catriona},
  year = 2025,
  month = jul,
  number = {arXiv:2507.02136},
  eprint = {2507.02136},
  primaryclass = {astro-ph},
  publisher = {arXiv},
  doi = {10.48550/arXiv.2507.02136},
  urldate = {2025-07-04},
  archiveprefix = {arXiv}
}

@article{buchner_x-ray_2014,
  title = {X-Ray Spectral Modelling of the {{AGN}} Obscuring Region in the {{CDFS}}: {{Bayesian}} Model Selection and Catalogue},
  shorttitle = {X-Ray Spectral Modelling of the {{AGN}} Obscuring Region in the {{CDFS}}},
  author = {Buchner, J. and Georgakakis, A. and Nandra, K. and Hsu, L. and Rangel, C. and Brightman, M. and Merloni, A. and Salvato, M. and Donley, J. and Kocevski, D.},
  year = 2014,
  month = apr,
  journal = {Astron. Astrophys.},
  volume = {564},
  pages = {A125},
  publisher = {EDP Sciences},
  issn = {0004-6361, 1432-0746},
  doi = {10.1051/0004-6361/201322971},
  urldate = {2026-01-29},
  copyright = {\copyright{} ESO, 2014},
  langid = {english}
}

@article{burkner_efficient_2021,
  title = {Efficient Leave-One-out Cross-Validation for {{Bayesian}} Non-Factorized Normal and {{Student}}-t Models},
  author = {B{\"u}rkner, Paul-Christian and Gabry, Jonah and Vehtari, Aki},
  year = 2021,
  month = jun,
  journal = {Comput. Stat.},
  volume = {36},
  number = {2},
  pages = {1243--1261},
  issn = {1613-9658},
  doi = {10.1007/s00180-020-01045-4},
  urldate = {2026-10-28},
  langid = {english}
}

@misc{cadieux_transmission_2024,
  title = {Transmission {{Spectroscopy}} of the {{Habitable Zone Exoplanet LHS}} 1140 b with {{JWST}}/{{NIRISS}}},
  author = {Cadieux, Charles and Doyon, Ren{\'e} and MacDonald, Ryan J. and Turbet, Martin and Artigau, {\'E}tienne and Lim, Olivia and Radica, Michael and Fauchez, Thomas J. and Salhi, Salma and Dang, Lisa and Albert, Lo{\"i}c and Coulombe, Louis-Philippe and Cowan, Nicolas B. and Lafreni{\`e}re, David and L'Heureux, Alexandrine and Piaulet, Caroline and Benneke, Bj{\"o}rn and Cloutier, Ryan and Charnay, Benjamin and Cook, Neil J. and {Fournier-Tondreau}, Marylou and Plotnykov, Mykhaylo and Valencia, Diana},
  year = 2024,
  month = jun,
  number = {arXiv:2406.15136},
  eprint = {2406.15136},
  primaryclass = {astro-ph},
  publisher = {arXiv},
  urldate = {2024-06-24},
  archiveprefix = {arXiv}
}

@article{collins_astroimagej_2017,
  title = {{{AstroImageJ}}: {{Image Processing}} and {{Photometric Extraction}} for {{Ultra}}-Precise {{Astronomical Light Curves}}},
  shorttitle = {{{AstroImageJ}}},
  author = {Collins, Karen A. and Kielkopf, John F. and Stassun, Keivan G. and Hessman, Frederic V.},
  year = 2017,
  month = feb,
  journal = {Astron. J.},
  volume = {153},
  pages = {77},
  issn = {0004-6256},
  doi = {10.3847/1538-3881/153/2/77},
  urldate = {2021-02-20}
}

@article{dressing_occurrence_2013,
  title = {The {{Occurrence Rate}} of {{Small Planets}} around {{Small Stars}}},
  author = {Dressing, Courtney D. and Charbonneau, David},
  year = 2013,
  month = apr,
  journal = {Astrophys. J.},
  volume = {767},
  number = {1},
  pages = {95},
  doi = {10.1088/0004-637X/767/1/95},
  urldate = {2019-08-14},
  langid = {english}
}

@article{dressing_occurrence_2015,
  title = {The {{Occurrence}} of {{Potentially Habitable Planets Orbiting M Dwarfs Estimated}} from the {{Full Kepler Dataset}} and an {{Empirical Measurement}} of the {{Detection Sensitivity}}},
  author = {Dressing, Courtney D. and Charbonneau, David},
  year = 2015,
  month = jul,
  journal = {Astrophys. J.},
  volume = {807},
  pages = {45},
  issn = {0004-637X},
  doi = {10.1088/0004-637X/807/1/45},
  urldate = {2019-01-18}
}

@article{feroz_multinest_2009,
  title = {{{MULTINEST}}: An Efficient and Robust {{Bayesian}} Inference Tool for Cosmology and Particle Physics},
  shorttitle = {{{MULTINEST}}},
  author = {Feroz, F. and Hobson, M. P. and Bridges, M.},
  year = 2009,
  month = oct,
  journal = {Mon. Not. R. Astron. Soc.},
  volume = {398},
  pages = {1601--1614},
  publisher = {OUP},
  issn = {0035-8711},
  doi = {10.1111/j.1365-2966.2009.14548.x},
  urldate = {2026-01-29}
}

@misc{foreman-mackey_cornerpy_2016,
  title = {Corner.Py: Corner.Py v1.0.2},
  shorttitle = {Corner.Py},
  author = {{Foreman-Mackey}, Dan and Vousden, Will and {Price-Whelan}, Adrian and Pitkin, Matt and Zabalza, Victor and Ryan, Geoffrey and Emily and Smith, Michael and Ashton, Gregory and Cruz, Kelle and Kerzendorf, Wolfgang and Caswell, Thomas A. and Hoyer, Stephan and Barbary, Kyle and Czekala, Ian and Hogg, David W. and Brewer, Brendon J.},
  year = 2016,
  month = feb,
  doi = {10.5281/zenodo.45906},
  urldate = {2022-03-24},
  howpublished = {Zenodo}
}

@article{foreman-mackey_exoplanet_2021,
  title = {Exoplanet: {{Gradient}}-Based Probabilistic Inference for Exoplanet Data \& Other Astronomical Time Series},
  shorttitle = {Exoplanet},
  author = {{Foreman-Mackey}, Daniel and Luger, Rodrigo and Agol, Eric and Barclay, Thomas and Bouma, Luke and Brandt, Timothy and Czekala, Ian and David, Trevor and Dong, Jiayin and Gilbert, Emily and Gordon, Tyler and Hedges, Christina and Hey, Daniel and Morris, Brett and {Price-Whelan}, Adrian and Savel, Arjun},
  year = 2021,
  month = jun,
  journal = {J. Open Source Softw.},
  volume = {6},
  pages = {3285},
  doi = {10.21105/joss.03285},
  urldate = {2021-08-12}
}

@misc{foreman-mackey_exoplanet-devexoplanet_2021,
  title = {Exoplanet-Dev/Exoplanet v0.4.4},
  author = {{Foreman-Mackey}, Daniel and Savel, Arjun and Luger, Rodrigo and Czekala, Ian and Agol, Eric and {Price-Whelan}, Adrian and Gilbert, Emily and Brandt, Timothy D. and Barclay, Tom and Bouma, Luke},
  year = 2021,
  month = jan,
  doi = {10.5281/zenodo.1998447}
}

@article{foreman-mackey_fast_2017,
  title = {Fast and Scalable {{Gaussian}} Process Modeling with Applications to Astronomical Time Series},
  author = {{Foreman-Mackey}, Daniel and Agol, Eric and Ambikasaran, Sivaram and Angus, Ruth},
  year = 2017,
  month = nov,
  journal = {Astron. J.},
  volume = {154},
  number = {6},
  eprint = {1703.09710},
  pages = {220},
  issn = {1538-3881},
  doi = {10.3847/1538-3881/aa9332},
  urldate = {2020-06-05},
  archiveprefix = {arXiv}
}

@article{foreman-mackey_scalable_2018,
  title = {Scalable Backpropagation for Gaussian Processes Using Celerite},
  author = {{Foreman-Mackey}, D.},
  year = 2018,
  month = feb,
  journal = {Res. Notes Am. Astron. Soc.},
  volume = {2},
  number = {1},
  pages = {31},
  doi = {10.3847/2515-5172/aaaf6c}
}

@misc{gillon_first_2025,
  title = {First {{JWST}} Thermal Phase Curves of Temperate Terrestrial Exoplanets Reveal No Thick Atmosphere around {{TRAPPIST-1}} b and c},
  author = {Gillon, Micha{\"e}l and Ducrot, Elsa and Bell, Taylor J. and Huang, Ziyu and Lincowski, Andrew and Lyu, Xintong and Maurel, Alice and Revol, Alexandre and Agol, Eric and Bolmont, Emeline and Dong, Chuanfei and Fauchez, Thomas J. and Koll, Daniel D. B. and Leconte, J{\'e}r{\'e}my and Meadows, Victoria S. and Selsis, Franck and Turbet, Martin and Charnay, Benjamin and Delre, Laetita and Demory, Brice-Olivier and Householder, Aaron and Zieba, Sebastian and Berardo, David and Dyrek, Achr{\`e}ne and Edwards, Billy and de Wit, Julien and Greene, Thomas P. and Hu, Renyu and Iro, Nicolas and Kreidberg, Laura and Lagage, Pierre-Olivier and {Lustig-Yaeger}, Jacob and Iyer, Aishwarya},
  year = 2025,
  month = sep,
  number = {arXiv:2509.02128},
  eprint = {2509.02128},
  primaryclass = {astro-ph},
  publisher = {arXiv},
  doi = {10.48550/arXiv.2509.02128},
  urldate = {2025-09-03},
  archiveprefix = {arXiv}
}

@article{huehnerhoff_astrophysical_2016,
  title = {Astrophysical {{Research Consortium Telescope Imaging Camera}} ({{ARCTIC}}) Facility Optical Imager for the {{Apache Point Observatory}} 3.5m Telescope},
  author = {Huehnerhoff, Joseph and Ketzeback, William and Bradley, Alaina and Dembicky, Jack and Doughty, Caitlin and Hawley, Suzanne and Johnson, Courtney and Klaene, Mark and Leon, Ed and McMillan, Russet and Owen, Russell and Sayres, Conor and Sheen, Tyler and Shugart, Alysha},
  year = 2016,
  month = aug,
  volume = {9908},
  pages = {99085H},
  doi = {10.1117/12.2234214},
  urldate = {2021-02-21}
}

@article{hunter_matplotlib:_2007,
  title = {Matplotlib: {{A 2D Graphics Environment}}},
  shorttitle = {Matplotlib},
  author = {Hunter, J. D.},
  year = 2007,
  month = may,
  journal = {Comput. Sci. Eng.},
  volume = {9},
  number = {3},
  pages = {90--95},
  issn = {1521-9615},
  doi = {10.1109/MCSE.2007.55}
}

@article{kanodia_toi-5205b_2023,
  title = {{{TOI-5205}}b: {{A Short}}-Period {{Jovian Planet Transiting}} a {{Mid-M Dwarf}}},
  shorttitle = {{{TOI-5205}}b},
  author = {Kanodia, Shubham and Mahadevan, Suvrath and {Libby-Roberts}, Jessica and Stefansson, Gudmundur and Ca{\~n}as, Caleb I. and Piette, Anjali A. A. and Boss, Alan and Teske, Johanna and Chambers, John and Zeimann, Greg and Monson, Andrew and Robertson, Paul and Ninan, Joe P. and Lin, Andrea S. J. and Bender, Chad F. and Cochran, William D. and Diddams, Scott A. and Gupta, Arvind F. and Halverson, Samuel and Hawley, Suzanne and Kobulnicky, Henry A. and Metcalf, Andrew J. and Parker, Brock A. and Powers, Luke and Ramsey, Lawrence W. and Roy, Arpita and Schwab, Christian and Swaby, Tera N. and Terrien, Ryan C. and Wisniewski, John},
  year = 2023,
  month = mar,
  journal = {Astron. J.},
  volume = {165},
  pages = {120},
  issn = {0004-6256},
  doi = {10.3847/1538-3881/acabce},
  urldate = {2023-03-02}
}

@article{kempton_reflective_2023,
  title = {A Reflective, Metal-Rich Atmosphere for {{GJ}} 1214b from Its {{JWST}} Phase Curve},
  author = {Kempton, Eliza M.-R. and Zhang, Michael and Bean, Jacob L. and Steinrueck, Maria E. and Piette, Anjali A. A. and Parmentier, Vivien and Malsky, Isaac and Roman, Michael T. and Rauscher, Emily and Gao, Peter and Bell, Taylor J. and Xue, Qiao and Taylor, Jake and Savel, Arjun B. and Arnold, Kenneth E. and Nixon, Matthew C. and Stevenson, Kevin B. and Mansfield, Megan and Kendrew, Sarah and Zieba, Sebastian and Ducrot, Elsa and Dyrek, Achr{\`e}ne and Lagage, Pierre-Olivier and Stassun, Keivan G. and Henry, Gregory W. and Barman, Travis and Lupu, Roxana and Malik, Matej and Kataria, Tiffany and Ih, Jegug and Fu, Guangwei and Welbanks, Luis and McGill, Peter},
  year = 2023,
  month = may,
  journal = {Nature},
  eprint = {2305.06240},
  primaryclass = {astro-ph},
  issn = {0028-0836, 1476-4687},
  doi = {10.1038/s41586-023-06159-5},
  urldate = {2023-05-11},
  archiveprefix = {arXiv}
}

@article{kipping_efficient_2013,
  title = {Efficient, Uninformative Sampling of Limb Darkening Coefficients for Two-Parameter Laws},
  author = {Kipping, D. M.},
  year = 2013,
  month = nov,
  journal = {Mon. Not. R. Astron. Soc.},
  volume = {435},
  pages = {2152--2160},
  doi = {10.1093/mnras/stt1435}
}

@article{kipping_parametrizing_2013,
  title = {Parametrizing the Exoplanet Eccentricity Distribution with the {{Beta}} Distribution},
  author = {Kipping, David M.},
  year = 2013,
  month = sep,
  journal = {Mon. Not. R. Astron. Soc. Lett.},
  volume = {434},
  number = {1},
  eprint = {1306.4982},
  primaryclass = {astro-ph.EP},
  pages = {L51-L55},
  issn = {1745-3933, 1745-3925},
  doi = {10.1093/mnrasl/slt075},
  urldate = {2026-10-27},
  archiveprefix = {arXiv}
}

@article{kumar_arviz_2019,
  title = {{{ArviZ}} a Unified Library for Exploratory Analysis of {{Bayesian}} Models in {{Python}}},
  author = {Kumar, Ravin and Carroll, Colin and Hartikainen, Ari and Martin, Osvaldo A.},
  year = 2019,
  journal = {J. Open Source Softw.},
  doi = {10.21105/joss.01143}
}

@article{lomb_least-squares_1976,
  title = {Least-Squares Frequency Analysis of Unequally Spaced Data},
  author = {Lomb, N. R.},
  year = 1976,
  month = feb,
  journal = {Astrophys. Space Sci.},
  volume = {39},
  pages = {447--462},
  issn = {0004-640X},
  doi = {10.1007/BF00648343},
  urldate = {2020-04-24}
}

@article{luger_starry_2019,
  title = {Starry: {{Analytic Occultation Light Curves}}},
  shorttitle = {Starry},
  author = {Luger, Rodrigo and Agol, Eric and {Foreman-Mackey}, Daniel and Fleming, David P. and {Lustig-Yaeger}, Jacob and Deitrick, Russell},
  year = 2019,
  month = feb,
  journal = {Astron. J.},
  volume = {157},
  pages = {64},
  issn = {0004-6256},
  doi = {10.3847/1538-3881/aae8e5},
  urldate = {2021-07-28}
}

@inproceedings{mckinney-proc-scipy-2010,
  title = {Data {{Structures}} for {{Statistical Computing}} in {{Python}}},
  booktitle = {Proc. 9th {{Python Sci}}. {{Conf}}.},
  author = {McKinney, Wes},
  editor = {{van der Walt}, St{\'e}fan and Millman, Jarrod},
  year = 2010,
  pages = {56--61},
  doi = {10.25080/Majora-92bf1922-00a}
}

@book{oliphant_numpy:_2006,
  title = {{{NumPy}}: {{A}} Guide to {{NumPy}}},
  author = {Oliphant, Travis},
  year = 2006
}

@article{oliphant_python_2007,
  title = {Python for {{Scientific Computing}}},
  author = {Oliphant, T. E.},
  year = 2007,
  month = may,
  journal = {Comput. Sci. Eng.},
  volume = {9},
  number = {3},
  pages = {10--20},
  issn = {1521-9615},
  doi = {10.1109/MCSE.2007.58}
}

@article{perez_ipython:_2007,
  title = {{{IPython}}: A {{System}} for {{Interactive Scientific Computing}}},
  author = {P{\'e}rez, Fernando and Granger, Brian E.},
  year = 2007,
  month = may,
  journal = {Comput. Sci. Eng.},
  volume = {9},
  number = {3},
  pages = {21--29},
  issn = {1521-9615},
  doi = {10.1109/MCSE.2007.53}
}

@article{robitaille_astropy:_2013,
  title = {Astropy: {{A}} Community {{Python}} Package for Astronomy},
  shorttitle = {Astropy},
  author = {Robitaille, Thomas P. and Tollerud, Erik J. and Greenfield, Perry and Droettboom, Michael and Bray, Erik and Aldcroft, Tom and Davis, Matt and Ginsburg, Adam and {Price-Whelan}, Adrian M. and Kerzendorf, Wolfgang E. and Conley, Alexander and Crighton, Neil and Barbary, Kyle and Muna, Demitri and Ferguson, Henry and Grollier, Fr{\'e}d{\'e}ric and Parikh, Madhura M. and Nair, Prasanth H. and G{\"u}nther, Hans M. and Deil, Christoph and Woillez, Julien and Conseil, Simon and Kramer, Roban and Turner, James E. H. and Singer, Leo and Fox, Ryan and Weaver, Benjamin A. and Zabalza, Victor and Edwards, Zachary I. and Bostroem, K. Azalee and Burke, D. J. and Casey, Andrew R. and Crawford, Steven M. and Dencheva, Nadia and Ely, Justin and Jenness, Tim and Labrie, Kathleen and Lim, Pey Lian and Pierfederici, Francesco and Pontzen, Andrew and Ptak, Andy and Refsdal, Brian and Servillat, Mathieu and Streicher, Ole},
  year = 2013,
  month = oct,
  journal = {Astron. Astrophys.},
  volume = {558},
  pages = {A33},
  issn = {0004-6361, 1432-0746},
  doi = {10.1051/0004-6361/201322068},
  urldate = {2019-01-16},
  copyright = {\copyright{} ESO, 2013},
  langid = {english}
}

@article{salvatier_probabilistic_2016,
  title = {Probabilistic Programming in {{Python}} Using {{PyMC3}}},
  author = {Salvatier, John and Wiecki, Thomas V and Fonnesbeck, Christopher},
  year = 2016,
  journal = {PeerJ Comput. Sci.},
  volume = {2},
  pages = {e55},
  publisher = {PeerJ Inc.}
}

@article{scargle_studies_1982,
  title = {Studies in Astronomical Time Series Analysis. {{II}} - {{Statistical}} Aspects of Spectral Analysis of Unevenly Spaced Data},
  author = {Scargle, J. D.},
  year = 1982,
  month = dec,
  journal = {Astrophys. J.},
  volume = {263},
  pages = {835--853},
  issn = {0004-637X},
  doi = {10.1086/160554},
  urldate = {2020-04-24}
}

@article{sivula_uncertainty_2025,
  title = {Uncertainty in {{Bayesian Leave-One-Out Cross-Validation Based Model Comparison}}},
  author = {Sivula, Tuomas and Magnusson, M{\aa}ns and Matamoros, Asael Alonzo and Vehtari, Aki},
  year = 2025,
  month = jan,
  journal = {Bayesian Anal.},
  volume = {-1},
  number = {-1},
  pages = {1--31},
  publisher = {International Society for Bayesian Analysis},
  issn = {1936-0975, 1931-6690},
  doi = {10.1214/25-BA1569},
  urldate = {2026-10-19},
  langid = {english}
}

@article{stefansson_toward_2017,
  title = {Toward {{Space}}-like {{Photometric Precision}} from the {{Ground}} with {{Beam}}-Shaping {{Diffusers}}},
  author = {Stefansson, Gudmundur and Mahadevan, Suvrath and Hebb, Leslie and Wisniewski, John and Huehnerhoff, Joseph and Morris, Brett and Halverson, Sam and Zhao, Ming and Wright, Jason and O'rourke, Joseph and Knutson, Heather and Hawley, Suzanne and Kanodia, Shubham and Li, Yiting and Hagen, Lea M. Z. and Liu, Leo J. and Beatty, Thomas and Bender, Chad and Robertson, Paul and Dembicky, Jack and Gray, Candace and Ketzeback, William and McMillan, Russet and Rudyk, Theodore},
  year = 2017,
  month = oct,
  journal = {Astrophys. J.},
  volume = {848},
  pages = {9},
  issn = {0004-637X},
  doi = {10.3847/1538-4357/aa88aa},
  urldate = {2021-03-26}
}

@article{the_theano_development_team_theano_2016,
  title = {Theano: {{A Python}} Framework for Fast Computation of Mathematical Expressions},
  shorttitle = {Theano},
  author = {{The Theano Development Team} and {Al-Rfou}, Rami and Alain, Guillaume and Almahairi, Amjad and Angermueller, Christof and Bahdanau, Dzmitry and Ballas, Nicolas and Bastien, Fr{\'e}d{\'e}ric and Bayer, Justin and Belikov, Anatoly and Belopolsky, Alexander and Bengio, Yoshua and Bergeron, Arnaud and Bergstra, James and Bisson, Valentin and Bleecher Snyder, Josh and Bouchard, Nicolas and {Boulanger-Lewandowski}, Nicolas and Bouthillier, Xavier and {de Br{\'e}bisson}, Alexandre and Breuleux, Olivier and Carrier, Pierre-Luc and Cho, Kyunghyun and Chorowski, Jan and Christiano, Paul and Cooijmans, Tim and C{\^o}t{\'e}, Marc-Alexandre and C{\^o}t{\'e}, Myriam and Courville, Aaron and Dauphin, Yann N. and Delalleau, Olivier and Demouth, Julien and Desjardins, Guillaume and Dieleman, Sander and Dinh, Laurent and Ducoffe, M{\'e}lanie and Dumoulin, Vincent and Ebrahimi Kahou, Samira and Erhan, Dumitru and Fan, Ziye and Firat, Orhan and Germain, Mathieu and Glorot, Xavier and Goodfellow, Ian and Graham, Matt and Gulcehre, Caglar and Hamel, Philippe and Harlouchet, Iban and Heng, Jean-Philippe and Hidasi, Bal{\'a}zs and Honari, Sina and Jain, Arjun and Jean, S{\'e}bastien and Jia, Kai and Korobov, Mikhail and Kulkarni, Vivek and Lamb, Alex and Lamblin, Pascal and Larsen, Eric and Laurent, C{\'e}sar and Lee, Sean and Lefrancois, Simon and Lemieux, Simon and L{\'e}onard, Nicholas and Lin, Zhouhan and Livezey, Jesse A. and Lorenz, Cory and Lowin, Jeremiah and Ma, Qianli and Manzagol, Pierre-Antoine and Mastropietro, Olivier and McGibbon, Robert T. and Memisevic, Roland and {van Merri{\"e}nboer}, Bart and Michalski, Vincent and Mirza, Mehdi and Orlandi, Alberto and Pal, Christopher and Pascanu, Razvan and Pezeshki, Mohammad and Raffel, Colin and Renshaw, Daniel and Rocklin, Matthew and Romero, Adriana and Roth, Markus and Sadowski, Peter and Salvatier, John and Savard, Fran{\c c}ois and Schl{\"u}ter, Jan and Schulman, John and Schwartz, Gabriel and Vlad Serban, Iulian and Serdyuk, Dmitriy and Shabanian, Samira and Simon, {\'E}tienne and Spieckermann, Sigurd and Ramana Subramanyam, S. and Sygnowski, Jakub and Tanguay, J{\'e}r{\'e}mie and {van Tulder}, Gijs and Turian, Joseph and Urban, Sebastian and Vincent, Pascal and Visin, Francesco and {de Vries}, Harm and {Warde-Farley}, David and Webb, Dustin J. and Willson, Matthew and Xu, Kelvin and Xue, Lijun and Yao, Li and Zhang, Saizheng and Zhang, Ying},
  year = 2016,
  month = may,
  journal = {arXiv e-prints},
  pages = {arXiv:1605.02688},
  urldate = {2021-07-28}
}

@article{vehtari_practical_2017,
  title = {Practical {{Bayesian}} Model Evaluation Using Leave-One-out Cross-Validation and {{WAIC}}},
  author = {Vehtari, Aki and Gelman, Andrew and Gabry, Jonah},
  year = 2017,
  month = sep,
  journal = {Stat. Comput.},
  volume = {27},
  number = {5},
  pages = {1413--1432},
  issn = {1573-1375},
  doi = {10.1007/s11222-016-9696-4},
  urldate = {2026-09-05},
  langid = {english}
}

@article{vehtari_rank-normalization_2021,
  title = {Rank-Normalization, Folding, and Localization: {{An}} Improved {{R}}-Hat for Assessing Convergence of {{MCMC}} (with {{Discussion}})},
  shorttitle = {Rank-Normalization, Folding, and Localization},
  author = {Vehtari, Aki and Gelman, Andrew and Simpson, Daniel and Carpenter, Bob and B{\"u}rkner, Paul-Christian},
  year = 2021,
  month = jun,
  journal = {Bayesian Anal.},
  volume = {16},
  pages = {667--718},
  doi = {10.1214/20-BA1221},
  urldate = {2026-09-05}
}

@article{virtanen_scipy_2020,
  title = {{{SciPy}} 1.0: {{Fundamental Algorithms}} for {{Scientific Computing}} in {{Python}}},
  author = {Virtanen, Pauli and Gommers, Ralf and Oliphant, Travis E. and Haberland, Matt and Reddy, Tyler and Cournapeau, David and Burovski, Evgeni and Peterson, Pearu and Weckesser, Warren and Bright, Jonathan and {van der Walt}, St{\'e}fan J. and Brett, Matthew and Wilson, Joshua and Jarrod Millman, K. and Mayorov, Nikolay and Nelson, Andrew R. J. and Jones, Eric and Kern, Robert and Larson, Eric and Carey, {\relax CJ} and Polat, {\.I}lhan and Feng, Yu and Moore, Eric W. and {Vand erPlas}, Jake and Laxalde, Denis and Perktold, Josef and Cimrman, Robert and Henriksen, Ian and Quintero, E. A. and Harris, Charles R and Archibald, Anne M. and Ribeiro, Ant{\^o}nio H. and Pedregosa, Fabian and {van Mulbregt}, Paul and Contributors, SciPy 1. 0},
  year = 2020,
  journal = {Nat. Methods},
  volume = {17},
  pages = {261--272},
  doi = {10.1038/s41592-019-0686-2}
}

@article{zechmeister_generalised_2009,
  title = {The Generalised {{Lomb-Scargle}} Periodogram. {{A}} New Formalism for the Floating-Mean and {{Keplerian}} Periodograms},
  author = {Zechmeister, M. and K{\"u}rster, M.},
  year = 2009,
  month = mar,
  journal = {Astron. Astrophys.},
  volume = {496},
  pages = {577--584},
  issn = {0004-6361},
  doi = {10.1051/0004-6361:200811296},
  urldate = {2020-04-24}
}

@software{bushouse_jwst,
  author       = {{Bushouse}, Howard and {Eisenhamer}, Jonathan and {Dencheva}, Nadia and
                  {Davies}, James and {Greenfield}, Perry and {Morrison}, Jane and
                  {Hodge}, Phil and {Simon}, Bernie and {Grumm}, David and
                  {Droettboom}, Michael and {Slavich}, Edward and {Sosey}, Megan and
                  {Pauly}, Tyler and {Miller}, Todd and {Jedrzejewski}, Robert and
                  {Hack}, Warren and {Davis}, David and {Crawford}, Steven and
                  {Law}, David and {Gordon}, Karl and {Regan}, Michael and
                  {Cara}, Mihai and {MacDonald}, Ken and {Bradley}, Larry and
                  {Shanahan}, Clare and {Jamieson}, William and {Teodoro}, Mairan and
                  {Williams}, Thomas and {Pena-Guerrero}, Maria and {Graham}, Brett and
                  {Molter}, Edward and {Brandt}, Timothy and {Hayes}, Christian and
                  {Cooper}, Rachel and {Clarke}, Melanie and {Filippazzo}, Joseph},
  title        = {{JWST Calibration Pipeline}},
  version      = {3.0.0},
  publisher    = {Zenodo},

  doi          = {10.5281/zenodo.6984365},
  url          = {https://doi.org/10.5281/zenodo.6984365}
}

@article{pymc2023,
  title = {{PyMC}: A Modern and Comprehensive Probabilistic Programming Framework in {P}ython},
  author = {Oriol Abril-Pla and Virgile Andreani and Colin Carroll and Larry Dong and Christopher J. Fonnesbeck and Maxim Kochurov and Ravin Kumar and Junpeng Lao and Christian C. Luhmann and Osvaldo A. Martin and Michael Osthege and Ricardo Vieira and Thomas Wiecki and Robert Zinkov },
  journal = {{PeerJ} Computer Science},
  volume = {9},
  number = {e1516},
  doi = {10.7717/peerj-cs.1516},
  year = {2023}
}

@ARTICLE{bandfield15,
  author  = {{Bandfield}, Joshua L. and {Hayne}, Paul O. and {Williams}, Jean-Pierre and
             {Greenhagen}, Benjamin T. and {Paige}, David A.},
  title   = {Lunar surface roughness derived from LRO Diviner Radiometer observations},
  journal = {Icarus},
  year    = 2015,
  volume  = 248,
  pages   = {357-372},
  doi     = {10.1016/j.icarus.2014.11.009}
}

@ARTICLE{ranjan17,
       author = {{Ranjan}, Sukrit and {Sasselov}, Dimitar D.},
        title = "{Constraints on the Early Terrestrial Surface UV Environment Relevant to Prebiotic Chemistry}",
      journal = {Astrobiology},
         year = 2017,
        month = mar,
       volume = {17},
       number = {3},
        pages = {169-204},
          doi = {10.1089/ast.2016.1519},
archivePrefix = {arXiv},
       eprint = {1610.06223},
 primaryClass = {astro-ph.EP},
       adsurl = {https://ui.adsabs.harvard.edu/abs/2017AsBio..17..169R}
}

@ARTICLE{Agol2016,
       author = {{Agol}, Eric and {Deck}, Katherine},
        title = "{Transit Timing to First Order in Eccentricity}",
      journal = {\apj},
         year = 2016,
        month = feb,
       volume = {818},
       number = {2},
          eid = {177},
        pages = {177},
          doi = {10.3847/0004-637X/818/2/177},
archivePrefix = {arXiv},
       eprint = {1509.01623},
 primaryClass = {astro-ph.EP},
       adsurl = {https://ui.adsabs.harvard.edu/abs/2016ApJ...818..177A}
}

@ARTICLE{Spencer2000,
       author = {{Spencer}, John R. and {Rathbun}, Julie A. and {Travis}, Larry D. and {Tamppari}, Leslie K. and {Barnard}, Laura and {Martin}, Terry Z. and {McEwen}, Alfred S.},
        title = "{Io's Thermal Emission from the Galileo Photopolarimeter- Radiometer}",
      journal = {Science},
         year = 2000,
        month = may,
       volume = {288},
       number = {5469},
        pages = {1198-1201},
          doi = {10.1126/science.288.5469.1198},
       adsurl = {https://ui.adsabs.harvard.edu/abs/2000Sci...288.1198S}
}

@ARTICLE{Veeder2004,
       author = {{Veeder}, Glenn J. and {Matson}, Dennis L. and {Johnson}, Torrence V. and {Davies}, Ashley G. and {Blaney}, Diana L.},
        title = "{The polar contribution to the heat flow of Io}",
      journal = {\icarus},
         year = 2004,
        month = may,
       volume = {169},
       number = {1},
        pages = {264-270},
          doi = {10.1016/j.icarus.2003.11.016},
       adsurl = {https://ui.adsabs.harvard.edu/abs/2004Icar..169..264V}
}

@ARTICLE{deKleer2019,
       author = {{de Kleer}, Katherine and {Nimmo}, Francis and {Kite}, Edwin},
        title = "{Variability in Io's Volcanism on Timescales of Periodic Orbital Changes}",
      journal = {\grl},
         year = 2019,
        month = jun,
       volume = {46},
       number = {12},
        pages = {6327-6332},
          doi = {10.1029/2019GL082691},
       adsurl = {https://ui.adsabs.harvard.edu/abs/2019GeoRL..46.6327D}
}

@ARTICLE{Eggleton1998,
       author = {{Eggleton}, Peter P. and {Kiseleva}, Ludmila G. and {Hut}, Piet},
        title = "{The Equilibrium Tide Model for Tidal Friction}",
      journal = {\apj},
         year = 1998,
        month = may,
       volume = {499},
       number = {2},
        pages = {853-870},
          doi = {10.1086/305670},
archivePrefix = {arXiv},
       eprint = {astro-ph/9801246},
 primaryClass = {astro-ph},
       adsurl = {https://ui.adsabs.harvard.edu/abs/1998ApJ...499..853E}
}

@ARTICLE{Seligman2024,
       author = {{Seligman}, Darryl Z. and {Feinstein}, Adina D. and {Lai}, Dong and {Welbanks}, Luis and {Taylor}, Aster G. and {Becker}, Juliette and {Adams}, Fred C. and {Morgan}, Marvin and {Bergner}, Jennifer B.},
        title = "{Potential Melting of Extrasolar Planets by Tidal Dissipation}",
      journal = {\apj},
         year = 2024,
        month = jan,
       volume = {961},
       number = {1},
          eid = {22},
        pages = {22},
          doi = {10.3847/1538-4357/ad0b82},
archivePrefix = {arXiv},
       eprint = {2311.01187},
 primaryClass = {astro-ph.EP},
       adsurl = {https://ui.adsabs.harvard.edu/abs/2024ApJ...961...22S}
}

@ARTICLE{Veeder2012,
       author = {{Veeder}, Glenn J. and {Davies}, Ashley Gerard and {Matson}, Dennis L. and {Johnson}, Torrence V. and {Williams}, David A. and {Radebaugh}, Jani},
        title = "{Io: Volcanic thermal sources and global heat flow}",
      journal = {\icarus},
         year = 2012,
        month = jun,
       volume = {219},
       number = {2},
        pages = {701-722},
          doi = {10.1016/j.icarus.2012.04.004},
       adsurl = {https://ui.adsabs.harvard.edu/abs/2012Icar..219..701V}
}

@ARTICLE{Lu2023,
       author = {{Lu}, Tiger and {Rein}, Hanno and {Tamayo}, Daniel and {Hadden}, Sam and {Mardling}, Rosemary and {Millholland}, Sarah C. and {Laughlin}, Gregory},
        title = "{Self-consistent Spin, Tidal, and Dynamical Equations of Motion in the REBOUNDx Framework}",
      journal = {\apj},
         year = 2023,
        month = may,
       volume = {948},
       number = {1},
          eid = {41},
        pages = {41},
          doi = {10.3847/1538-4357/acc06d},
archivePrefix = {arXiv},
       eprint = {2303.00006},
 primaryClass = {astro-ph.EP},
       adsurl = {https://ui.adsabs.harvard.edu/abs/2023ApJ...948...41L}
}

@ARTICLE{Greklek-McKeon2025a,
       author = {{Greklek-McKeon}, Michael and {Vissapragada}, Shreyas and {Knutson}, Heather A. and {Fukui}, Akihiko and {Saidel}, Morgan and {Gomez Barrientos}, Jonathan and {Levine}, W. Garrett and {Behmard}, Aida and {Batygin}, Konstantin and {Chachan}, Yayaati and {Vasisht}, Gautam and {Hu}, Renyu and {Cloutier}, Ryan and {Latham}, David and {L{\'o}pez-Morales}, Mercedes and {Vanderburg}, Andrew and {Heffner}, Carolyn and {Nied}, Paul and {Milburn}, Jennifer and {Wilson}, Isaac and {Roderick}, Diana and {Koviak}, Kathleen and {Barlow}, Tom and {Stone}, John F. and {Kiman}, Rocio and {Korth}, Judith and {de Leon}, Jerome P. and {Fukuda}, Izuru and {Hayashi}, Yuya and {Ikoma}, Masahiro and {Ikuta}, Kai and {Isogai}, Keisuke and {Kawai}, Yugo and {Kawauchi}, Kiyoe and {Kusakabe}, Nobuhiko and {Livingston}, John H. and {Mori}, Mayuko and {Narita}, Norio and {Tamura}, Motohide and {Watanabe}, Noriharu and {Fern{\'a}ndez-Rodr{\'\i}guez}, Gareb},
        title = "{Tidally Heated Sub-Neptunes, Refined Planetary Compositions, and Confirmation of a Third Planet in the TOI-1266 System}",
      journal = {\aj},
         year = 2025,
        month = jun,
       volume = {169},
       number = {6},
          eid = {292},
        pages = {292},
          doi = {10.3847/1538-3881/adc0fe},
archivePrefix = {arXiv},
       eprint = {2409.16374},
 primaryClass = {astro-ph.EP},
       adsurl = {https://ui.adsabs.harvard.edu/abs/2025AJ....169..292G}
}

@ARTICLE{Greklek-McKeon2025b,
       author = {{Greklek-McKeon}, Michael and {Knutson}, Heather A. and {Levine}, W. Garrett and {Hu}, Renyu and {Saidel}, Morgan and {Gomez Barrientos}, Jonathan and {Batygin}, Konstantin and {Benneke}, Bj{\"o}rn},
        title = "{Updated Mass, Eccentricity, and Tidal Heating Constraints for the Earth-sized Planet LP 791-18 d}",
      journal = {\aj},
         year = 2025,
        month = aug,
       volume = {170},
       number = {2},
          eid = {65},
        pages = {65},
          doi = {10.3847/1538-3881/addf2d},
archivePrefix = {arXiv},
       eprint = {2501.18700},
 primaryClass = {astro-ph.EP},
       adsurl = {https://ui.adsabs.harvard.edu/abs/2025AJ....170...65G}
}

@ARTICLE{Goldreich1966,
       author = {{Goldreich}, Peter and {Soter}, Steven},
        title = "{Q in the Solar System}",
      journal = {\icarus},
         year = 1966,
        month = jan,
       volume = {5},
       number = {1},
        pages = {375-389},
          doi = {10.1016/0019-1035(66)90051-0},
       adsurl = {https://ui.adsabs.harvard.edu/abs/1966Icar....5..375G}
}

@ARTICLE{Peterson2023,
       author = {{Peterson}, Merrin S. and {Benneke}, Bj{\"o}rn and {Collins}, Karen and {Piaulet}, Caroline and {Crossfield}, Ian J.~M. and {Ali-Dib}, Mohamad and {Christiansen}, Jessie L. and {Gagn{\'e}}, Jonathan and {Faherty}, Jackie and {Kite}, Edwin and {Dressing}, Courtney and {Charbonneau}, David and {Murgas}, Felipe and {Cointepas}, Marion and {Almenara}, Jose Manuel and {Bonfils}, Xavier and {Kane}, Stephen and {Werner}, Michael W. and {Gorjian}, Varoujan and {Roy}, Pierre-Alexis and {Shporer}, Avi and {Pozuelos}, Francisco J. and {Socia}, Quentin Jay and {Cloutier}, Ryan and {Dietrich}, Jeremy and {Irwin}, Jonathan and {Weiss}, Lauren and {Waalkes}, William and {Berta-Thomson}, Zach and {Evans}, Thomas and {Apai}, Daniel and {Parviainen}, Hannu and {Pall{\'e}}, Enric and {Narita}, Norio and {Howard}, Andrew W. and {Dragomir}, Diana and {Barkaoui}, Khalid and {Gillon}, Micha{\"e}l and {Jehin}, Emmanuel and {Ducrot}, Elsa and {Benkhaldoun}, Zouhair and {Fukui}, Akihiko and {Mori}, Mayuko and {Nishiumi}, Taku and {Kawauchi}, Kiyoe and {Ricker}, George and {Latham}, David W. and {Winn}, Joshua N. and {Seager}, Sara and {Isaacson}, Howard and {Bixel}, Alex and {Gibbs}, Aidan and {Jenkins}, Jon M. and {Smith}, Jeffrey C. and {Chavez}, Jose Perez and {Rackham}, Benjamin V. and {Henning}, Thomas and {Gabor}, Paul and {Chen}, Wen-Ping and {Espinoza}, Nestor and {Jensen}, Eric L.~N. and {Collins}, Kevin I. and {Schwarz}, Richard P. and {Conti}, Dennis M. and {Wang}, Gavin and {Kielkopf}, John F. and {Mao}, Shude and {Horne}, Keith and {Sefako}, Ramotholo and {Quinn}, Samuel N. and {Moldovan}, Dan and {Fausnaugh}, Michael and {F{\.z}{\.z}r{\'e}sz}, G{\'a}bor and {Barclay}, Thomas},
        title = "{A temperate Earth-sized planet with tidal heating transiting an M6 star}",
      journal = {\nat},
         year = 2023,
        month = may,
       volume = {617},
       number = {7962},
        pages = {701-705},
          doi = {10.1038/s41586-023-05934-8},
       adsurl = {https://ui.adsabs.harvard.edu/abs/2023Natur.617..701P}
}

@ARTICLE{Puranam2018,
       author = {{Puranam}, Abhijit and {Batygin}, Konstantin},
        title = "{Chaotic Excitation and Tidal Damping in the GJ 876 System}",
      journal = {\aj},
         year = 2018,
        month = apr,
       volume = {155},
       number = {4},
          eid = {157},
        pages = {157},
          doi = {10.3847/1538-3881/aab09f},
archivePrefix = {arXiv},
       eprint = {1802.08385},
 primaryClass = {astro-ph.EP},
       adsurl = {https://ui.adsabs.harvard.edu/abs/2018AJ....155..157P}
}

@ARTICLE{Zuniga-Fernandez2025,
       author = {{Z{\'u}{\~n}iga-Fern{\'a}ndez}, S. and {Pozuelos}, F.~J. and {D{\'e}vora-Pajares}, M. and {Cuello}, N. and {Greklek-McKeon}, M. and {Stassun}, K.~G. and {Van Grootel}, V. and {Rojas-Ayala}, B. and {Korth}, J. and {G{\"u}nther}, M.~N. and {Burgasser}, A.~J. and {Hsu}, C. and {Rackham}, B.~V. and {Barkaoui}, K. and {Timmermans}, M. and {Cadieux}, C. and {Alonso}, R. and {Strakhov}, I.~A. and {Howell}, S.~B. and {Littlefield}, C. and {Furlan}, E. and {Amado}, P.~J. and {Jenkins}, J.~M. and {Twicken}, J.~D. and {Sucerquia}, M. and {Davis}, Y.~T. and {Schanche}, N. and {Collins}, K.~A. and {Burdanov}, A. and {Davoudi}, F. and {Demory}, B.-O. and {Delrez}, L. and {Dransfield}, G. and {Ducrot}, E. and {Garcia}, L.~J. and {Gillon}, M. and {G{\'o}mez Maqueo Chew}, Y. and {Jan{\'o} Mu{\~n}oz}, C. and {Jehin}, E. and {Murray}, C.~A. and {Niraula}, P. and {Pedersen}, P.~P. and {Queloz}, D. and {Rebolo-L{\'o}pez}, R. and {Scott}, M.~G. and {Sebastian}, D. and {Hooton}, M.~J. and {Thompson}, S.~J. and {Triaud}, A.~H.~M.~J. and {de Wit}, J. and {Ghachoui}, M. and {Benkhaldoun}, Z. and {Doyon}, R. and {Lafreni{\`e}re}, D. and {Casanova}, V. and {Sota}, A. and {Plauchu-Frayn}, I. and {Khandelwal}, A. and {Zong Lang}, F. and {Schroffenegger}, U. and {Wampfler}, S. and {Lendl}, M. and {Schwarz}, R.~P. and {Murgas}, F. and {Palle}, E. and {Parviainen}, H.},
        title = "{Two warm Earth-sized exoplanets and an Earth-sized candidate in the M5V-M6V binary system TOI-2267}",
      journal = {\aap},
         year = 2025,
        month = oct,
       volume = {702},
          eid = {A85},
        pages = {A85},
          doi = {10.1051/0004-6361/202554419},
archivePrefix = {arXiv},
       eprint = {2508.14176},
 primaryClass = {astro-ph.EP},
       adsurl = {https://ui.adsabs.harvard.edu/abs/2025A&A...702A..85Z}
}

@ARTICLE{Tamayo2020,
       author = {{Tamayo}, Daniel and {Rein}, Hanno and {Shi}, Pengshuai and {Hernandez}, David M.},
        title = "{REBOUNDx: a library for adding conservative and dissipative forces to otherwise symplectic N-body integrations}",
      journal = {\mnras},
         year = 2020,
        month = jan,
       volume = {491},
       number = {2},
        pages = {2885-2901},
          doi = {10.1093/mnras/stz2870},
archivePrefix = {arXiv},
       eprint = {1908.05634},
 primaryClass = {astro-ph.EP},
       adsurl = {https://ui.adsabs.harvard.edu/abs/2020MNRAS.491.2885T}
}

@ARTICLE{he21_scattering,
       author = {{He}, Quanfu and {Fang}, Zheng and {Shoshanim}, Ofir and {Brown}, Steven S. and {Rudich}, Yinon},
        title = "{Scattering and absorption cross sections of atmospheric gases in the ultraviolet-visible wavelength range (307-725 nm)}",
      journal = {Atmospheric Chemistry \& Physics},
         year = 2021,
        month = oct,
       volume = {21},
       number = {19},
        pages = {14927-14940},
          doi = {10.5194/acp-21-14927-2021},
       adsurl = {https://ui.adsabs.harvard.edu/abs/2021ACP....2114927H}
}

@ARTICLE{palmer75,
       author = {{Palmer}, K.~F. and {Williams}, D.},
        title = "{Optical constants of sulfuric acid; application to the clouds of Venus?}",
      journal = {\ao},
         year = 1975,
        month = jan,
       volume = {14},
       number = {1},
        pages = {208-219},
          doi = {10.1364/AO.14.000208},
       adsurl = {https://ui.adsabs.harvard.edu/abs/1975ApOpt..14..208P}
}

@ARTICLE{terragni25,
       author = {{Terragni}, J. and {Gordon}, I.~E. and {Adkins}, E.~M. and {Boulet}, C. and {Campargue}, A. and {Chistikov}, D. and {Finenko}, A. and {Finkenzeller}, H. and {Fleurbaey}, H. and {Hargreaves}, R.~J. and {Hanson}, R.~K. and {Hartmann}, J.-M. and {Klingberg}, A. and {Kohler}, E. and {Koroleva}, A.~O. and {Mondelain}, D. and {Piccioni}, G. and {Stefani}, S. and {Strand}, C.~L. and {Tran}, H. and {Turbet}, M. and {Vigasin}, A. and {Vitali}, F. and {Volkamer}, R. and {Wei}, C.},
        title = "{Collision induced absorption in HITRAN2024: Enhanced and improved data for atmospheric and planetary studies}",
      journal = {\jqsrt},
         year = 2025,
        month = dec,
       volume = {347},
          eid = {109631},
        pages = {109631},
          doi = {10.1016/j.jqsrt.2025.109631},
       adsurl = {https://ui.adsabs.harvard.edu/abs/2025JQSRT.34709631T}
}

@ARTICLE{rozitis11,
       author = {{Rozitis}, B. and {Green}, S.~F.},
        title = "{Directional characteristics of thermal-infrared beaming from atmosphereless planetary surfaces - a new thermophysical model}",
      journal = {\mnras},
         year = 2011,
        month = aug,
       volume = {415},
       number = {3},
        pages = {2042-2062},
          doi = {10.1111/j.1365-2966.2011.18718.x},
       adsurl = {https://ui.adsabs.harvard.edu/abs/2011MNRAS.415.2042R}
}

@ARTICLE{lagerros98_thermal,
       author = {{Lagerros}, Johan S.~V.},
        title = "{Thermal physics of asteroids. IV. Thermal infrared beaming}",
      journal = {\aap},
         year = 1998,
        month = apr,
       volume = {332},
        pages = {1123-1132},
       adsurl = {https://ui.adsabs.harvard.edu/abs/1998A&A...332.1123L}
}

@ARTICLE{spencer90_rough,
       author = {{Spencer}, J.~R.},
        title = "{A rough-surface thermophysical model for airless planets}",
      journal = {\icarus},
         year = 1990,
        month = jan,
       volume = {83},
       number = {1},
        pages = {27-38},
          doi = {10.1016/0019-1035(90)90004-S},
       adsurl = {https://ui.adsabs.harvard.edu/abs/1990Icar...83...27S}
}

@article{wolff09,
author = {Wolff, M. J. and Smith, M. D. and Clancy, R. T. and Arvidson, R. and Kahre, M. and Seelos IV, F. and Murchie, S. and Savijärvi, H.},
title = {Wavelength dependence of dust aerosol single scattering albedo as observed by the Compact Reconnaissance Imaging Spectrometer},
journal = {Journal of Geophysical Research: Planets},
volume = {114},
number = {E2},
pages = {},
doi = {https://doi.org/10.1029/2009JE003350},
url = {https://agupubs.onlinelibrary.wiley.com/doi/abs/10.1029/2009JE003350},
eprint = {https://agupubs.onlinelibrary.wiley.com/doi/pdf/10.1029/2009JE003350},
year = {2009}
}

@ARTICLE{deguine20_complex,
       author = {{Deguine}, Alexandre and {Petitprez}, Denis and {Clarisse}, Lieven and {Gu{\dj}mundsson}, Sn{\ae}varr and {Outes}, Valeria and {Villarosa}, Gustavo and {Herbin}, Herv{\'e}},
        title = "{Complex refractive index of volcanic ash aerosol in the infrared, visible, and ultraviolet}",
      journal = {\ao},
         year = 2020,
        month = feb,
       volume = {59},
       number = {4},
        pages = {884},
          doi = {10.1364/AO.59.000884},
       adsurl = {https://ui.adsabs.harvard.edu/abs/2020ApOpt..59..884D}
}

@ARTICLE{khare84_optical,
       author = {{Khare}, B.~N. and {Sagan}, C. and {Arakawa}, E.~T. and {Suits}, F. and {Callcott}, T.~A. and {Williams}, M.~W.},
        title = "{Optical constants of organic tholins produced in a simulated Titanian atmosphere: From soft x-ray to microwave frequencies}",
      journal = {\icarus},
         year = 1984,
        month = oct,
       volume = {60},
       number = {1},
        pages = {127-137},
          doi = {10.1016/0019-1035(84)90142-8},
       adsurl = {https://ui.adsabs.harvard.edu/abs/1984Icar...60..127K}
}

@ARTICLE{bennett25_hst,
       author = {{Bennett}, Katherine A. and {Sing}, David K. and {Stevenson}, Kevin B. and {Wakeford}, Hannah R. and {Rustamkulov}, Zafar and {Allen}, Natalie H. and {Lothringer}, Joshua D. and {MacDonald}, Ryan J. and {Mayne}, Nathan J. and {Fu}, Guangwei},
        title = "{An HST Transmission Spectrum of the Closest M Dwarf Transiting Rocky Planet LTT 1445Ab}",
      journal = {\aj},
         year = 2025,
        month = feb,
       volume = {169},
       number = {2},
          eid = {111},
        pages = {111},
          doi = {10.3847/1538-3881/ad9dd1},
archivePrefix = {arXiv},
       eprint = {2410.11054},
 primaryClass = {astro-ph.EP},
       adsurl = {https://ui.adsabs.harvard.edu/abs/2025AJ....169..111B}
}

@INCOLLECTION{winn10,
       author = {{Winn}, J.~N.},
        title = "{Exoplanet Transits and Occultations}",
    booktitle = {Exoplanets},
         year = 2010,
       editor = {{Seager}, S.},
        pages = {55-77},
          doi = {10.48550/arXiv.1001.2010},
       adsurl = {https://ui.adsabs.harvard.edu/abs/2010exop.book...55W}
}

@ARTICLE{carter23,
       author = {{Carter}, Aarynn L. and {Hinkley}, Sasha and {Kammerer}, Jens and {Skemer}, Andrew and {Biller}, Beth A. and {Leisenring}, Jarron M. and {Millar-Blanchaer}, Maxwell A. and {Petrus}, Simon and {Stone}, Jordan M. and {Ward-Duong}, Kimberly and {Wang}, Jason J. and {Girard}, Julien H. and {Hines}, Dean C. and {Perrin}, Marshall D. and {Pueyo}, Laurent and {Balmer}, William O. and {Bonavita}, Mariangela and {Bonnefoy}, Mickael and {Chauvin}, Gael and {Choquet}, Elodie and {Christiaens}, Valentin and {Danielski}, Camilla and {Kennedy}, Grant M. and {Matthews}, Elisabeth C. and {Miles}, Brittany E. and {Patapis}, Polychronis and {Ray}, Shrishmoy and {Rickman}, Emily and {Sallum}, Steph and {Stapelfeldt}, Karl R. and {Whiteford}, Niall and {Zhou}, Yifan and {Absil}, Olivier and {Boccaletti}, Anthony and {Booth}, Mark and {Bowler}, Brendan P. and {Chen}, Christine H. and {Currie}, Thayne and {Fortney}, Jonathan J. and {Grady}, Carol A. and {Greebaum}, Alexandra Z. and {Henning}, Thomas and {Hoch}, Kielan K.~W. and {Janson}, Markus and {Kalas}, Paul and {Kenworthy}, Matthew A. and {Kervella}, Pierre and {Kraus}, Adam L. and {Lagage}, Pierre-Olivier and {Liu}, Michael C. and {Macintosh}, Bruce and {Marino}, Sebastian and {Marley}, Mark S. and {Marois}, Christian and {Matthews}, Brenda C. and {Mawet}, Dimitri and {McElwain}, Michael W. and {Metchev}, Stanimir and {Meyer}, Michael R. and {Molliere}, Paul and {Moran}, Sarah E. and {Morley}, Caroline V. and {Mukherjee}, Sagnick and {Pantin}, Eric and {Quirrenbach}, Andreas and {Rebollido}, Isabel and {Ren}, Bin B. and {Schneider}, Glenn and {Vasist}, Malavika and {Worthen}, Kadin and {Wyatt}, Mark C. and {Briesemeister}, Zackery W. and {Bryan}, Marta L. and {Calissendorff}, Per and {Cantalloube}, Faustine and {Cugno}, Gabriele and {De Furio}, Matthew and {Dupuy}, Trent J. and {Factor}, Samuel M. and {Faherty}, Jacqueline K. and {Fitzgerald}, Michael P. and {Franson}, Kyle and {Gonzales}, Eileen C. and {Hood}, Callie E. and {Howe}, Alex R. and {Kuzuhara}, Masayuki and {Lagrange}, Anne-Marie and {Lawson}, Kellen and {Lazzoni}, Cecilia and {Lew}, Ben W.~P. and {Liu}, Pengyu and {Llop-Sayson}, Jorge and {Lloyd}, James P. and {Martinez}, Raquel A. and {Mazoyer}, Johan and {Palma-Bifani}, Paulina and {Quanz}, Sascha P. and {Redai}, Jea Adams and {Samland}, Matthias and {Schlieder}, Joshua E. and {Tamura}, Motohide and {Tan}, Xianyu and {Uyama}, Taichi and {Vigan}, Arthur and {Vos}, Johanna M. and {Wagner}, Kevin and {Wolff}, Schuyler G. and {Ygouf}, Marie and {Zhang}, Xi and {Zhang}, Keming and {Zhang}, Zhoujian},
        title = "{The JWST Early Release Science Program for Direct Observations of Exoplanetary Systems I: High-contrast Imaging of the Exoplanet HIP 65426 b from 2 to 16 {\ensuremath{\mu}}m}",
      journal = {\apjl},
         year = 2023,
        month = jul,
       volume = {951},
       number = {1},
          eid = {L20},
        pages = {L20},
          doi = {10.3847/2041-8213/acd93e},
archivePrefix = {arXiv},
       eprint = {2208.14990},
 primaryClass = {astro-ph.EP},
       adsurl = {https://ui.adsabs.harvard.edu/abs/2023ApJ...951L..20C}
}

@ARTICLE{ji25_cosmic,
       author = {{Ji}, Xuan and {Chatterjee}, Richard D. and {Coy}, Brandon Park and {Kite}, Edwin},
        title = "{The Cosmic Shoreline Revisited: A Metric for Atmospheric Retention Informed by Hydrodynamic Escape}",
      journal = {\apj},
         year = 2025,
        month = oct,
       volume = {992},
       number = {2},
          eid = {198},
        pages = {198},
          doi = {10.3847/1538-4357/adfe69},
archivePrefix = {arXiv},
       eprint = {2504.19872},
 primaryClass = {astro-ph.EP},
       adsurl = {https://ui.adsabs.harvard.edu/abs/2025ApJ...992..198J}
}

@ARTICLE{mandel02,
       author = {{Mandel}, Kaisey and {Agol}, Eric},
        title = "{Analytic Light Curves for Planetary Transit Searches}",
      journal = {\apjl},
         year = 2002,
        month = dec,
       volume = {580},
       number = {2},
        pages = {L171-L175},
          doi = {10.1086/345520},
archivePrefix = {arXiv},
       eprint = {astro-ph/0210099},
 primaryClass = {astro-ph},
       adsurl = {https://ui.adsabs.harvard.edu/abs/2002ApJ...580L.171M}
}

@ARTICLE{2015arXiv150702646V,
       author = {{Vehtari}, Aki and {Simpson}, Daniel and {Gelman}, Andrew and {Yao}, Yuling and {Gabry}, Jonah},
        title = "{Pareto Smoothed Importance Sampling}",
      journal = {arXiv e-prints},
         year = 2015,
        month = jul,
          eid = {arXiv:1507.02646},
        pages = {arXiv:1507.02646},
          doi = {10.48550/arXiv.1507.02646},
archivePrefix = {arXiv},
       eprint = {1507.02646},
 primaryClass = {stat.CO},
       adsurl = {https://ui.adsabs.harvard.edu/abs/2015arXiv150702646V}
}

@ARTICLE{wirth26_analytic,
       author = {{Wirth}, Christopher P. and {Powell}, Diana and {Wordsworth}, Robin},
        title = "{Analytic Modeling of Tidally Locked Rocky Planet Atmospheres across Dynamical Regimes}",
      journal = {\apj},
         year = 2026,
        month = mar,
       volume = {999},
       number = {2},
          eid = {153},
        pages = {153},
          doi = {10.3847/1538-4357/ae3da9},
archivePrefix = {arXiv},
       eprint = {2601.12556},
 primaryClass = {astro-ph.EP},
       adsurl = {https://ui.adsabs.harvard.edu/abs/2026ApJ...999..153W}
}

@ARTICLE{dyrek24,
       author = {{Dyrek}, A. and {Ducrot}, E. and {Lagage}, P.-O. and {Tremblin}, P. and {Kendrew}, S. and {Bouwman}, J. and {Bouffet}, R.},
        title = "{Transiting exoplanets with the Mid-InfraRed Instrument on board JWST: From simulations to observations}",
      journal = {\aap},
         year = 2024,
        month = mar,
       volume = {683},
          eid = {A212},
        pages = {A212},
          doi = {10.1051/0004-6361/202347127},
archivePrefix = {arXiv},
       eprint = {2403.00676},
 primaryClass = {astro-ph.EP},
       adsurl = {https://ui.adsabs.harvard.edu/abs/2024A&A...683A.212D}
}

@ARTICLE{morrison23,
       author = {{Morrison}, Jane E. and {Dicken}, Daniel and {Argyriou}, Ioannis and {Ressler}, Michael E. and {Gordon}, Karl D. and {Regan}, Michael W. and {Cracraft}, Misty and {Rieke}, George H. and {Engesser}, Michael and {Alberts}, Stacey and et al.},
        title = "{JWST MIRI Flight Performance: Detector Effects and Data Reduction Algorithms}",
      journal = {\pasp},
         year = 2023,
        month = jul,
       volume = {135},
       number = {1049},
          eid = {075004},
        pages = {075004},
          doi = {10.1088/1538-3873/acdea6},
archivePrefix = {arXiv},
       eprint = {2308.16327},
 primaryClass = {astro-ph.IM},
       adsurl = {https://ui.adsabs.harvard.edu/abs/2023PASP..135g5004M}
}

@ARTICLE{steinrueck25,
       author = {{Steinrueck}, Maria E. and {Parmentier}, Vivien and {Kreidberg}, Laura and {Gao}, Peter and {Kempton}, Eliza M.-R. and {Zhang}, Michael and {Stevenson}, Kevin B. and {Malsky}, Isaac and {Roman}, Michael T. and {Rauscher}, Emily and et al.},
        title = "{The Radiative Effects of Photochemical Hazes on the Atmospheric Circulation and Phase Curves of Sub-Neptunes}",
      journal = {\apj},
         year = 2025,
        month = may,
       volume = {985},
       number = {1},
          eid = {98},
        pages = {98},
          doi = {10.3847/1538-4357/adc99a},
archivePrefix = {arXiv},
       eprint = {2503.22642},
 primaryClass = {astro-ph.EP},
       adsurl = {https://ui.adsabs.harvard.edu/abs/2025ApJ...985...98S}
}

@ARTICLE{ragozzine07,
       author = {{Ragozzine}, D. and {Brown}, M.~E.},
        title = "{Candidate Members and Age Estimate of the Family of Kuiper Belt Object 2003 EL61}",
      journal = {\aj},
         year = 2007,
        month = dec,
       volume = {134},
       number = {6},
        pages = {2160-2167},
          doi = {10.1086/522334},
archivePrefix = {arXiv},
       eprint = {0709.0328},
 primaryClass = {astro-ph},
       adsurl = {https://ui.adsabs.harvard.edu/abs/2007AJ....134.2160R}
}

@ARTICLE{henderson94_nearsurface1,
       author = {{Henderson}, B.~G. and {Jakosky}, B.~M.},
        title = "{Near-surface thermal gradients and their effects on mid-infrared emission spectra of planetary surfaces}",
      journal = {\jgr},
         year = 1994,
        month = sep,
       volume = {99},
       number = {E9},
        pages = {19063-19074},
          doi = {10.1029/94JE01861},
       adsurl = {https://ui.adsabs.harvard.edu/abs/1994JGR....9919063H}
}

@ARTICLE{henderson97_nearsurface2,
       author = {{Henderson}, Bradley G. and {Jakosky}, Bruce M.},
        title = "{Near-surface thermal gradients and mid-IR emission spectra: A new model including scattering and application to real data}",
      journal = {\jgr},
         year = 1997,
        month = mar,
       volume = {102},
       number = {E3},
        pages = {6567-6580},
          doi = {10.1029/96JE03781},
       adsurl = {https://ui.adsabs.harvard.edu/abs/1997JGR...102.6567H}
}

@BOOK{murray_dermott99,
       author = {{Murray}, Carl D. and {Dermott}, Stanley F.},
        title = "{Solar System Dynamics}",
         year = 1999,
          doi = {10.1017/CBO9781139174817},
       adsurl = {https://ui.adsabs.harvard.edu/abs/1999ssd..book.....M}
}

@ARTICLE{yee26,
       author = {{Yee}, Samuel W. and {Dziire}, Tarisai and {Collins}, Karen A. and {Bouma}, Luke G. and {Tamayo}, Daniel and {Petigura}, Erik A. and {Bieryla}, Allyson and {Latham}, David W. and {Quinn}, Samuel N. and {Winn}, Joshua N. and et al.},
        title = "{TOI-3686: A Hot Jupiter and Sub-Neptune In A 2:1 Mean-Motion Resonance}",
      journal = {arXiv e-prints},
         year = 2026,
        month = oct,
          eid = {arXiv:2610.28533},
        pages = {arXiv:2610.28533},
          doi = {10.48550/arXiv.2610.28533},
archivePrefix = {arXiv},
       eprint = {2610.28533},
 primaryClass = {astro-ph.EP},
       adsurl = {https://ui.adsabs.harvard.edu/abs/2026arXiv261028533Y}
}

@ARTICLE{lithwick12,
       author = {{Lithwick}, Yoram and {Xie}, Jiwei and {Wu}, Yanqin},
        title = "{Extracting Planet Mass and Eccentricity from TTV Data}",
      journal = {\apj},
         year = 2012,
        month = dec,
       volume = {761},
       number = {2},
          eid = {122},
        pages = {122},
          doi = {10.1088/0004-637X/761/2/122},
archivePrefix = {arXiv},
       eprint = {1207.4192},
 primaryClass = {astro-ph.EP},
       adsurl = {https://ui.adsabs.harvard.edu/abs/2012ApJ...761..122L}
}

@ARTICLE{zieba26_lhs3844,
       author = {{Zieba}, Sebastian and {Kreidberg}, Laura and {Coy}, Brandon P. and {Bello-Arufe}, Aaron and {Paragas}, Kimberly and {Lyu}, Xintong and {Hu}, Renyu and {Iyer}, Aishwarya and {Kite}, Edwin S. and {Koll}, Daniel D.~B. and {Wohlfarth}, Kay and {Whittaker}, Emerson and {Knutson}, Heather and {Wordsworth}, Robin and {Morley}, Caroline and {Schaefer}, Laura},
        title = "{The dark and featureless surface of rocky exoplanet LHS 3844 b from JWST mid-infrared spectroscopy}",
      journal = {Nature Astronomy},
         year = 2026,
        month = aug,
       volume = {10},
        pages = {1129-1144},
          doi = {10.1038/s41550-026-02860-3},
archivePrefix = {arXiv},
       eprint = {2605.00100},
 primaryClass = {astro-ph.EP},
       adsurl = {https://ui.adsabs.harvard.edu/abs/2026NatAs..10.1129Z}
}

@ARTICLE{eastman2013,
       author = {{Eastman}, Jason and {Gaudi}, B. Scott and {Agol}, Eric},
        title = "{EXOFAST: A Fast Exoplanetary Fitting Suite in IDL}",
      journal = {\pasp},
         year = 2013,
        month = jan,
       volume = {125},
       number = {923},
        pages = {83},
          doi = {10.1086/669497},
archivePrefix = {arXiv},
       eprint = {1206.5798},
 primaryClass = {astro-ph.IM},
       adsurl = {https://ui.adsabs.harvard.edu/abs/2013PASP..125...83E}
}

@misc{charbonneau07_dynamics,
      title={The Dynamics-Based Approach to Studying Terrestrial Exoplanets}, 
      author={David Charbonneau and Drake Deming},
      year={2007},
      eprint={0706.1047},
      archivePrefix={arXiv},
      primaryClass={astro-ph},
      url={https://arxiv.org/abs/0706.1047}, 
}

@MISC{espinoza_dd9234,
       author = {{Espinoza}, Nestor and {Diamond-Lowe}, Hannah},
        title = "{Rocky Worlds DDT: JWST Observations of LTT 1445 A c}",
 howpublished = {JWST Proposal. Cycle 3, ID. \#9234},
         year = 2025,
        month = jan,
        pages = {9234},
       adsurl = {https://ui.adsabs.harvard.edu/abs/2025jwst.prop.9234E}
}

@ARTICLE{maurel25_constraints,
       author = {{Maurel}, Alice and {Turbet}, Martin and {Ducrot}, Elsa and {Leconte}, J{\'e}r{\'e}my and {Chaverot}, Guillaume and {Milcareck}, Gwenael and {Revol}, Alexandre and {Charnay}, Benjamin and {Fauchez}, Thomas J. and {Gillon}, Micha{\"e}l and et al.},
        title = "{Constraints on the possible atmospheres on TRAPPIST-1 b: insights from 3D climate modeling}",
      journal = {\aap},
         year = 2025,
        month = sep,
       volume = {701},
          eid = {A193},
        pages = {A193},
          doi = {10.1051/0004-6361/202554243},
archivePrefix = {arXiv},
       eprint = {2509.02120},
 primaryClass = {astro-ph.EP},
       adsurl = {https://ui.adsabs.harvard.edu/abs/2025A&A...701A.193M}
}

@ARTICLE{lyu26_subsurface,
       author = {{Lyu}, Xintong and {Koll}, Daniel D.~B.},
        title = "{Impact of Subsurface Temperature Gradients on Emission Spectra of Airless Exoplanets: The Solid-state Greenhouse and Anti-greenhouse}",
      journal = {\apj},
         year = 2026,
        month = apr,
       volume = {1000},
       number = {2},
          eid = {274},
        pages = {274},
          doi = {10.3847/1538-4357/ae4a9f},
archivePrefix = {arXiv},
       eprint = {2510.22932},
 primaryClass = {astro-ph.EP},
       adsurl = {https://ui.adsabs.harvard.edu/abs/2026ApJ..1000..274L}
}

@ARTICLE{kreidberg25_first,
       author = {{Kreidberg}, Laura and {Stevenson}, Kevin B.},
        title = "{A first look at rocky exoplanets with JWST}",
      journal = {Proceedings of the National Academy of Science},
         year = 2025,
        month = sep,
       volume = {122},
       number = {39},
          eid = {e2416190122},
        pages = {e2416190122},
          doi = {10.1073/pnas.2416190122},
archivePrefix = {arXiv},
       eprint = {2507.00933},
 primaryClass = {astro-ph.EP},
       adsurl = {https://ui.adsabs.harvard.edu/abs/2025PNAS..12216190K}
}

@article{Grant2024_exoticld,
  author  = {Grant, David and Wakeford, Hannah R.},
  title   = {{ExoTiC-LD}: thirty seconds to stellar limb-darkening coefficients},
  journal = {Journal of Open Source Software},
  year    = {2024},
  volume  = {9},
  number  = {100},
  pages   = {6816},
  doi     = {10.21105/joss.06816}
}

@article{Kostogryz2023_mpsatlas,
  author  = {Kostogryz, N. and Shapiro, A. I. and Witzke, V. and Grant, D. and
             Wakeford, H. R. and Stevenson, K. B. and Solanki, S. K. and Gizon, L.},
  title   = {{MPS-ATLAS} Library of Stellar Model Atmospheres and Spectra},
  journal = {Research Notes of the AAS},
  year    = {2023},
  volume  = {7},
  number  = {3},
  pages   = {39},
  doi     = {10.3847/2515-5172/acc180}
}

@ARTICLE{august25_hot,
       author = {{August}, Prune C. and {Buchhave}, Lars A. and {Diamond-Lowe}, Hannah and {Mendon{\c{c}}a}, Jo{\~a}o M. and {Gressier}, Am{\'e}lie and {Rathcke}, Alexander D. and {Allen}, Natalie H. and {Fortune}, Mark and {Jones}, Kathryn D. and {Meier Vald{\'e}s}, Erik A. and {Demory}, Brice-Olivier and {Espinoza}, N{\'e}stor and {Fisher}, Chloe E. and {Gibson}, Neale P. and {Heng}, Kevin and {Hoeijmakers}, Jens and {Hooton}, Matthew J. and {Kitzmann}, Daniel and {Prinoth}, Bibiana and {Eastman}, Jason D. and {Barnes}, Rory},
        title = "{Hot Rocks Survey I: A possible shallow eclipse for LHS 1478 b}",
      journal = {\aap},
         year = 2025,
       volume = {695},
        pages = {A171},
          doi = {10.1051/0004-6361/202452611},
       eprint = {2410.11048}
}

@ARTICLE{batalha19_picaso,
       author = {{Batalha}, Natasha E. and {Marley}, Mark S. and {Lewis}, Nikole K. and {Fortney}, Jonathan J.},
        title = "{Exoplanet Reflected-light Spectroscopy with PICASO}",
      journal = {\apj},
         year = 2019,
        month = jun,
       volume = {878},
       number = {1},
          eid = {70},
        pages = {70},
          doi = {10.3847/1538-4357/ab1b51},
archivePrefix = {arXiv},
       eprint = {1904.09355},
 primaryClass = {astro-ph.EP},
       adsurl = {https://ui.adsabs.harvard.edu/abs/2019ApJ...878...70B}
}

@ARTICLE{connors2025,
       author = {{Connors}, Nicholas J. and {Monaghan}, Christopher and {Benneke}, Bj{\"o}rn and {Dang}, Lisa},
        title = "{Uniform Reanalysis of JWST MIRI 15 {\ensuremath{\mu}}m Exoplanet Eclipse Observations Using Frame-normalized Principal Component Analysis}",
      journal = {\apjl},
         year = 2025,
        month = aug,
       volume = {989},
       number = {1},
          eid = {L11},
        pages = {L11},
          doi = {10.3847/2041-8213/adee0d},
archivePrefix = {arXiv},
       eprint = {2507.02052},
 primaryClass = {astro-ph.EP},
       adsurl = {https://ui.adsabs.harvard.edu/abs/2025ApJ...989L..11C}
}

@article{wogan25_photochem,
doi = {10.3847/PSJ/ae0e1c},
url = {https://doi.org/10.3847/PSJ/ae0e1c},
year = {2025},
month = {nov},
publisher = {The American Astronomical Society},
volume = {6},
number = {11},
pages = {256},
author = {Wogan, Nicholas F. and Batalha, Natasha E. and Zahnle, Kevin and Krissansen-Totton, Joshua and Catling, David C. and Wolf, Eric T. and Robinson, Tyler D. and Meadows, Victoria and Arney, Giada and Domagal-Goldman, Shawn},
title = {The Open-source Photochem Code: A General Chemical and Climate Model for Interpreting (Exo)Planet Observations},
journal = {The Planetary Science Journal}
}

@ARTICLE{oddo23_characterization,
       author = {{Oddo}, Dominic and {Dragomir}, Diana and {Brandeker}, Alexis and {Osborn}, Hugh P. and {Collins}, Karen and {Stassun}, Keivan G. and {Astudillo-Defru}, Nicola and {Bieryla}, Allyson and {Howell}, Steve B. and {Ciardi}, David R. and {Quinn}, Samuel and {Almenara}, Jose M. and {Brice{\~n}o}, C{\'e}sar and {Collins}, Kevin I. and {Col{\'o}n}, Knicole D. and {Conti}, Dennis M. and {Crouzet}, Nicolas and {Furlan}, Elise and {Gan}, Tianjun and {Gnilka}, Crystal L. and {Goeke}, Robert F. and {Gonzales}, Erica and {Harris}, Mallory and {Jenkins}, Jon M. and {Jensen}, Eric L.~N. and {Latham}, David and {Law}, Nicholas and {Lund}, Michael B. and {Mann}, Andrew W. and {Massey}, Bob and {Murgas}, Felipe and {Ricker}, George and {Relles}, Howard M. and {Rowden}, Pamela and {Schwarz}, Richard P. and {Schlieder}, Joshua and {Shporer}, Avi and {Seager}, Sara and {Srdoc}, Gregor and {Torres}, Guillermo and {Twicken}, Joseph D. and {Vanderspek}, Roland and {Winn}, Joshua N. and {Ziegler}, Carl},
        title = "{Characterization of a Set of Small Planets with TESS and CHEOPS and an Analysis of Photometric Performance}",
      journal = {\aj},
         year = 2023,
        month = mar,
       volume = {165},
       number = {3},
          eid = {134},
        pages = {134},
          doi = {10.3847/1538-3881/acb4e3},
archivePrefix = {arXiv},
       eprint = {2301.08162},
 primaryClass = {astro-ph.EP},
       adsurl = {https://ui.adsabs.harvard.edu/abs/2023AJ....165..134O}
}

@ARTICLE{monaghan26_uniform,
       author = {{Monaghan}, Christopher and {Benneke}, Bj{\"o}rn and {Connors}, Nicholas J. and {Coulombe}, Louis-Philippe and {Roy}, Pierre-Alexis},
        title = "{Uniform Reinterpretation of Rocky Exoplanet Secondary Eclipse Observations and the Impact of Stellar and Orbital Uncertainties}",
      journal = {\aj},
         year = 2026,
        month = may,
       volume = {171},
       number = {5},
          eid = {319},
        pages = {319},
          doi = {10.3847/1538-3881/ae5ba5},
archivePrefix = {arXiv},
       eprint = {2604.15421},
 primaryClass = {astro-ph.EP},
       adsurl = {https://ui.adsabs.harvard.edu/abs/2026AJ....171..319M}
}

@ARTICLE{malik19_self,
       author = {{Malik}, Matej and {Kitzmann}, Daniel and {Mendon{\c{c}}a}, Jo{\~a}o M. and {Grimm}, Simon L. and {Marleau}, Gabriel-Dominique and {Linder}, Esther F. and {Tsai}, Shang-Min and {Heng}, Kevin},
        title = "{Self-luminous and Irradiated Exoplanetary Atmospheres Explored with HELIOS}",
      journal = {\aj},
         year = 2019,
        month = may,
       volume = {157},
       number = {5},
          eid = {170},
        pages = {170},
          doi = {10.3847/1538-3881/ab1084},
archivePrefix = {arXiv},
       eprint = {1903.06794},
 primaryClass = {astro-ph.EP},
       adsurl = {https://ui.adsabs.harvard.edu/abs/2019AJ....157..170M}
}

@ARTICLE{malik17_helios,
       author = {{Malik}, Matej and {Grosheintz}, Luc and {Mendon{\c{c}}a}, Jo{\~a}o M. and {Grimm}, Simon L. and {Lavie}, Baptiste and {Kitzmann}, Daniel and {Tsai}, Shang-Min and {Burrows}, Adam and {Kreidberg}, Laura and {Bedell}, Megan and {Bean}, Jacob L. and {Stevenson}, Kevin B. and {Heng}, Kevin},
        title = "{HELIOS: An Open-source, GPU-accelerated Radiative Transfer Code for Self-consistent Exoplanetary Atmospheres}",
      journal = {\aj},
         year = 2017,
        month = feb,
       volume = {153},
       number = {2},
          eid = {56},
        pages = {56},
          doi = {10.3847/1538-3881/153/2/56},
archivePrefix = {arXiv},
       eprint = {1606.05474},
 primaryClass = {astro-ph.EP},
       adsurl = {https://ui.adsabs.harvard.edu/abs/2017AJ....153...56M}
}

@software{Foreman-Mackey2024,
       author = {{Foreman-Mackey}, Daniel and {Yu}, Weixiang and {Yadav}, Sachin and {Reynolds Becker}, McCoy and {Caplar}, Neven and {Huppenkothen}, Daniela and {Killestein}, Thomas and {Tronsgaard}, Ren{\'e} and {Rashid}, Theo and {Schmerler}, Steve},
        title = "{dfm/tinygp: The tiniest of Gaussian Process libraries}",
         year = 2024,
        month = jan,
          eid = {10.5281/zenodo.10463641},
          doi = {10.5281/zenodo.10463641},
      version = {v0.3.0},
    publisher = {Zenodo},
       adsurl = {https://ui.adsabs.harvard.edu/abs/2024zndo..10463641F}
}

@software{jaxoplanet,
  author       = {Soichiro Hattori and
                  Lionel Garcia and
                  Catriona Murray and
                  Jiayin Dong and
                  Shashank Dholakia and
                  David Degen and
                  Daniel Foreman-Mackey},
  title        = {{exoplanet-dev/jaxoplanet: Astronomical time series analysis with JAX}},
  month        = mar,
  year         = 2024,
  publisher    = {Zenodo},
  version      = {v0.0.2},
  doi          = {10.5281/zenodo.10736936},
  url          = {https://doi.org/10.5281/zenodo.10736936}
}

@ARTICLE{fortune25_lhs1140c,
       author = {{Fortune}, Mark and {Gibson}, Neale P. and {Diamond-Lowe}, Hannah and {Mendon{\c{c}}a}, Jo{\~a}o M. and {Gressier}, Am{\'e}lie and {Kitzmann}, Daniel and {Allen}, Natalie H. and {August}, Prune C. and {Ih}, Jegug and {Meier Vald{\'e}s}, Erik and {Zgraggen}, Merlin and {Buchhave}, Lars A. and {Demory}, Brice-Olivier and {Espinoza}, N{\'e}stor and {Heng}, Kevin and {Jones}, Kathryn and {Rathcke}, Alexander D.},
        title = "{Hot Rocks Survey: III. A deep eclipse for LHS 1140c and a new Gaussian process method to account for correlated noise in individual pixels}",
      journal = {\aap},
         year = 2025,
        month = sep,
       volume = {701},
          eid = {A25},
        pages = {A25},
          doi = {10.1051/0004-6361/202554198},
archivePrefix = {arXiv},
       eprint = {2505.22186},
 primaryClass = {astro-ph.EP},
       adsurl = {https://ui.adsabs.harvard.edu/abs/2025A&A...701A..25F}
}

@ARTICLE{meier2025_toi1468b,
       author = {{Meier Vald{\'e}s}, E.~A. and {Demory}, B. -O. and {Diamond-Lowe}, H. and {Mendon{\c{c}}a}, J.~M. and {August}, P.~C. and {Fortune}, M. and {Allen}, N.~H. and {Kitzmann}, D. and {Gressier}, A. and {Hooton}, M. and {Jones}, K.~D. and {Buchhave}, L.~A. and {Espinoza}, N. and {Fisher}, C.~E. and {Gibson}, N.~P. and {Heng}, K. and {Hoeijmakers}, J. and {Prinoth}, B. and {Rathcke}, A.~D. and {Eastman}, J.~D.},
        title = "{Hot Rocks Survey: II. The thermal emission of TOI-1468 b reveals a bare hot rock}",
      journal = {\aap},
         year = 2025,
        month = jun,
       volume = {698},
          eid = {A68},
        pages = {A68},
          doi = {10.1051/0004-6361/202453449},
archivePrefix = {arXiv},
       eprint = {2503.19772},
 primaryClass = {astro-ph.EP},
       adsurl = {https://ui.adsabs.harvard.edu/abs/2025A&A...698A..68M}
}

@ARTICLE{france16,
       author = {{France}, Kevin and {Loyd}, R.~O. Parke and {Youngblood}, Allison and {Brown}, Alexander and {Schneider}, P. Christian and {Hawley}, Suzanne L. and {Froning}, Cynthia S. and {Linsky}, Jeffrey L. and {Roberge}, Aki and {Buccino}, Andrea P. and {Davenport}, James R.~A. and {Fontenla}, Juan M. and {Kaltenegger}, Lisa and {Kowalski}, Adam F. and {Mauas}, Pablo J.~D. and {Miguel}, Yamila and {Redfield}, Seth and {Rugheimer}, Sarah and {Tian}, Feng and {Vieytes}, Mariela C. and {Walkowicz}, Lucianne M. and {Weisenburger}, Kolby L.},
        title = "{The MUSCLES Treasury Survey. I. Motivation and Overview}",
      journal = {\apj},
         year = 2016,
        month = apr,
       volume = {820},
       number = {2},
          eid = {89},
        pages = {89},
          doi = {10.3847/0004-637X/820/2/89},
archivePrefix = {arXiv},
       eprint = {1602.09142},
 primaryClass = {astro-ph.SR},
       adsurl = {https://ui.adsabs.harvard.edu/abs/2016ApJ...820...89F}
}

@article{tian09,
doi = {10.1088/0004-637X/703/1/905},
url = {https://dx.doi.org/10.1088/0004-637X/703/1/905},
year = {2009},
month = {sep},
publisher = {The American Astronomical Society},
volume = {703},
number = {1},
pages = {905},
author = {Tian, Feng},
title = {THERMAL ESCAPE FROM SUPER EARTH ATMOSPHERES IN THE HABITABLE ZONES OF M STARS},
journal = {The Astrophysical Journal}
}

@ARTICLE{chatterjee24,
       author = {{Chatterjee}, Richard D. and {Pierrehumbert}, Raymond T.},
        title = "{Novel Physics of Escaping Secondary Atmospheres May Shape the Cosmic Shoreline}",
      journal = {arXiv e-prints},
         year = 2024,
        month = dec,
          eid = {arXiv:2412.05188},
        pages = {arXiv:2412.05188},
          doi = {10.48550/arXiv.2412.05188},
archivePrefix = {arXiv},
       eprint = {2412.05188},
 primaryClass = {astro-ph.EP},
       adsurl = {https://ui.adsabs.harvard.edu/abs/2024arXiv241205188C}
}

@article{kass95,
author = {{Kass}, Robert E. and {Raftery}, Adrian E. },
title = {Bayes Factors},
journal = {Journal of the American Statistical Association},
volume = {90},
number = {430},
pages = {773--795},
year = {1995},
publisher = {ASA Website},
doi = {10.1080/01621459.1995.10476572},


URL = { 
    
    
        https://www.tandfonline.com/doi/abs/10.1080/01621459.1995.10476572
    

},
eprint = { 
    
    
        https://www.tandfonline.com/doi/pdf/10.1080/01621459.1995.10476572
    

}

}

@ARTICLE{pass25,
       author = {{Pass}, Emily K and {Charbonneau}, David and {Vanderburg}, Andrew},
        title = "{The Receding Cosmic Shoreline of Mid-to-Late M Dwarfs: Measurements of Active Lifetimes Worsen Challenges for Atmosphere Retention by Rocky Exoplanets}",
      journal = {arXiv e-prints},
         year = 2025,
        month = apr,
          eid = {arXiv:2504.01182},
        pages = {arXiv:2504.01182},
          doi = {10.48550/arXiv.2504.01182},
archivePrefix = {arXiv},
       eprint = {2504.01182},
 primaryClass = {astro-ph.EP},
       adsurl = {https://ui.adsabs.harvard.edu/abs/2025arXiv250401182P}
}

@ARTICLE{fauchez25,
       author = {{Fauchez}, Thomas J. and {Rackham}, Benjamin V. and {Ducrot}, Elsa and {Stevenson}, Kevin B. and {de Wit}, Julien},
        title = "{Stellar Models Also Limit Exoplanet Atmosphere Studies in Emission}",
      journal = {arXiv e-prints},
         year = 2025,
        month = feb,
          eid = {arXiv:2502.19585},
        pages = {arXiv:2502.19585},
          doi = {10.48550/arXiv.2502.19585},
archivePrefix = {arXiv},
       eprint = {2502.19585},
 primaryClass = {astro-ph.EP},
       adsurl = {https://ui.adsabs.harvard.edu/abs/2025arXiv250219585F}
}

@ARTICLE{paragas25,
       author = {{Paragas}, Kimberly and {Knutson}, Heather A. and {Hu}, Renyu and {Ehlmann}, Bethany L. and {Alemanno}, Giulia and {Helbert}, J{\"o}rn and {Maturilli}, Alessandro and {Zhang}, Michael and {Iyer}, Aishwarya and {Rossman}, George},
        title = "{A New Spectral Library for Modeling the Surfaces of Hot, Rocky Exoplanets}",
      journal = {\apj},
         year = 2025,
        month = mar,
       volume = {981},
       number = {2},
          eid = {130},
        pages = {130},
          doi = {10.3847/1538-4357/ada9eb},
archivePrefix = {arXiv},
       eprint = {2502.04433},
 primaryClass = {astro-ph.EP},
       adsurl = {https://ui.adsabs.harvard.edu/abs/2025ApJ...981..130P}
}

@ARTICLE{shields16,
       author = {{Shields}, Aomawa L. and {Ballard}, Sarah and {Johnson}, John Asher},
        title = "{The habitability of planets orbiting M-dwarf stars}",
      journal = {\physrep},
         year = 2016,
        month = dec,
       volume = {663},
        pages = {1},
          doi = {10.1016/j.physrep.2016.10.003},
archivePrefix = {arXiv},
       eprint = {1610.05765},
 primaryClass = {astro-ph.EP},
       adsurl = {https://ui.adsabs.harvard.edu/abs/2016PhR...663....1S}
}

@ARTICLE{millerricci09,
       author = {{Miller-Ricci}, Eliza and {Seager}, Sara and {Sasselov}, Dimitar},
        title = "{The Atmospheric Signatures of Super-Earths: How to Distinguish Between Hydrogen-Rich and Hydrogen-Poor Atmospheres}",
      journal = {\apj},
         year = 2009,
        month = jan,
       volume = {690},
       number = {2},
        pages = {1056-1067},
          doi = {10.1088/0004-637X/690/2/1056},
archivePrefix = {arXiv},
       eprint = {0808.1902},
 primaryClass = {astro-ph},
       adsurl = {https://ui.adsabs.harvard.edu/abs/2009ApJ...690.1056M}
}

@ARTICLE{deming09,
       author = {{Deming}, D. and {Seager}, S. and {Winn}, J. and {Miller-Ricci}, E. and {Clampin}, M. and {Lindler}, D. and {Greene}, T. and {Charbonneau}, D. and {Laughlin}, G. and {Ricker}, G. and {Latham}, D. and {Ennico}, K.},
        title = "{Discovery and Characterization of Transiting Super Earths Using an All-Sky Transit Survey and Follow-up by the James Webb Space Telescope}",
      journal = {\pasp},
         year = 2009,
        month = sep,
       volume = {121},
       number = {883},
        pages = {952},
          doi = {10.1086/605913},
archivePrefix = {arXiv},
       eprint = {0903.4880},
 primaryClass = {astro-ph.EP},
       adsurl = {https://ui.adsabs.harvard.edu/abs/2009PASP..121..952D}
}

@article{greene23,
  title={Thermal emission from the Earth-sized exoplanet TRAPPIST-1 b using JWST},
  author={Greene, Thomas P and Bell, Taylor J and Ducrot, Elsa and Dyrek, Achr{\`e}ne and Lagage, Pierre-Olivier and Fortney, Jonathan J},
  journal={Nature},
  volume={618},
  number={7963},
  pages={39--42},
  year={2023},
  publisher={Nature Publishing Group UK London}
}

@ARTICLE{koll22,
       author = {{Koll}, Daniel D.~B.},
        title = "{A Scaling for Atmospheric Heat Redistribution on Tidally Locked Rocky Planets}",
      journal = {\apj},
         year = 2022,
        month = jan,
       volume = {924},
       number = {2},
          eid = {134},
        pages = {134},
          doi = {10.3847/1538-4357/ac3b48},
archivePrefix = {arXiv},
       eprint = {1907.13145},
 primaryClass = {astro-ph.EP},
       adsurl = {https://ui.adsabs.harvard.edu/abs/2022ApJ...924..134K}
}

@ARTICLE{mansfield19,
       author = {{Mansfield}, Megan and {Kite}, Edwin S. and {Hu}, Renyu and {Koll}, Daniel D.~B. and {Malik}, Matej and {Bean}, Jacob L. and {Kempton}, Eliza M. -R.},
        title = "{Identifying Atmospheres on Rocky Exoplanets through Inferred High Albedo}",
      journal = {\apj},
         year = 2019,
        month = dec,
       volume = {886},
       number = {2},
          eid = {141},
        pages = {141},
          doi = {10.3847/1538-4357/ab4c90},
archivePrefix = {arXiv},
       eprint = {1907.13150},
 primaryClass = {astro-ph.EP},
       adsurl = {https://ui.adsabs.harvard.edu/abs/2019ApJ...886..141M}
}

@ARTICLE{kreidberg19,
       author = {{Kreidberg}, Laura and {Koll}, Daniel D.~B. and {Morley}, Caroline and {Hu}, Renyu and {Schaefer}, Laura and {Deming}, Drake and {Stevenson}, Kevin B. and {Dittmann}, Jason and {Vanderburg}, Andrew and {Berardo}, David and {Guo}, Xueying and {Stassun}, Keivan and {Crossfield}, Ian and {Charbonneau}, David and {Latham}, David W. and {Loeb}, Abraham and {Ricker}, George and {Seager}, Sara and {Vanderspek}, Roland},
        title = "{Absence of a thick atmosphere on the terrestrial exoplanet LHS 3844b}",
      journal = {\nat},
         year = 2019,
        month = aug,
       volume = {573},
       number = {7772},
        pages = {87-90},
          doi = {10.1038/s41586-019-1497-4},
archivePrefix = {arXiv},
       eprint = {1908.06834},
 primaryClass = {astro-ph.EP},
       adsurl = {https://ui.adsabs.harvard.edu/abs/2019Natur.573...87K}
}

@ARTICLE{whittaker22,
       author = {{Whittaker}, Emily A. and {Malik}, Matej and {Ih}, Jegug and {Kempton}, Eliza M. -R. and {Mansfield}, Megan and {Bean}, Jacob L. and {Kite}, Edwin S. and {Koll}, Daniel D.~B. and {Cronin}, Timothy W. and {Hu}, Renyu},
        title = "{The Detectability of Rocky Planet Surface and Atmosphere Composition with the JWST: The Case of LHS 3844b}",
      journal = {\aj},
         year = 2022,
        month = dec,
       volume = {164},
       number = {6},
          eid = {258},
        pages = {258},
          doi = {10.3847/1538-3881/ac9ab3},
archivePrefix = {arXiv},
       eprint = {2207.08889},
 primaryClass = {astro-ph.EP},
       adsurl = {https://ui.adsabs.harvard.edu/abs/2022AJ....164..258W}
}

@ARTICLE{moran23,
       author = {{Moran}, Sarah E. and {Stevenson}, Kevin B. and {Sing}, David K. and {MacDonald}, Ryan J. and {Kirk}, James and {Lustig-Yaeger}, Jacob and {Peacock}, Sarah and {Mayorga}, L.~C. and {Bennett}, Katherine A. and {L{\'o}pez-Morales}, Mercedes and {May}, E.~M. and {Rustamkulov}, Zafar and {Valenti}, Jeff A. and {Adams Redai}, J{\'e}a I. and {Alam}, Munazza K. and {Batalha}, Natasha E. and {Fu}, Guangwei and {Gonzalez-Quiles}, Junellie and {Highland}, Alicia N. and {Kruse}, Ethan and {Lothringer}, Joshua D. and {Ortiz Ceballos}, Kevin N. and {Sotzen}, Kristin S. and {Wakeford}, Hannah R.},
        title = "{High Tide or Riptide on the Cosmic Shoreline? A Water-Rich Atmosphere or Stellar Contamination for the Warm Super-Earth GJ\raisebox{-0.5ex}\textasciitilde486b from JWST Observations}",
      journal = {arXiv e-prints},
         year = 2023,
        month = may,
          eid = {arXiv:2305.00868},
        pages = {arXiv:2305.00868},
          doi = {10.48550/arXiv.2305.00868},
archivePrefix = {arXiv},
       eprint = {2305.00868},
 primaryClass = {astro-ph.EP},
       adsurl = {https://ui.adsabs.harvard.edu/abs/2023arXiv230500868M}
}

@ARTICLE{rackham18,
       author = {{Rackham}, Benjamin V. and {Apai}, D{\'a}niel and {Giampapa}, Mark S.},
        title = "{The Transit Light Source Effect: False Spectral Features and Incorrect Densities for M-dwarf Transiting Planets}",
      journal = {\apj},
         year = 2018,
        month = feb,
       volume = {853},
       number = {2},
          eid = {122},
        pages = {122},
          doi = {10.3847/1538-4357/aaa08c},
archivePrefix = {arXiv},
       eprint = {1711.05691},
 primaryClass = {astro-ph.EP},
       adsurl = {https://ui.adsabs.harvard.edu/abs/2018ApJ...853..122R}
}

@ARTICLE{zahnle17,
       author = {{Zahnle}, Kevin J. and {Catling}, David C.},
        title = "{The Cosmic Shoreline: The Evidence that Escape Determines which Planets Have Atmospheres, and what this May Mean for Proxima Centauri B}",
      journal = {\apj},
         year = 2017,
        month = jul,
       volume = {843},
       number = {2},
          eid = {122},
        pages = {122},
          doi = {10.3847/1538-4357/aa7846},
archivePrefix = {arXiv},
       eprint = {1702.03386},
 primaryClass = {astro-ph.EP},
       adsurl = {https://ui.adsabs.harvard.edu/abs/2017ApJ...843..122Z}
}

@ARTICLE{malik19,
       author = {{Malik}, Matej and {Kempton}, Eliza M. -R. and {Koll}, Daniel D.~B. and {Mansfield}, Megan and {Bean}, Jacob L. and {Kite}, Edwin},
        title = "{Analyzing Atmospheric Temperature Profiles and Spectra of M Dwarf Rocky Planets}",
      journal = {\apj},
         year = 2019,
        month = dec,
       volume = {886},
       number = {2},
          eid = {142},
        pages = {142},
          doi = {10.3847/1538-4357/ab4a05},
archivePrefix = {arXiv},
       eprint = {1907.13135},
 primaryClass = {astro-ph.EP},
       adsurl = {https://ui.adsabs.harvard.edu/abs/2019ApJ...886..142M}
}

@article{zieba23,
  title={No thick carbon dioxide atmosphere on the rocky exoplanet TRAPPIST-1 c},
  author={Zieba, Sebastian and Kreidberg, Laura and Ducrot, Elsa and Gillon, Micha{\"e}l and Morley, Caroline and Schaefer, Laura and Tamburo, Patrick and Koll, Daniel DB and Lyu, Xintong and Acu{\~n}a, Lorena and others},
  journal={Nature},
  pages={1--4},
  year={2023},
  publisher={Nature Publishing Group UK London}
}

@article{zhang24,
  title={GJ 367b is a dark, hot, airless sub-Earth},
  author={Zhang, Michael and Hu, Renyu and Inglis, Julie and Dai, Fei and Bean, Jacob L and Knutson, Heather A and Lam, Kristine and Goffo, Elisa and Gandolfi, Davide},
  journal={The Astrophysical Journal Letters},
  volume={961},
  number={2},
  pages={L44},
  year={2024},
  publisher={IOP Publishing}
}

@article{iyer23,
  title={The SPHINX M-dwarf spectral grid. I. Benchmarking new model atmospheres to derive fundamental M-dwarf properties},
  author={Iyer, Aishwarya R and Line, Michael R and Muirhead, Philip S and Fortney, Jonathan J and Gharib-Nezhad, Ehsan},
  journal={The Astrophysical Journal},
  volume={944},
  number={1},
  pages={41},
  year={2023},
  publisher={IOP Publishing}
}

@article{cowan11,
  title={The statistics of albedo and heat recirculation on hot exoplanets},
  author={Cowan, Nicolas B and Agol, Eric},
  journal={The Astrophysical Journal},
  volume={729},
  number={1},
  pages={54},
  year={2011},
  publisher={IOP Publishing}
}

@article{jackson08,
  title={Tidal heating of terrestrial extrasolar planets and implications for their habitability},
  author={Jackson, Brian and Barnes, Rory and Greenberg, Richard},
  journal={Monthly Notices of the Royal Astronomical Society},
  volume={391},
  number={1},
  pages={237--245},
  year={2008},
  publisher={Blackwell Publishing Ltd Oxford, UK}
}

@article{xue24,
  title={JWST Thermal Emission of the Terrestrial Exoplanet GJ 1132b},
  author={Xue, Qiao and Bean, Jacob L and Zhang, Michael and Mahajan, Alexandra and Ih, Jegug and Eastman, Jason D and Lunine, Jonathan and Mansfield, Megan Weiner and Coy, Brandon Park and Kempton, Eliza M-R and others},
  journal={The Astrophysical Journal Letters},
  volume={973},
  number={1},
  pages={L8},
  year={2024},
  publisher={IOP Publishing}
}

@article{mansfield24,
  title={No Thick Atmosphere on the Terrestrial Exoplanet Gl 486b},
  author={Weiner Mansfield, Megan and Xue, Qiao and Zhang, Michael and Mahajan, Alexandra S and Ih, Jegug and Koll, Daniel and Bean, Jacob L and Coy, Brandon Park and Eastman, Jason D and Kempton, Eliza M-R and others},
  journal={The Astrophysical Journal Letters},
  volume={975},
  number={1},
  pages={L22},
  year={2024},
  publisher={IOP Publishing}
}

@article{diamond24,
  title={High-energy spectra of LTT 1445A and GJ 486 reveal flares and activity},
  author={Diamond-Lowe, H and King, GW and Youngblood, A and Brown, A and Howard, WS and Winters, JG and Wilson, DJ and France, K and Mendon{\c{c}}a, JM and Buchhave, LA and others},
  journal={Astronomy \& Astrophysics},
  volume={689},
  pages={A48},
  year={2024},
  publisher={EDP Sciences}
}

@article{hammond2024,
  title={Reliable Detections of Atmospheres on Rocky Exoplanets with Photometric JWST Phase Curves},
  author={Hammond, Mark and Guimond, Claire Marie and Lichtenberg, Tim and Nicholls, Harrison and Fisher, Chloe and Luque, Rafael and Meier, Tobias G and Taylor, Jake and Changeat, Quentin and Dang, Lisa and others},
  journal={The Astrophysical Journal Letters},
  volume={978},
  number={2},
  pages={L40},
  year={2025},
  publisher={IOP Publishing}
}

@article{speagle20,
  title={dynesty: a dynamic nested sampling package for estimating Bayesian posteriors and evidences},
  author={Speagle, Joshua S},
  journal={Monthly Notices of the Royal Astronomical Society},
  volume={493},
  number={3},
  pages={3132--3158},
  year={2020},
  publisher={Oxford University Press}
}

@ARTICLE{Koll2019,
       author = {{Koll}, Daniel D.~B.},
        title = "{A Scaling Theory for Atmospheric Heat Redistribution on Rocky Exoplanets}",
      journal = {arXiv e-prints},
         year = 2019,
        month = jul,
          eid = {arXiv:1907.13145v1},
        pages = {arXiv:1907.13145v1},
archivePrefix = {arXiv},
       eprint = {1907.13145v1},
 primaryClass = {astro-ph.EP},
          doi = {10.48550/arXiv.1907.13145}
}

@ARTICLE{allen25_hot,
       author = {{Allen}, Natalie H. and {Espinoza}, N{\'e}stor and {Diamond-Lowe}, Hannah and {Mendon{\c{c}}a}, Jo{\~a}o M. and {Demory}, Brice-Olivier and {Gressier}, Am{\'e}lie and {Ih}, Jegug and {Fortune}, Mark and {August}, Prune C. and {Holmberg}, M{\r{a}}ns and {Meier Vald{\'e}s}, Erik and {Zgraggen}, Merlin and {Buchhave}, Lars A. and {Burgasser}, Adam J. and {Fisher}, Chloe and {Gibson}, Neale P. and {Heng}, Kevin and {Hoeijmakers}, Jens and {Kitzmann}, Daniel and {Prinoth}, Bibiana and {Rathcke}, Alexander D. and {Morris}, Brett M.},
        title = "{Hot Rocks Survey. IV. Emission from LTT 3780 b Is Consistent with a Bare Rock}",
      journal = {\aj},
         year = 2025,
        month = oct,
       volume = {170},
        pages = {240},
          doi = {10.3847/1538-3881/adfc51},
archivePrefix = {arXiv},
       eprint = {2508.14210},
 primaryClass = {astro-ph.EP}
}

@ARTICLE{august26_atmospheric,
       author = {{August}, Prune C. and {Wordsworth}, Robin and {Huffman}, Mikayla and {Brain}, David and {Buchhave}, Lars A.},
        title = "{Atmospheric Collapse and Reinflation through Impacts for Terrestrial Planets around M Dwarfs}",
      journal = {\apjl},
         year = 2026,
        month = jan,
       volume = {996},
        pages = {L26},
          doi = {10.3847/2041-8213/ae2a27},
archivePrefix = {arXiv},
       eprint = {2510.25896},
 primaryClass = {astro-ph.EP}
}

@ARTICLE{batalha26_jwst,
       author = {{Batalha}, Natasha E. and {Wallack}, Nicole L. and {Gordon}, Tyler A. and {Wogan}, Nicholas F. and {Bennett}, Katherine A. and {Adams Redai}, Jea and {L{\'o}pez-Morales}, Mercedes and {Teske}, Johanna and {Valenti}, Jeff and {Alam}, Munazza K. and {Alderson}, Lili and {Aguichine}, Artyom and {Batalha}, Natalie M. and {Gagnebin}, Anna and {Gao}, Peter and {Meech}, Annabella and {Moran}, Sarah E. and {Wakeford}, Hannah R. and {Wolfgang}, Angie},
        title = "{JWST COMPASS: The 3--5 $\mu$ m Transmission Spectrum of LTT 1445 A b}",
      journal = {\aj},
         year = 2026,
        month = jul,
       volume = {172},
        pages = {52},
          doi = {10.3847/1538-3881/ae6c26},
archivePrefix = {arXiv},
       eprint = {2605.15003},
 primaryClass = {astro-ph.EP}
}

@ARTICLE{bell22_eureka,
       author = {{Bell}, Taylor J. and {Ahrer}, Eva-Maria and {Brande}, Jonathan and {Carter}, Aarynn L. and {Feinstein}, Adina D. and {Guzman Caloca}, Giannina and {Mansfield}, Megan and {Zieba}, Sebastian and {Piaulet}, Caroline and {Benneke}, Bj{\"o}rn and {Filippazzo}, Joseph and {May}, Erin M. and {Roy}, Pierre-Alexis and {Kreidberg}, Laura and {Stevenson}, Kevin B.},
        title = "{Eureka!: An End-to-End Pipeline for JWST Time-Series Observations}",
      journal = {The Journal of Open Source Software},
         year = 2022,
        month = nov,
       volume = {7},
       number = {79},
        pages = {4503},
          doi = {10.21105/joss.04503},
archivePrefix = {arXiv},
       eprint = {2207.03585},
 primaryClass = {astro-ph.IM}
}

@ARTICLE{brown22_x,
       author = {{Brown}, Alexander and {Froning}, Cynthia S. and {Youngblood}, Allison and {France}, Kevin and {Wilson}, David J. and {Duvvuri}, Girish and {Miguel}, Yamila and {Diamond-Lowe}, Hannah},
        title = "{X-Ray Emission from the Exoplanet Hosting LTT 1445 Triple Star System}",
      journal = {\aj},
         year = 2022,
        month = nov,
       volume = {164},
        pages = {206},
          doi = {10.3847/1538-3881/ac8f25},
archivePrefix = {arXiv},
       eprint = {2209.11314},
 primaryClass = {astro-ph.SR}
}

@ARTICLE{connors26_gj,
       author = {{Connors}, Nicholas J. and {Monaghan}, Christopher and {Benneke}, Bj{\"o}rn and {Dang}, Lisa and {Roy}, Pierre-Alexis},
        title = "{GJ 3929 b as the First Complete Rocky Worlds DDT Data Set}",
      journal = {\apjl},
         year = 2026,
        month = jul,
       volume = {1006},
        pages = {L6},
          doi = {10.3847/2041-8213/ae8018},
archivePrefix = {arXiv},
       eprint = {2606.07511},
 primaryClass = {astro-ph.EP}
}

@ARTICLE{diamondlowe23_ground,
       author = {{Diamond-Lowe}, Hannah and {Mendon{\c{c}}a}, Jo{\~a}o M. and {Charbonneau}, David and {Buchhave}, Lars A.},
        title = "{Ground-based Optical Transmission Spectroscopy of the Nearby Terrestrial Exoplanet LTT 1445Ab}",
      journal = {\aj},
         year = 2023,
        month = apr,
       volume = {165},
        pages = {169},
          doi = {10.3847/1538-3881/acbf39},
archivePrefix = {arXiv},
       eprint = {2210.11809},
 primaryClass = {astro-ph.EP}
}

@ARTICLE{ducrot25_combined,
       author = {{Ducrot}, Elsa and {Lagage}, Pierre-Olivier and {Min}, Michiel and {Gillon}, Micha{\"e}l and {Bell}, Taylor J. and {Tremblin}, Pascal and {Greene}, Thomas and {Dyrek}, Achr{\`e}ne and {Bouwman}, Jeroen and {Waters}, Rens and {G{\"u}del}, Manuel and {Henning}, Thomas and {Vandenbussche}, Bart and {Absil}, Olivier and {Barrado}, David and {Boccaletti}, Anthony and {Coulais}, Alain and {Decin}, Leen and {Edwards}, Billy and {Gastaud}, Ren{\'e} and {Glasse}, Alistair and {Kendrew}, Sarah and {Olofsson}, Goran and {Patapis}, Polychronis and {Pye}, John and {Rouan}, Daniel and {Whiteford}, Niall and {Argyriou}, Ioannis and {Cossou}, Christophe and {Glauser}, Adrian M. and {Krause}, Oliver and {Lahuis}, Fred and {Royer}, Pierre and {Scheithauer}, Silvia and {Colina}, Luis and {van Dishoeck}, Ewine F. and {Ostlin}, G{\"o}ran and {Ray}, Tom P. and {Wright}, Gillian},
        title = "{Combined analysis of the 12.8 and 15 $\mu$m JWST/MIRI eclipse observations of TRAPPIST-1 b}",
      journal = {Nature Astronomy},
         year = 2025,
        month = mar,
       volume = {9},
       number = {3},
        pages = {358-369},
          doi = {10.1038/s41550-024-02428-z},
archivePrefix = {arXiv},
       eprint = {2412.11627},
 primaryClass = {astro-ph.EP}
}

@ARTICLE{holmberg26_hot,
       author = {{Holmberg}, M{\r{a}}ns and {Diamond-Lowe}, Hannah and {Mendon{\c{c}}a}, Jo{\~a}o M. and {Kitzmann}, Daniel and {Espinoza}, N{\'e}stor and {Allen}, Natalie H. and {August}, Prune C. and {Fortune}, Mark and {Gressier}, Am{\'e}lie and {Ih}, Jegug and {Vald{\'e}s}, Erik Meier and {Zgraggen}, Merlin and {Buchhave}, Lars A. and {Demory}, Brice-Olivier and {Fisher}, Chloe and {Gibson}, Neale P. and {Heng}, Kevin and {Prinoth}, Bibiana and {Burgasser}, Adam J.},
        title = "{Hot Rocks Survey. V. Secondary Eclipse Photometry of GJ 3473 b with JWST/MIRI}",
      journal = {\aj},
         year = 2026,
        month = apr,
       volume = {171},
        pages = {251},
          doi = {10.3847/1538-3881/ae4c45},
archivePrefix = {arXiv},
       eprint = {2604.02332},
 primaryClass = {astro-ph.EP}
}

@ARTICLE{ih23_constraining,
       author = {{Ih}, Jegug and {Kempton}, Eliza M.-R. and {Whittaker}, Emily A. and {Lessard}, Madeline},
        title = "{Constraining the Thickness of TRAPPIST-1 b's Atmosphere from Its JWST Secondary Eclipse Observation at 15 $\mu$m}",
      journal = {\apjl},
         year = 2023,
        month = jul,
       volume = {952},
        pages = {L4},
          doi = {10.3847/2041-8213/ace03b},
archivePrefix = {arXiv},
       eprint = {2305.10414},
 primaryClass = {astro-ph.EP}
}

@ARTICLE{ih26_do,
       author = {{Ih}, Jegug and {Kempton}, Eliza M.-R. and {Diamond-Lowe}, Hannah and {Krissansen-Totton}, Joshua and {Mansfield}, Megan Weiner and {Xue}, Qiao and {Wogan}, Nicholas and {Nixon}, Matthew C. and {Hord}, Benjamin J.},
        title = "{Do Rocky Planets around M Stars Have Atmospheres? A Statistical Approach to the Cosmic Shoreline}",
      journal = {\aj},
         year = 2026,
        month = jul,
       volume = {172},
        pages = {59},
          doi = {10.3847/1538-3881/ae71c1},
archivePrefix = {arXiv},
       eprint = {2508.08253},
 primaryClass = {astro-ph.EP}
}

@ARTICLE{koll19_identifying,
       author = {{Koll}, Daniel D. B. and {Malik}, Matej and {Mansfield}, Megan and {Kempton}, Eliza M.-R. and {Kite}, Edwin and {Abbot}, Dorian and {Bean}, Jacob L.},
        title = "{Identifying Candidate Atmospheres on Rocky M Dwarf Planets via Eclipse Photometry}",
      journal = {\apj},
         year = 2019,
        month = dec,
       volume = {886},
        pages = {140},
          doi = {10.3847/1538-4357/ab4c91},
archivePrefix = {arXiv},
       eprint = {1907.13138},
 primaryClass = {astro-ph.EP}
}

@ARTICLE{lavie23_planetary,
       author = {{Lavie}, B. and {Bouchy}, F. and {Lovis}, C. and {Zapatero Osorio}, M. and {Deline}, A. and {Barros}, S. and {Figueira}, P. and {Sozzetti}, A. and {Gonz{\'a}lez Hern{\'a}ndez}, J. I. and {Lillo-Box}, J. and {Rodrigues}, J. and {Mehner}, A. and {Damasso}, M. and {Adibekyan}, V. and {Alibert}, Y. and {Allende Prieto}, C. and {Cristiani}, S. and {D'Odorico}, V. and {Di Marcantonio}, P. and {Ehrenreich}, D. and {G{\'e}nova Santos}, R. and {Lo Curto}, G. and {Martins}, C. J. A. P. and {Micela}, G. and {Molaro}, P. and {Nunes}, N. and {Palle}, E. and {Pepe}, F. and {Poretti}, E. and {Rebolo}, R. and {Santos}, N. and {Sousa}, S. and {Su{\'a}rez Mascare{\~n}o}, A. and {Tabrenero}, H. and {Udry}, S.},
        title = "{Planetary system around LTT 1445A unveiled by ESPRESSO: Multiple planets in a triple M-dwarf system}",
      journal = {\aap},
         year = 2023,
        month = may,
       volume = {673},
        pages = {A69},
          doi = {10.1051/0004-6361/202143007}
}

@ARTICLE{parkcoy25_population,
       author = {{Park Coy}, Brandon and {Ih}, Jegug and {Kite}, Edwin S. and {Koll}, Daniel D. B. and {Tenthoff}, Moritz and {Bean}, Jacob L. and {Weiner Mansfield}, Megan and {Zhang}, Michael and {Xue}, Qiao and {Kempton}, Eliza M.-R. and {Wohlfarth}, Kay and {Hu}, Renyu and {Lyu}, Xintong and {W{\"o}hler}, Christian},
        title = "{Population-level Hypothesis Testing with Rocky Planet Emission Data: A Tentative Trend in the Brightness Temperatures of M-Earths}",
      journal = {\apj},
         year = 2025,
        month = jul,
       volume = {987},
        pages = {22},
          doi = {10.3847/1538-4357/add3f7},
archivePrefix = {arXiv},
       eprint = {2412.06573},
 primaryClass = {astro-ph.EP}
}

@ARTICLE{pass23_hst,
       author = {{Pass}, Emily K. and {Winters}, Jennifer G. and {Charbonneau}, David and {Balkanski}, Aurelia and {Lewis}, Nikole and {Lally}, Maura and {Bean}, Jacob L. and {Cloutier}, Ryan and {Eastman}, Jason D.},
        title = "{HST/WFC3 Light Curve Supports a Terrestrial Composition for the Closest Exoplanet to Transit an M Dwarf}",
      journal = {\aj},
         year = 2023,
        month = oct,
       volume = {166},
        pages = {171},
          doi = {10.3847/1538-3881/acf561},
archivePrefix = {arXiv},
       eprint = {2307.02970},
 primaryClass = {astro-ph.EP}
}

@ARTICLE{redfield24_report,
       author = {{Redfield}, Seth and {Batalha}, Natasha and {Benneke}, Bj{\"o}rn and {Biller}, Beth and {Espinoza}, Nestor and {France}, Kevin and {Konopacky}, Quinn and {Kreidberg}, Laura and {Rauscher}, Emily and {Sing}, David},
        title = "{Report of the Working Group on Strategic Exoplanet Initiatives with HST and JWST}",
      journal = {arXiv e-prints},
         year = 2024,
        month = apr,
        pages = {arXiv:2404.02932},
archivePrefix = {arXiv},
       eprint = {2404.02932},
 primaryClass = {astro-ph.IM}
}

@ARTICLE{wachiraphan25_the,
       author = {{Wachiraphan}, Patcharapol and {Berta-Thompson}, Zachory K. and {Diamond-Lowe}, Hannah and {Winters}, Jennifer G. and {Murray}, Catriona and {Zhang}, Michael and {Xue}, Qiao and {Morley}, Caroline V. and {Rosario-Franco}, Marialis and {Duvvuri}, Girish M.},
        title = "{The Thermal Emission Spectrum of the Nearby Rocky Exoplanet LTT 1445A b from JWST MIRI/LRS}",
      journal = {\aj},
         year = 2025,
        month = jun,
       volume = {169},
        pages = {311},
          doi = {10.3847/1538-3881/adc990},
archivePrefix = {arXiv},
       eprint = {2410.10987},
 primaryClass = {astro-ph.EP}
}

@ARTICLE{winters19_three,
       author = {{Winters}, Jennifer G. and {Medina}, Amber A. and {Irwin}, Jonathan M. and {Charbonneau}, David and {Astudillo-Defru}, Nicola and {Horch}, Elliott P. and {Eastman}, Jason D. and {Vrijmoet}, Eliot Halley and {Henry}, Todd J. and {Diamond-Lowe}, Hannah and {Winston}, Elaine and {Barclay}, Thomas and {Bonfils}, Xavier and {Ricker}, George R. and {Vanderspek}, Roland and {Latham}, David W. and {Seager}, Sara and {Winn}, Joshua N. and {Jenkins}, Jon M. and {Udry}, St{\'e}phane and {Twicken}, Joseph D. and {Teske}, Johanna K. and {Tenenbaum}, Peter and {Pepe}, Francesco and {Murgas}, Felipe and {Muirhead}, Philip S. and {Mink}, Jessica and {Lovis}, Christophe and {Levine}, Alan M. and {L{\'e}pine}, S{\'e}bastien and {Jao}, Wei-Chun and {Henze}, Christopher E. and {Fur{\'e}sz}, G{\'a}bor and {Forveille}, Thierry and {Figueira}, Pedro and {Esquerdo}, Gilbert A. and {Dressing}, Courtney D. and {D{\'i}az}, Rodrigo F. and {Delfosse}, Xavier and {Burke}, Christopher J. and {Bouchy}, Fran{\c{c}}ois and {Berlind}, Perry and {Almenara}, Jose-Manuel},
        title = "{Three Red Suns in the Sky: A Transiting, Terrestrial Planet in a Triple M-dwarf System at 6.9 pc}",
      journal = {\aj},
         year = 2019,
        month = oct,
       volume = {158},
        pages = {152},
          doi = {10.3847/1538-3881/ab364d},
archivePrefix = {arXiv},
       eprint = {1906.10147},
 primaryClass = {astro-ph.EP}
}

@ARTICLE{winters22_a,
       author = {{Winters}, Jennifer G. and {Cloutier}, Ryan and {Medina}, Amber A. and {Irwin}, Jonathan M. and {Charbonneau}, David and {Astudillo-Defru}, Nicola and {Bonfils}, Xavier and {Howard}, Andrew W. and {Isaacson}, Howard and {Bean}, Jacob L. and {Seifahrt}, Andreas and {Teske}, Johanna K. and {Eastman}, Jason D. and {Twicken}, Joseph D. and {Collins}, Karen A. and {Jensen}, Eric L. N. and {Quinn}, Samuel N. and {Payne}, Matthew J. and {Kristiansen}, Martti H. and {Spencer}, Alton and {Vanderburg}, Andrew and {Zechmeister}, Mathias and {Weiss}, Lauren M. and {Wang}, Sharon Xuesong and {Wang}, Gavin and {Udry}, St{\'e}phane and {Terentev}, Ivan A. and {St{\"u}rmer}, Julian and {Stef{\'a}nsson}, Gudmundur and {Shporer}, Avi and {Shectman}, Stephen and {Sefako}, Ramotholo and {Schwengeler}, Hans Martin and {Schwarz}, Richard P. and {Scarsdale}, Nicholas and {Rubenzahl}, Ryan A. and {Roy}, Arpita and {Rosenthal}, Lee J. and {Robertson}, Paul and {Petigura}, Erik A. and {Pepe}, Francesco and {Omohundro}, Mark and {Murphy}, Joseph M. Akana and {Murgas}, Felipe and {Mo{\v{c}}nik}, Teo and {Montet}, Benjamin T. and {Mennickent}, Ronald and {Mayo}, Andrew W. and {Massey}, Bob and {Lubin}, Jack and {Lovis}, Christophe and {Lewin}, Pablo and {Kasper}, David and {Kane}, Stephen R. and {Jenkins}, Jon M. and {Huber}, Daniel and {Horne}, Keith and {Hill}, Michelle L. and {Gorrini}, Paula and {Giacalone}, Steven and {Fulton}, Benjamin and {Forveille}, Thierry and {Figueira}, Pedro and {Fetherolf}, Tara and {Dressing}, Courtney and {D{\'i}az}, Rodrigo F. and {Delfosse}, Xavier and {Dalba}, Paul A. and {Dai}, Fei and {Cort{\'e}s}, C. C. and {Crossfield}, Ian J. M. and {Crane}, Jeffrey D. and {Conti}, Dennis M. and {Collins}, Kevin I. and {Chontos}, Ashley and {Butler}, R. Paul and {Brown}, Peyton and {Brady}, Madison and {Behmard}, Aida and {Beard}, Corey and {Batalha}, Natalie M. and {Almenara}, Jose-Manuel},
        title = "{A Second Planet Transiting LTT 1445A and a Determination of the Masses of Both Worlds}",
      journal = {\aj},
         year = 2022,
        month = apr,
       volume = {163},
        pages = {168},
          doi = {10.3847/1538-3881/ac50a9}
}

@ARTICLE{wogan26_a,
       author = {{Wogan}, Nicholas F. and {Batalha}, Natasha E. and {Ih}, Jegug and {Lustig-Yaeger}, Jacob and {Stevenson}, Kevin B.},
        title = "{A Climate-constrained Bayesian Inverse Method for JWST Rocky Exoplanet Eclipse Spectra: Case Study of LTT 1445A b}",
      journal = {\apjl},
         year = 2026,
        month = aug,
       volume = {1007},
        pages = {L46},
          doi = {10.3847/2041-8213/ae9305},
archivePrefix = {arXiv},
       eprint = {2605.14997},
 primaryClass = {astro-ph.EP}
}

@ARTICLE{xue25_the,
       author = {{Xue}, Qiao and {Zhang}, Michael and {Park Coy}, Brandon and {Brady}, Madison and {Ji}, Xuan and {Bean}, Jacob L. and {Radica}, Michael and {Seifahrt}, Andreas and {St{\"u}rmer}, Julian and {Luque}, Rafael and {Basant}, Ritvik and {Brown}, Nina and {Das}, Tanya and {Kasper}, David and {Piaulet-Ghorayeb}, Caroline and {Kempton}, Eliza M.-R. and {Kite}, Edwin},
        title = "{The JWST Rocky Worlds DDT Program Reveals GJ 3929b to Likely Be a Bare Rock}",
      journal = {\apjl},
         year = 2025,
        month = dec,
       volume = {995},
        pages = {L52},
          doi = {10.3847/2041-8213/ae2098},
archivePrefix = {arXiv},
       eprint = {2508.12516},
 primaryClass = {astro-ph.EP}
}

@ARTICLE{zgraggen26_hot,
       author = {{Zgraggen}, Merlin and {Demory}, Brice-Olivier and {Diamond-Lowe}, Hannah and {Mendon{\c{c}}a}, Jo{\~a}o M. and {Vald{\'e}s}, Erik Meier and {Fortune}, Mark and {Jones}, Kathryn D. and {Kitzmann}, Daniel and {Smith}, Millie and {Allen}, Natalie H. and {August}, Prune C. and {Gressier}, Am{\'e}lie and {Holmberg}, M{\r{a}}ns and {Ih}, Jegug and {Buchhave}, Lars A. and {Espinoza}, N{\'e}stor and {Gibson}, Neale P. and {Heng}, Kevin},
        title = "{Hot Rocks Survey VI: An anomalously hot dayside for the rocky planet GJ 357 b}",
      journal = {arXiv e-prints},
         year = 2026,
        month = sep,
        pages = {arXiv:2609.30552},
archivePrefix = {arXiv},
       eprint = {2609.30552},
 primaryClass = {astro-ph.EP}
}

@ARTICLE{juliet,
       author = {{Espinoza}, N{\'e}stor and {Kossakowski}, Diana and {Brahm}, Rafael},
        title = "{juliet: a versatile modelling tool for transiting and non-transiting exoplanetary systems}",
      journal = {\mnras},
         year = 2019,
        month = dec,
       volume = {490},
       number = {2},
        pages = {2262-2283},
          doi = {10.1093/mnras/stz2688},
archivePrefix = {arXiv},
       eprint = {1812.08549},
 primaryClass = {astro-ph.EP},
       adsurl = {https://ui.adsabs.harvard.edu/abs/2019MNRAS.490.2262E}
}

@article{celerite,
	doi = {10.3847/1538-3881/aa9332},
	url = {https://doi.org/10.3847%2F1538-3881%2Faa9332},
	year = 2017,
	month = {nov},
	publisher = {American Astronomical Society},
	volume = {154},
	number = {6},
	pages = {220},
	author = {Daniel Foreman-Mackey and Eric Agol and Sivaram Ambikasaran and Ruth Angus},
	title = {Fast and Scalable Gaussian Process Modeling with Applications to Astronomical Time Series},
	journal = {The Astronomical Journal}
}

@ARTICLE{Coulombe_2024,
       author = {{Coulombe}, Louis-Philippe and {Roy}, Pierre-Alexis and {Benneke}, Bj{\"o}rn},
        title = "{Biases in Exoplanet Transmission Spectra Introduced by Limb-darkening Parametrization}",
      journal = {\aj},
         year = 2024,
        month = nov,
       volume = {168},
       number = {5},
          eid = {227},
        pages = {227},
          doi = {10.3847/1538-3881/ad7aef},
archivePrefix = {arXiv},
       eprint = {2409.03812},
 primaryClass = {astro-ph.EP},
       adsurl = {https://ui.adsabs.harvard.edu/abs/2024AJ....168..227C}
}

@INPROCEEDINGS{Stefansson2018_diffuserSPIE,
       author = {{Stefansson}, Gudmundur and {Mahadevan}, Suvrath and {Wisniewski}, John and {Li}, Yiting and {Hebb}, Leslie and {Morris}, Brett and {Halverson}, Samuel and {Monson}, Andrew and {Robertson}, Paul},
        title = "{Extreme precision photometry from the ground with beam-shaping diffusers for K2, TESS, and beyond}",
    booktitle = {Ground-based and Airborne Instrumentation for Astronomy VII},
         year = 2018,
       editor = {{Evans}, Christopher J. and {Simard}, Luc and {Takami}, Hideki},
       series = {Society of Photo-Optical Instrumentation Engineers (SPIE) Conference Series},
       volume = {10702},
        month = aug,
          eid = {1070250},
        pages = {1070250},
          doi = {10.1117/12.2312833},
archivePrefix = {arXiv},
       eprint = {1808.02187},
 primaryClass = {astro-ph.IM},
       adsurl = {https://ui.adsabs.harvard.edu/abs/2018SPIE10702E..50S}
}

@ARTICLE{Osborn2015_scintillation,
       author = {{Osborn}, J. and {F{\"o}hring}, D. and {Dhillon}, V.~S. and {Wilson}, R.~W.},
        title = "{Atmospheric scintillation in astronomical photometry}",
      journal = {\mnras},
         year = 2015,
        month = sep,
       volume = {452},
       number = {2},
        pages = {1707-1716},
          doi = {10.1093/mnras/stv1400},
archivePrefix = {arXiv},
       eprint = {1506.06921},
 primaryClass = {astro-ph.IM},
       adsurl = {https://ui.adsabs.harvard.edu/abs/2015MNRAS.452.1707O}
}

@ARTICLE{Young1967_scintillation,
       author = {{Young}, Andrew T.},
        title = "{Photometric error analysis. VI. Confirmation of Reiger's theory of scintillation}",
      journal = {\aj},
         year = 1967,
        month = aug,
       volume = {72},
        pages = {747},
          doi = {10.1086/110303},
       adsurl = {https://ui.adsabs.harvard.edu/abs/1967AJ.....72..747Y}
}
\bibliographystyle{aasjournalv7}



\end{document}